\documentclass[10pt,twocolumn,superscriptaddress,floats,showpacs,nobalancelastpage,longbibliography,prl]{revtex4-1}
\pdfoutput=1

\usepackage{array,longtable}
\newcolumntype{L}{>{\tiny $}p{0.33\columnwidth}<{$}}
\newcolumntype{M}{>{\scriptsize $}p{0.33\columnwidth}<{$}}
\newcolumntype{N}{>{\scriptsize $}p{0.43\columnwidth}<{$}}
\usepackage{amsmath}
\usepackage{amssymb}
\usepackage{amsfonts}
\usepackage{graphicx}
\usepackage{hyperref}
\usepackage{tabularx}

\usepackage{float}
\usepackage{graphics}
\usepackage[caption=false]{subfig}
\usepackage{times}

\allowdisplaybreaks[1]

\newif\ifhyper
\hypertrue
\ifhyper
\hypersetup{
   citecolor = {green},
   colorlinks = {true}, 
   urlcolor = {blue} 
}
\fi

\begin{document}

\title{SU$(3)_1$ Chiral Spin Liquid on the Square Lattice: a View from Symmetric PEPS}

\author{Ji-Yao Chen}
\affiliation{Max-Planck-Institut f\"ur Quantenoptik, Hans-Kopfermann-Stra{\ss}e 1, 85748 Garching, Germany}
\affiliation{Munich Center for Quantum Science and Technology, Schellingstra{\ss}e 4, 80799 M{\"u}nchen, Germany}

\author{Sylvain Capponi}
\affiliation{Laboratoire de Physique Th\'eorique, IRSAMC, Universit\'e de Toulouse, CNRS, UPS, 31062 Toulouse, France}

\author{Alexander Wietek}
\affiliation{Center for Computational Quantum Physics, Flatiron Institute, 162 5th Avenue, NY 10010, New York, USA}

\author{Matthieu Mambrini}
\affiliation{Laboratoire de Physique Th\'eorique, IRSAMC, Universit\'e de Toulouse, CNRS, UPS, 31062 Toulouse, France}

\author{Norbert Schuch}
\affiliation{Max-Planck-Institut f\"ur Quantenoptik, Hans-Kopfermann-Stra{\ss}e 1, 85748 Garching, Germany}
\affiliation{Munich Center for Quantum Science and Technology, Schellingstra{\ss}e 4, 80799 M{\"u}nchen, Germany}

\author{Didier Poilblanc}
\affiliation{Laboratoire de Physique Th\'eorique, IRSAMC, Universit\'e de Toulouse, CNRS, UPS, 31062 Toulouse, France}

\date{\today}

\begin{abstract}
Quantum spin liquids can be faithfully represented and efficiently
characterized within the framework of Projected Entangled Pair States
(PEPS). Guided by extensive exact diagonalization and density matrix
renormalization group calculations, we construct an optimized symmetric
PEPS for a SU$(3)_1$ chiral spin liquid on the square lattice. Characteristic features are revealed by the entanglement spectrum (ES) on an infinitely long cylinder. In all three $\mathbb{Z}_3$ sectors, the level counting of the linear dispersing modes is in full agreement with SU$(3)_1$ Wess-Zumino-Witten conformal field theory prediction. Special features in the ES are shown to be in correspondence with bulk anyonic correlations, indicating a fine structure in the holographic bulk-edge correspondence. Possible universal properties of topological SU$(N)_k$ chiral PEPS are discussed.

\end{abstract}
\maketitle

{\it Introduction} -- 
Quantum spin liquids are long-range entangled states of matter of two dimensional electronic spin systems~\cite{Misguich2005,Savary2016,Zhou2017}. Among the various classes~\cite{Wen2002}, spin liquids with broken time-reversal symmetry, i.e., chiral spin liquids (CSL)~\cite{Kalmeyer1987,WWZ1989}, exhibit chiral topological order~\cite{Wen1990}. Intimately related to Fractional Quantum Hall (FQH) states~\cite{Tsui1982}, CSL host both anyonic quasi-particles in the bulk~\cite{Halperin1984} and chiral gapless modes on the edge~\cite{Wen1991a}. It was early suggested that, in systems with enhanced SU($N$) symmetry, realizable with alkaline earth atoms loaded in optical lattices~\cite{Gorshkov2010}, CSL can naturally appear~\cite{Hermele2009}. Later on, many SU$(N)_1$ CSL with different $N$ were identified on the triangular lattice~\cite{Nataf2016}, while the original proposal on the square lattice~\cite{Hermele2009} remains controversial.

In recent years, Projected Entangled Pair States (PEPS)~\cite{Verstraete2004b} have progressively emerged as a powerful tool to study quantum spin liquids. As an ansatz, PEPS provide variational ground states competitive with other methods~\cite{Liao2017,Lee2019}, and equally importantly, offer a powerful framework to encode topological order~\cite{Schuch2010a} and construct non-chiral~\cite{Schuch2012,Chen2018a} and chiral -- both Abelian~\cite{Poilblanc2015} and non-Abelian~\cite{Chen2018b} -- SU(2) spin liquids. Generically, SU(2) CSL described by PEPS exhibit linearly dispersing chiral branches in the entanglement spectrum (ES) well described by Wess-Zumino-Witten (WZW) SU$(2)_k$ (with $k=1$ for Abelian CSL) conformal field theory (CFT) for one-dimensional edges~\cite{Franscesco1997}. However, to our knowledge, there is no known example of more general SU($N$) PEPS with unambiguous chiral edge modes. Thus it remains unclear whether symmetric PEPS can describe higher SU($N$) CSL \emph{faithfully}. In order to address these issues, we propose and investigate a frustrated SU(3) symmetric spin model on the square lattice with a symmetric PEPS ansatz, thereby taking the first step towards describing general SU$(N)_k$ CSL with PEPS.

{\it Model and exact diagonalization} --
On every site, we place a three-dimensional spin degree of freedom, which transforms as the fundamental representation of SU($3$). The Hamiltonian, defined on a square lattice, includes the most general SU(3)-symmetric short-range three-site interaction:
\begin{equation}
\label{eq:model}
\begin{split}
H &= J_1\sum_{\langle i,j\rangle}P_{ij} + J_2\sum_{\langle\langle k,l\rangle\rangle}P_{kl} \\
 & + J_{\rm R}\sum_{\triangle ijk}(P_{ijk} + P_{ijk}^{-1}) + iJ_{\rm I}\sum_{\triangle ijk}(P_{ijk} - P_{ijk}^{-1}),
\end{split}
\end{equation}
where the first (second) term corresponds to two-site permutations over all (next-)nearest-neighbor bonds, and the third and fourth terms are three-site (clockwise) permutations on all triangles of every plaquette. We have chosen $J_2=J_1/2$ so that the two-body part ($J_1$ and $J_2$) on the interacting triangular units becomes $S_{3}$ symmetric, hence mimicking the corresponding Hamiltonian on the triangular lattice~\footnote{The chiral spin liquid phase should also exist away from $J_2=J_1/2$, due to its gapped nature.} and further parameterized the amplitude of each term as: $J_1=2J_2=\frac{4}{3}\cos \theta \sin \phi, J_{\rm R}=\cos \theta\cos\phi, J_{\rm I}=\sin\theta$.

We have performed extensive exact diagonalization (ED) calculations~\cite{Laflorencie2004, Sandvik2010b} on various periodic $N_s$-site clusters to locate the CSL phase in parameter space. We expect (\ref{eq:model}) to host a SU$(3)_1$ CSL equivalent to the 221 Halperin FQH state~\cite{Halperin1983,Ardonne1999}, whose spectral signatures on small tori can be established precisely~\cite{WuTu2016,Sterdyniak2013}. A careful scan in $\theta$ and $\phi$ reveals that there is a small region where, for $N_s=3p$ $(p\in\mathbb{N}^{+})$, there are three low-lying singlets below the octet gap, reflecting the expected topological degeneracy of the CSL. For $N_s\ne 3p$, the low-energy quasi-degenerate manifold reflects perfectly the anyon content of the CSL. In both cases, the momenta of the low-energy states match the heuristic counting rules of the 221 Halperin FQH state with 0, 1 or 2 quasi-holes~\cite{Regnault2011,Bernevig2012,Estienne2012,Sterdyniak2013}. Finally, a $N_s=4\times 4$ cluster with open boundaries reveals the expected CFT edge spectrum (see Fig.~\ref{fig:ed} and supplemental material (SM)~\footnote{In the supplemental material, we provide several relevant details of this work, including basic knowledge of SU($3$) group~\cite{BarutRaczka1980}, ED study on various small clusters, DMRG study on finite cylinders, construction of SU($3$) symmetric PEPS, CTMRG and optimization procedure, additional data for ES, and topological excitation and correlation function of symmetric PEPS. In the end, we also list the nonzero elements of the tensors.}). In the following, we shall focus on angles $\theta=\phi=\frac{\pi}{4}$, for which clear evidence of a gapped CSL is found.~\footnote{Since our model possesses one SU(3) fermion per unit cell (equivalent to 1/3 filling), a trivial featureless gapped ground-state is impossible according to the Lieb-Schultz-Mattis theorem for SU(N) spin systems and its generalization to two dimensions~\cite{Lieb1961,Oshikawa2000,Hastings2005,Totsuka2017} and, hence, a uniform gapped phase has to be topological.}

\begin{figure}
	\centering
	\includegraphics[width=\columnwidth]{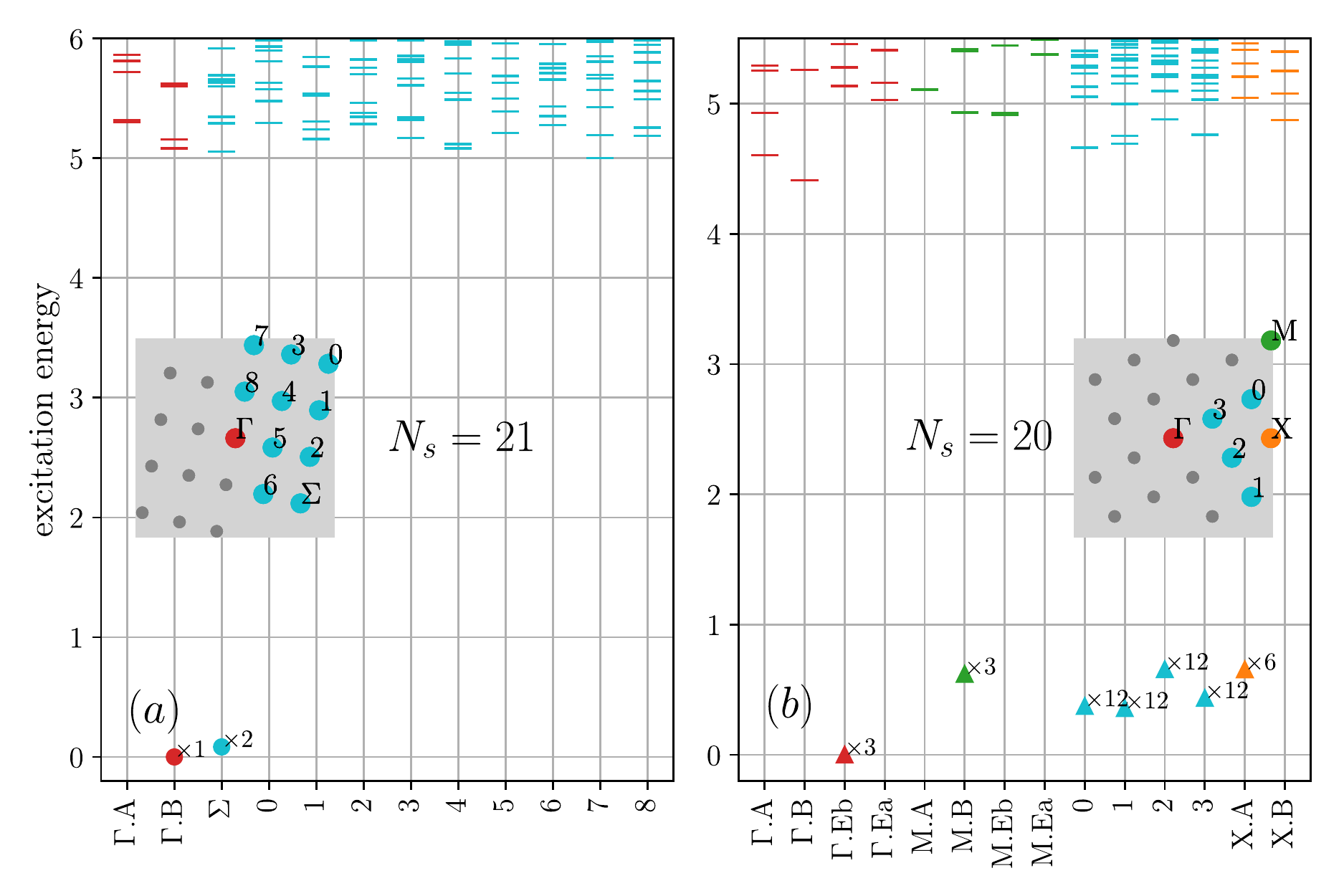}
	\caption{ED spectra obtained on periodic 21-site (a) and 20-site (b) tori. In the insets, available momenta in reciprocal space are shown. Dot, triangle and segment symbols correspond to 
	singlets, $\bf\overline{3}$ and other finite-dimensional SU(3) irreducible representations (irreps), respectively. }
	\label{fig:ed}
\end{figure}

\begin{figure}[tb]
\centering
	\includegraphics[width=0.95\columnwidth]{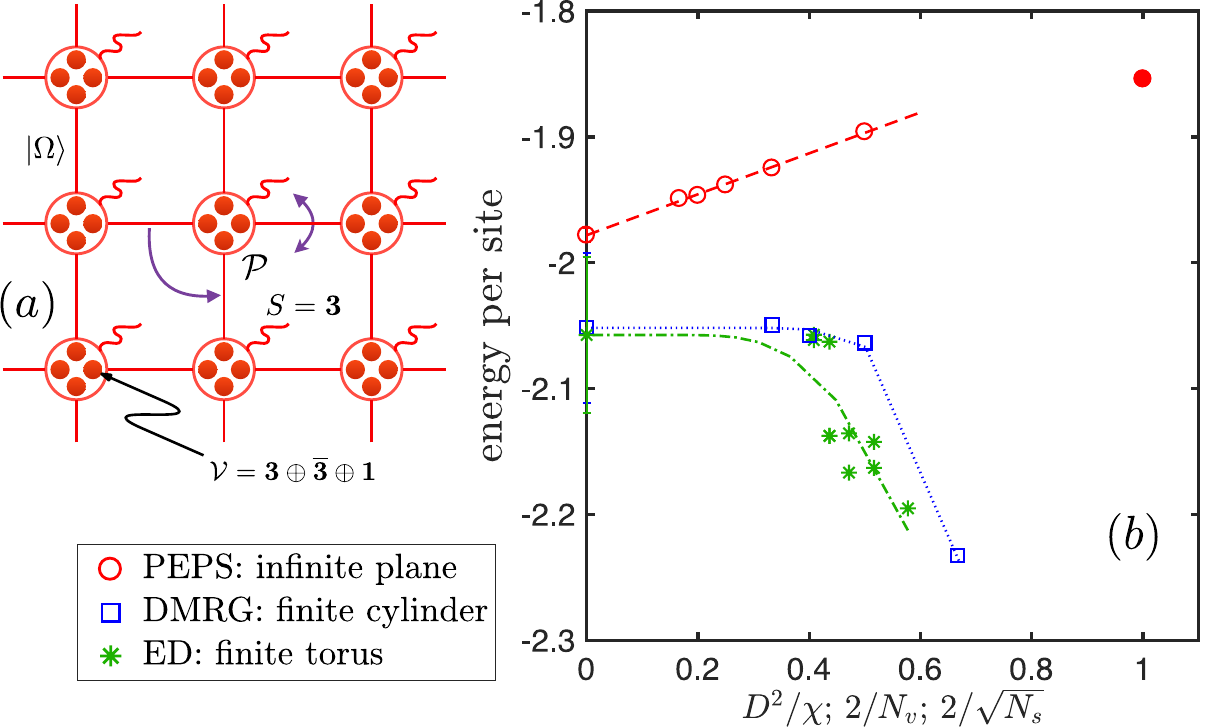}
\caption{(a) The symmetric PEPS construction. See the text for details. (b) Comparison of energy densities obtained by symmetric PEPS, DMRG and ED, plotted as a function of $D^2/\chi$, $2/N_v$ and $2/\sqrt{N_s}$, respectively. The PEPS energy is optimized at $\chi=D^2$ and extrapolated to $\chi\to\infty$ (red circles). Blue squares stand for DMRG data on several finite-width cylinders ($N_v=3,4,5,6$), and  the ED results on tori with $N_s=12,15,18,21,24$ sites and different geometries are indicated by stars. The dotted (dash-dotted) line is an exponential fit of the DMRG (ED) data to the thermodynamic limit.}
\label{fig:peps}
\end{figure}

{\it Symmetric PEPS ansatz} -- Representing the CSL in terms of a symmetric PEPS allows us not only to obtain short-range properties such as energy density efficiently, but also to reveal its topological properties by examining the entanglement structure~\cite{Li2008,Cirac2011}. This can be accomplished by using SU($3$)-symmetric tensors, analogously to the SU($2$) case~\cite{Mambrini2016}. The simplest virtual space available here is ${\cal V}={\bf 3}\oplus{\bf\overline 3}\oplus{\bf 1}$ such that (i) a symmetric maximally entangled SU($3$) singlet $|\Omega\rangle$ can be realized on every bond by pairing two neighboring virtual particles and (ii) four virtual particles around each site can be fused into the 3-dimensional physical spin with an on-site projector $\mathcal{P}$~\cite{Kurecic2019,Dong2018}, see Fig.~\ref{fig:peps}(a). The so-called bond dimension is thus $D=7$. In addition to the continuous rotation symmetry, full account of the  discrete $C_{\rm 4v}$ point group symmetry (shown as purple arrows in Fig.~\ref{fig:peps}(a)) can be taken~\cite{Mambrini2016}, and tensors can be classified according to the corresponding irreps. By linearly combining on-site projectors of two different irreps with opposite $\pm 1$ characters w.r.t. axis reflections, one can construct a complex PEPS ansatz breaking both parity (P) and time-reversal (T) symmetries while preserving PT, as required for a CSL ground state. Details about this construction, used for SU($2$) CSL~\cite{Poilblanc2015,Chen2018b}, are provided in the SM. For the ease of the following discussion, it is convenient to define the tensor $\mathcal{A}$ by absorbing the adjacent singlets on, e.g., the right and down bonds around each site into the on-site projector, forming an equivalent way to express the wavefunction.

The fact that center of the SU($N$) group is isomorphic to the $\mathbb{Z}_N$ group allows
one to associate a $\mathbb{Z}_3$ charge $Q=+1$ to the physical space of
the tensor $\mathcal{A}$, while the virtual space carries $\mathbb{Z}_3$
charges $Q=\{ +1,-1, 0\}$, i.e.\ it contains a regular representation of
$\mathbb{Z}_3$~\footnote{We note that, the requirement is that one can
recover a regular representation of $\mathbb{Z}_N$ after sufficient number
of blocking. The local tensor before blocking does not need to contain the
regular representation.}. Hence, the tensor $\mathcal{A}$ bears an important
$\mathbb{Z}_3$ gauge symmetry associated to local charge conservation:
$(Z\otimes Z^\dag\otimes Z \otimes Z^\dag) \circ\mathcal{A} = \omega \mathcal{A}$,
where the action on virtual indices reads as left, right, up and down, and
$\omega = \rm{e}^{i2\pi/3}$, $Z =
\rm{diag}(\omega,\omega,\omega,\omega^2,\omega^2,\omega^2,1)$ is the
representation of the $\mathbb{Z}_3$ generator in $\cal V$. This built-in
gauge symmetry is central to topological properties, such as topological
degeneracy on the torus and anyonic excitations~\cite{Schuch2010a}. 
Let us emphasize that the $\mathbb{Z}_3$ gauge symmetry naturally
appears from the physical SU(3) and point group symmetry, and is not a
symmetry we imposed ad hoc.

{\it Variational optimization} -- The best variational ground state is
obtained by taking the ansatz $\mathcal{P} =
\sum_{a=1}^{N_1}\lambda_1^{a}\mathcal{B}_1^a +
i\sum_{b=1}^{N_2}\lambda_2^{b}\mathcal{B}_2^b$, where $\mathcal{B}_{1}^a$
($\mathcal{B}_{2}^b$) transforms as $B_{1}$ ($B_2$) irrep of $C_{\rm 4v}$,
$N_1$ ($N_2$) is the number of linearly independent projectors in the
$\mathcal{B}_1$ ($\mathcal{B}_2$) class, and optimizing the (few)
variational parameters $\{\lambda_{1}^a,\lambda_{2}^b\}\in \mathbb{R}$
with a conjugate-gradient method~\cite{NoceWrig06}. For a given tensor the
energy is obtained via the corner transfer matrix renormalization group
(CTMRG) method, computing an effective environment of bond dimension
$\chi$ surrounding an active $2\times 2$ region embedded in the infinite
plane (so-called iPEPS)~\cite{Nishino1996, Orus2009, Corboz2014, Fishman2018}. The gradient
is then simply obtained by a finite difference
approach~\cite{Poilblanc2017a}. U(1) quantum number is also used
occasionally to speed up the
computation~\cite{Bauer2011,Singh2011,Weichselbaum2012}. The exact
contraction scheme corresponds to the limit $\chi\rightarrow\infty$.

To establish the relevance of our symmetric PEPS ansatz for the model
\eqref{eq:model}, we compare the PEPS energy density with that obtained by
ED on several different tori up to size $N_s=24$, and by the density matrix
renormalization group (DMRG) method~\cite{White1992,itensor} on various finite
cylinders. In DMRG, for each cylinder width $N_v$, we compute the ground
state energies at two different cylinder lengths to subtract the
contribution from the edges~\cite{Stoudenmire2012}. A detailed description
of the DMRG method and additional data can be found in the SM. As shown in
Fig.~\ref{fig:peps}(b), the PEPS energy density obtained on the infinite
plane turns out to lie close to the energy density in the thermodynamic
limit, estimated from a finite-size scaling of the ED and DMRG data.

\begin{figure}[b]
\centering
	\subfloat{\includegraphics[width=0.496\columnwidth]{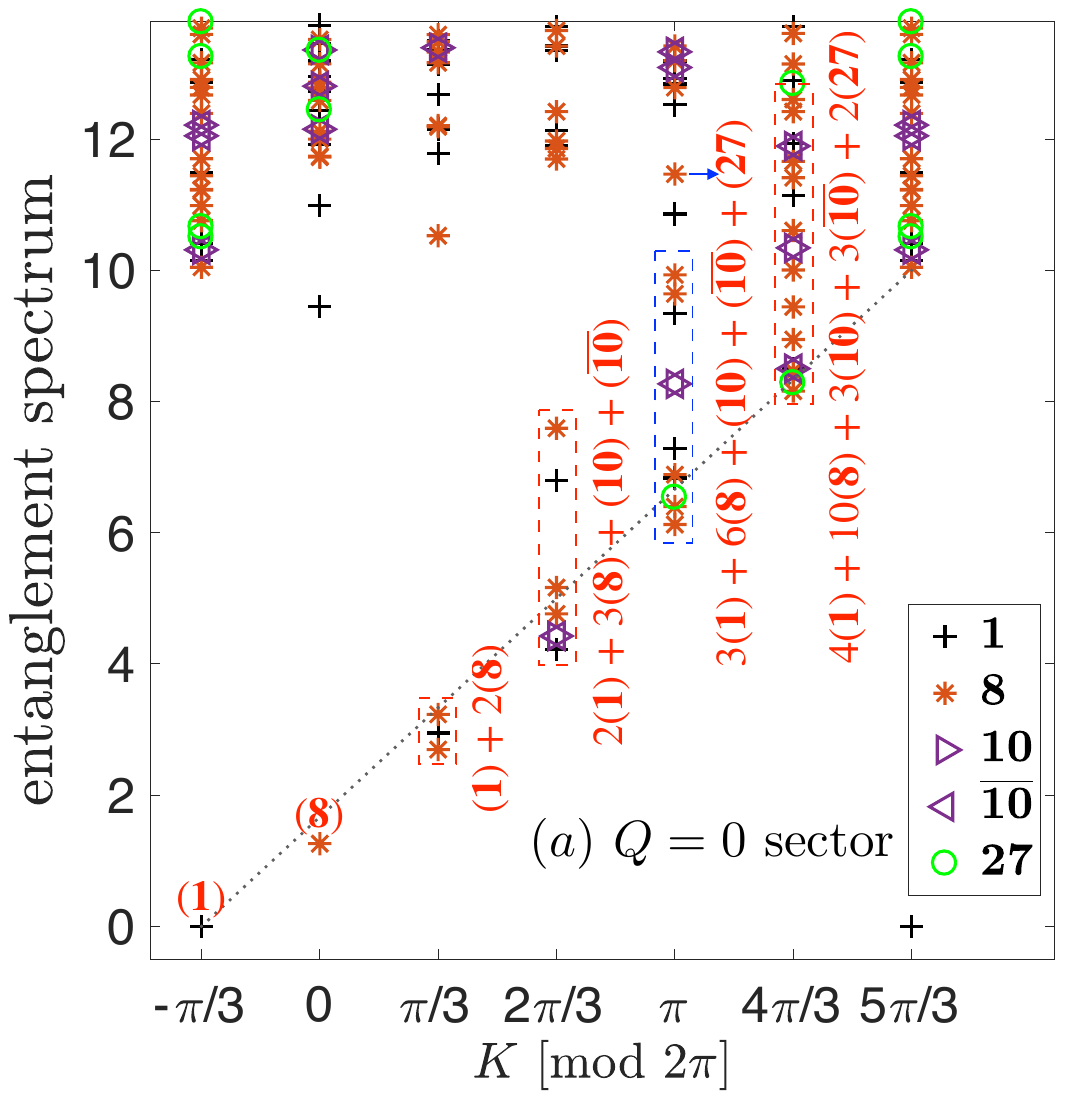}}
	\subfloat{\includegraphics[width=0.490\columnwidth]{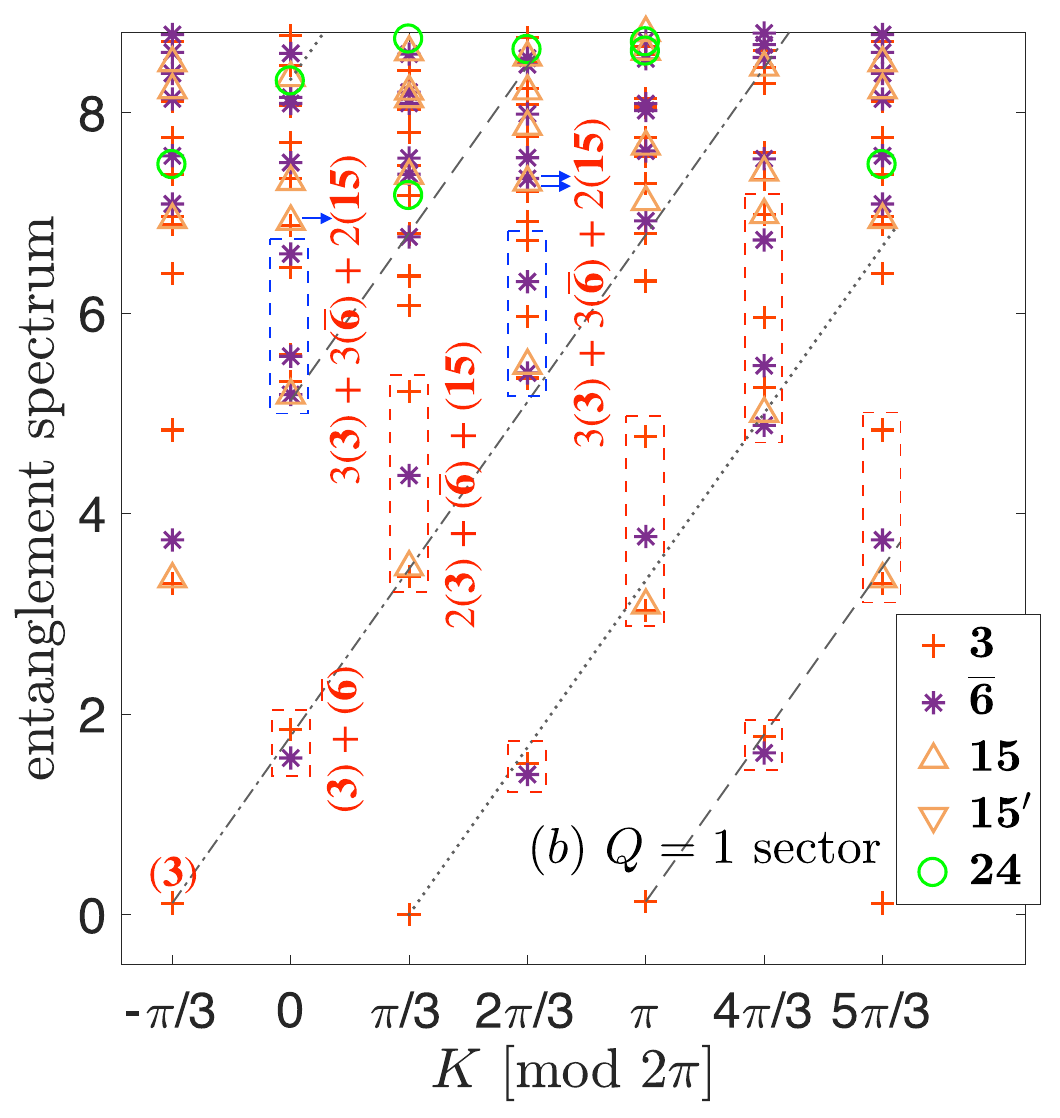}}
\caption{Entanglement spectrum on an infinitely long $N_v=6$ cylinder, computed with $\chi=343$, in the $Q=0$ (a) and $Q=+1$ (b) sectors. The spectrum in $Q=-1$ sector (not shown here) is found to be identical to that in $Q=+1$ sector but with conjugated SU($3$) irreps, and is shown in the SM.
For convenience, the lowest eigenvalue is subtracted in each plot. One chiral branch is seen in (a) starting at momentum $K_0=-\pi/3$ and three branches are seen in (b) starting at momenta $K_{\pm 1}=-\pi/3, \pi/3$ and $\pi$.
 In each sector, the irreps encircled by the red boxes (or the blue boxes and the arrows) agree with the level counting of the SU$(3)_1$ WZW CFT (shown on the plot vertically).
}
\label{fig:ES}
\end{figure}

{\it Entanglement spectrum} --
To get further insight into the nature of the CSL phase, we now explore the properties of our symmetric PEPS, where the $\mathbb{Z}_3$ gauge symmetry implies topological degeneracy on closed manifolds. On finite-width cylinders, quasi-degenerate ground states can be constructed by restricting the virtual boundary of PEPS to fixed $\mathbb{Z}_3$ charges $Q=0,\pm 1$, with or without inserting $\mathbb{Z}_3$ flux line through the cylinder. Here we shall focus on states without $\mathbb{Z}_3$ flux line, and briefly discuss those with flux line in the SM. The topological properties can be most easily obtained through a study of the entanglement spectrum, which is defined to be minus log of the spectrum of reduced density matrix (RDM) of subsystem, say the left half of a cylinder~\cite{Li2008}. For a PEPS on an infinite long cylinder, the RDM can be constructed from the leading eigenvector of the transfer operator~\cite{Cirac2011}. Since the onsite tensor carries charge $+1$, the cylinder width $N_v$ must be multiple of three. In our current setting with bond dimension $D=7$, exactly contracting the transfer operator is not feasible for large enough $N_v$. Instead we use the iPEPS environment tensors computed with CTMRG to construct the approximate leading eigenvector~\cite{Poilblanc2016}, where large enough environment dimension $\chi$ is needed to get converged results~\cite{Chen2018b}.  The constructed RDM is fully invariant under translation and SU($3$) rotations, which allows to block diagonalize it, introducing appropriate quantum numbers. In practice, we use the $\mathbb{Z}_3$ charges (associated to the $\mathbb{Z}_3$ gauge symmetry), two U(1) quantum numbers, and the momentum quantum number to do ED. The results with $N_v=6, \chi=343$ are shown in Fig.~\ref{fig:ES} for the three different charge sectors, i.e., $Q=0, \pm 1$.

Linearly dispersing chiral modes well separated from the high energy
continuum are seen with the same velocity, one mode in the $Q=0$ sector
and three modes in the $Q=+1$ sector. The $Q=\pm1$ sectors have
identical spectra: As both the bare tensor and the bond
$\vert\Omega\rangle$ are
PT-symmetric, so is the wavefunction, but after the reflection,  the
bonds are at the other side of the entanglement cut, and since the bonds exchange 
$\mathbf{3}\leftrightarrow\mathbf{\overline 3}$, this maps between the $Q=\pm1$ 
spectra.
Interestingly, for all different $\chi$ we have considered, the lowest level in the $Q=0$ sector appears at finite momentum $K_0=-\pi/3$, while the three branches in the $Q=\pm 1$ sectors start at momenta $K_{\pm 1}=-\pi/3, \pi/3$, and $\pi$. We believe the momentum shift is due to a quantum of magnetic flux trapped in the cylinder, and is an intrinsic property of the optimized PEPS, that constrains us to choose $N_v=6p$, $p$ integer
(For $N_v=3(2p+1)$, where $K_0$ and $K_{\pm 1}$ do not belong to the reciprocal space, see  SM).

Reconstructing the SU(3) irreps from the two U(1) quantum numbers (the Young tableaux for the relevant SU(3) irreps are provided in the SM), we found that the level contents follow the prediction of the Virasoro levels of the SU$(3)_1$ WZW CFT~\cite{Bouwknegt1996,Franscesco1997}. However, we observe a tripling of the branches in the $Q=\pm 1$ sectors that we shall discuss later.

{\it Bulk correlations } --
The above entanglement spectrum provides strong evidence of SU$(3)_1$
chiral topological order. However, it has been shown that in PEPS
describing chiral phases, certain bulk correlation lengths
computed from the transfer matrix spectrum
diverge~\cite{Wahl2013a,Dubail2015,Yang2015,Poilblanc2015,Poilblanc2016,Read2017,Poilblanc2017b,Hackenbroich2018,Chen2018b}.
Nevertheless, {\it a priori} it is not known which type of correlation is
quasi long-ranged, and how critical bulk correlations are related to the
observed chiral edge modes. Here we address this question with our
symmetric PEPS ansatz, where both the SU(3) symmetry and the associated
$\mathbb{Z}_3$ gauge symmetry will provide valuable insights.

\begin{figure}[!hbt]
\centering
\includegraphics[width=0.95\columnwidth]{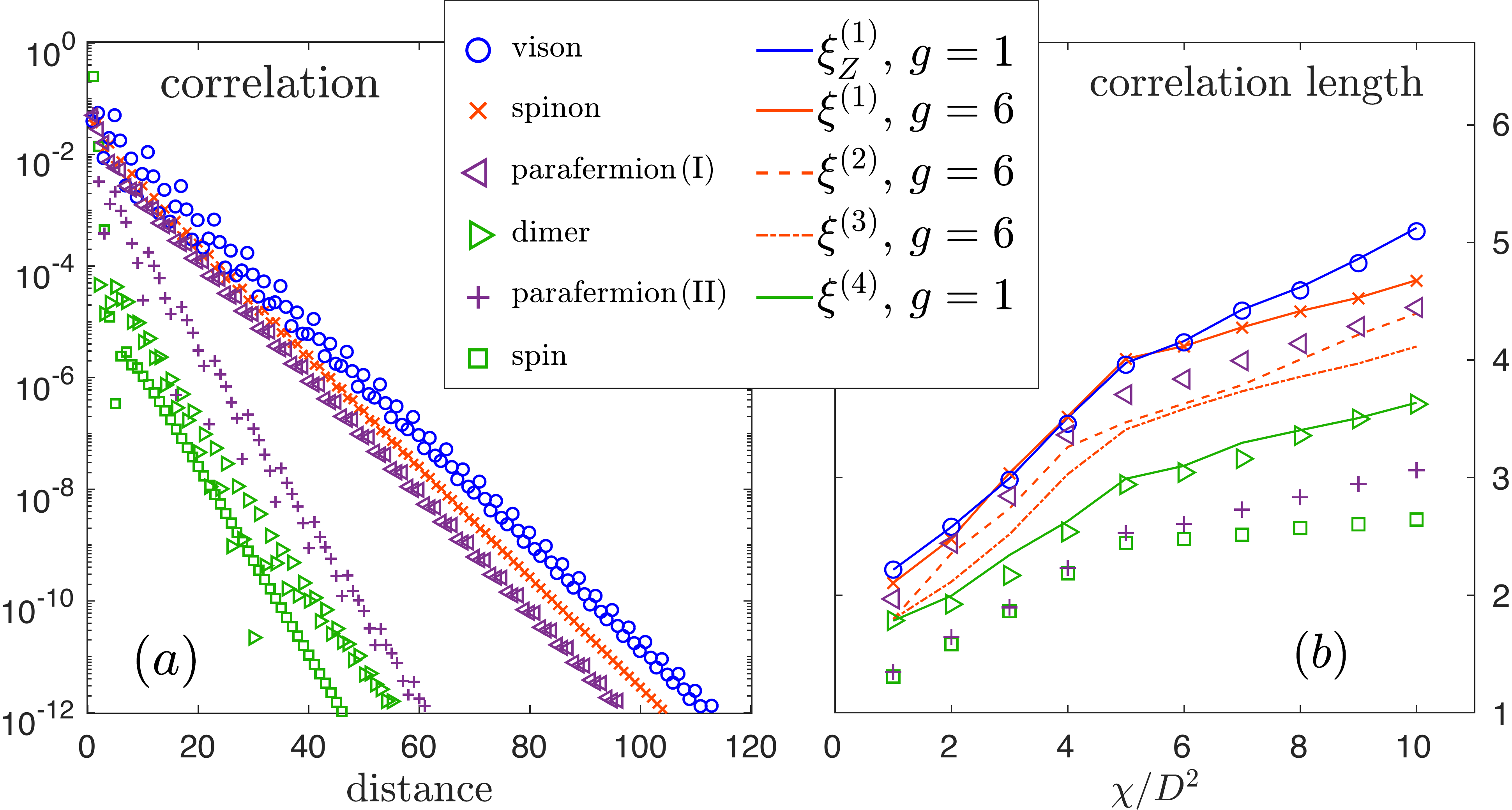}
\caption{Different bulk correlations in the optimized PEPS. From
the correlations versus distance (computed with $\chi=392$)
in (a), we extract the correlation lengths using
exponential fits, which are shown in (b) (using the same symbols), along with those
extracted from the transfer matrix spectrum with or without flux inserted
(shown as lines), with $g$ the degeneracy of the eigenvalue. Both
approaches agree for the spinon, vison and dimer correlation lengths which
show no saturation with increasing~$\chi$.} \label{fig:correlation}
\end{figure}

Within the PEPS methodology, correlation lengths of different types of
operators, including the anyonic type, can be obtained from two
complementary methods. On one hand, correlation functions of usual local
operators, e.g., spin-spin correlation $C_{\rm s}(d) =
\langle\mathbf{S}_i\cdot \mathbf{S}_{{i}+d\mathbf{e}_x}\rangle$ or dimer-dimer correlation
$C_{\rm d}(d) = \langle D^x_{i}D^x_{i+d\mathbf{e}_x}\rangle - \langle
D^x_{i}\rangle\langle D^x_{i+d\mathbf{e}_x}\rangle$, can be obtained
directly by applying the local operator on the physical indices. Here the
spin operators are the eight generators of su(3) algebra in fundamental
representation (see SM for explicit expression), and the dimer operator is
defined as $D^x_{i} = \mathbf{S}_i\cdot \mathbf{S}_{i+d\mathbf{e}_x}$. The
$\mathbb{Z}_3$ gauge symmetry enables to define topologically nontrivial
local excitations like spinon, vison and their bound state
~\cite{Schuch2010a,Jiang2015, Duivenvoorden2017}
. A local
excitation in the spinon sector can be created by applying an operator $X$ satisfying $XZ=\omega ZX$ on the virtual indices of the local tensor such that it carries zero $\mathbb{Z}_3$ charge
 instead of the original charge 1. Similarly, $X^2$ can create a charge
$-1$ spinon, since $X^2Z={\omega^2}ZX^2$. A pair of vison excitations can
be created by putting a string of $Z$ (or $Z^2$) operators on the virtual
level, whose end points correspond to the vison excitations. Parafermions,
bound states of a spinon and a vison, can be created by putting spinons at
the end points of the $Z$ string. All these real space correlations can be
obtained using the CTMRG environment tensors,  see SM for further details. The correlations of the different types are shown in Fig.~\ref{fig:correlation}(a), computed with $\chi=392$.

On the other hand, correlation lengths can also be extracted from the
spectrum of the transfer matrix of the environment tensors provided by
CTMRG (see SM), also termed
channel operator, whose eigenvalue degeneracies carry information about
the types of correlation.  Correlation lengths along horizontal and
vertical directions are found to be the same, as expected.  Denoting the
distinct transfer matrix eigenvalues as $t_a$ $(a=0,1,...)$
with $|t_0|>|t_1|>|t_2|>...$, it turns out $t_0$ is non-degenerate,
suggesting that there is no long-range order in the variational wave
function (confirming the ED results). The sub-leading eigenvalues $t_{a}$ $(a=1,2,3)$ are six-fold
degenerate, followed by a non-degenerate $t_4$. These eigenvalues give
direct access to series of correlation lengths $\xi^{(a)} =
-1/\mathrm{log}(|t_a/t_0|)$, which therefore carry the same degeneracies.
 We have also computed the correlation length with a $\pm 1$
$\mathbb{Z}_3$ flux by inserting a string of $Z$ (or $Z^2$) operators,
where the leading eigenvalue of the corresponding transfer matrix is denoted
as $t_{Z,1}$~\footnote{The transfer matrix with $Z$ virtual action on one
layer and $Z^2$ on the other layer has the same spectrum as the one
without flux, and the spectrum with $Z$ or $Z^2$ on only one layer share
the same absolute values but with different phase $\omega$ or $\omega^2$}.
From $t_{Z,1}$, which is non-degenerate, one can obtain the leading
correlation length in the flux sector $\xi^{(1)}_{Z} =
-1/\mathrm{log}(|t_{Z,1}/t_0|)$.

A summary of various correlation lengths versus $\chi$ from both methods
is shown in Fig.~\ref{fig:correlation}(b). We find that the largest one in
all sectors, $\xi^{(1)}_{Z}$, is equal to the correlation length found between a
pair of visons; it is non-degenerate, in agreement with the fact that
visons carry no spin.
In the sector without
flux, the leading correlation length $\xi^{(1)}$ is in perfect agreement
with the correlation length $\xi_{\mathrm{spinon}}$ extracted from placing
a spinon-antispinon pair.
Moreover, since PT symmetry maps spinons placed on reflected bonds to
anti-spinons, we expect the spinon correlations to have a degeneracy
structure
$\mathbf{3}\oplus\bf\overline{3}$ which indeed agrees with the six-fold
degeneracy in $\xi^{(1)}$, and is further supported by checking the U(1)
quantum numbers of the $t_1$ multiplet.
The U(1)
quantum numbers further suggest that $t_{2,3}$, which are also six-fold
degenerate, also carry SU(3) representation
$\mathbf{3}\oplus\bf\overline{3}$. Thus, $\xi^{(1,2,3)}$ all correspond to
spinon correlation length. This, in fact, is in correspondence with the
three linear dispersing branches in the ES in the $Q=\pm 1$ charged
sectors, as we shall discuss later. Further examining the correlation
length, we find $\xi^{(4)}$ is identical to dimer correlation length,
where non-degeneracy agrees with the fact that the dimer operator is invariant
under SU(3) rotation. Depending on the parafermion type,
the $\xi_{\mathrm{parafermion}}$ have different values, both of which are
smaller than the spinon correlation length. Interestingly, all of these
correlation lengths, except the spin correlation length
$\xi_{\mathrm{spin}}$, have no sign of saturation with increasing $\chi$,
in agreement with our expectation that the state is \emph{not} in the
$\mathbb{Z}_3$ quantum double phase.

\emph{Degeneracy structure of topological chiral PEPS} --
A remarkable feature of our results is the correspondence between the
leading four eigenvalues of the transfer matrix and the different sectors
in the ES: The $Q=0$ sector has one branch, while $Q=\pm 1$ each have three
almost degenerate branches.  This is in direct analogy to the
unique leading eigenvalue $t_0$ which has trivial spin, and the
approximate three-fold degeneracy of $t_1,t_2,t_3$, which have 
perfectly degenerate spins $\bf{3}$ and $\bf{\overline{3}}$, matching 
the perfect degeneracy between $Q=\pm1$. A similar correspondence between
(approximate) degeneracy of the (2D) transfer operator and of the ES branches was observed for chiral PEPS with SU$(2)_1$
counting, where it could be explained as arising from
the symmetry of the tensors, and subsequently used to remove the
degeneracy in the non-trivial sector in the vicinity of a (fine-tuned) 
perfectly degenerate point~\cite{Hackenbroich2018}. Furthermore, we checked that the same correspondence also holds in the PEPS description of non-Abelian SU$(2)_2$ CSL~\cite{Chen2018b}.
It is suggestive that such a correspondence in the (approximate) degeneracy structure is a
general feature of chiral PEPS and will also hold for general
SU$\mathrm(N)_k$ models; indeed, 
both ES and the eigenvalues $t_i$ are extracted from the
same objects, namely the left/right (or up/down) fixed points of the CTMRG
environment.  It would be interesting to see whether such a correspondence
holds in a context of general chiral models, and whether it could possibly
even be used to further characterize the precise nature of a chiral theory.

{\it Conclusion and outlook} --
In this work, we have proposed a model for a SU$(3)_1$ CSL on square lattice and unambiguously identified the relevant parameter space based on ED technique. Guided by ED and DMRG results, we have focused on constructing and optimizing a {\it symmetric} PEPS ansatz for the CSL, whose variational energy is remarkably good. For the first time, linear dispersing branches
in all three sectors of SU$(3)_1$ WZW CFT can be obtained with PEPS. A comparison between edge spectrum and bulk correlations reveal a fine structure in the bulk-edge correspondence, which will be tested in further study of SU$(N)_k$ PEPS CSL. Certain unresolved issues, e.g., the number of variationally degenerate ground states on torus, and anyon statistics of chiral topological order remain open, which we hope to uncover in the future.

{\it Acknowledgements} --
We acknowledge useful conversations with M.~Arildsen, C.~Delcamp, A.~Hackenbroich, M.~Iqbal, A.~Ludwig, G.~Sierra, and J.~Slingerland. JYC and NS acknowledge support by the European Union's Horizon 2020 programme through the ERC Starting Grant WASCOSYS (Grant No.~636201) and from the DFG (German Research Foundation) under Germany's Excellence Strategy (EXC-2111 -- 390814868). DP acknowledges support by the TNSTRONG ANR-16-CE30-0025  and TNTOP ANR-18-CE30-0026-01 grants awarded by the French Research Council.  This work was granted access to the HPC resources of CALMIP and GENCI supercomputing centers under the allocation 2017-P1231 and A0030500225, respectively, and computations have also been carried out on the TQO cluster of the Max-Planck-Institute of Quantum Optics.

\bibliography{bibliography}

\end{document}


\title{Supplementary Material for\\ ``SU$(3)_1$ Chiral Spin Liquid on the Square Lattice: a View from Symmetric PEPS''}

\author{Ji-Yao Chen}
\affiliation{Max-Planck-Institut f\"ur Quantenoptik, Hans-Kopfermann-Stra{\ss}e 1, 85748 Garching, Germany}
\affiliation{Munich Center for Quantum Science and Technology, Schellingstra{\ss}e 4, 80799 M{\"u}nchen, Germany}

\author{Sylvain Capponi}
\affiliation{Laboratoire de Physique Th\'eorique, IRSAMC, Universit\'e de Toulouse, CNRS, UPS, 31062 Toulouse, France}

\author{Alexander Wietek}
\affiliation{Center for Computational Quantum Physics, Flatiron Institute, 162 5th Avenue, NY 10010, New York, USA}

\author{Matthieu Mambrini}
\affiliation{Laboratoire de Physique Th\'eorique, IRSAMC, Universit\'e de Toulouse, CNRS, UPS, 31062 Toulouse, France}

\author{Norbert Schuch}
\affiliation{Max-Planck-Institut f\"ur Quantenoptik, Hans-Kopfermann-Stra{\ss}e 1, 85748 Garching, Germany}
\affiliation{Munich Center for Quantum Science and Technology, Schellingstra{\ss}e 4, 80799 M{\"u}nchen, Germany}

\author{Didier Poilblanc}
\affiliation{Laboratoire de Physique Th\'eorique, IRSAMC, Universit\'e de Toulouse, CNRS, UPS, 31062 Toulouse, France}

\date{\today}
\maketitle

To complement the main findings in the manuscript, we provide several relevant details in this supplementary material, organized as follows: basic knowledge about SU($3$) group and its irreducible representations (irreps) in Sec.~\ref{sec:SU3}, exact diagonalization study on various small clusters in Sec.~\ref{sec:ED}, DMRG study on finite cylinders in Sec.~\ref{sec:DMRG}, construction of SU(3) symmetric PEPS and the tensor classification scheme in Sec.~\ref{sec:PEPSansatz}, the specific corner transfer matrix renormalization group (CTMRG) method we use and the optimization procedure in Sec.~\ref{sec:CTMRG}, additional data for entanglement spectrum (ES) in Sec.~\ref{sec:ES}, topological excitations and correlation functions of symmetric PEPS in Sec.~\ref{sec:excitations}, and in the end we also list the nonzero elements of the tensors in Sec.~\ref{sec:tensorExpression}.

\section{Brief overview of SU($3$) irreps}
\label{sec:SU3}

Since the theory of SU($3$) group and its irreps can be found in many textbooks, e.g. Ref.~\onlinecite{BarutRaczka1980}, here we only list the relevant known results without derivation.

\begin{figure}[!h]
	\centering
	\includegraphics[width=0.2\textwidth]{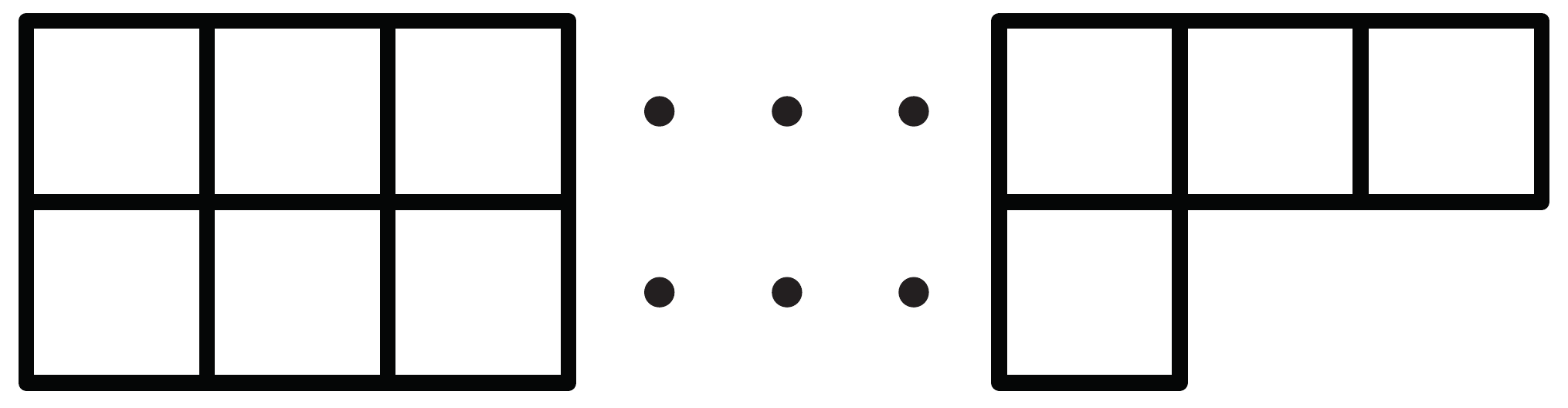}
	\caption{A generic Young tableau characterizing an irrep of SU($3$).}
	\label{fig:YT}
\end{figure}

As a special case of the general representation theory of SU($N$) group, to each irrep of SU($3$) we can associate a Young tableau containing a maximum of two rows (see Fig.~\ref{fig:YT}). Denoting by $p$ ($q$) the number of columns in the first (second) row, with $p\geq q$, the dimension of the corresponding irrep is $(1/2)(p+2)(q+1)(p-q+1)$.

Unlike the SU($2$) case where the states of a given multiplet are labeled by a unique U($1$) quantum number (eigenvalue of ${S}^z$) and related to each other by a unique ladder operator ${S}^{-}$ (or ${S}^{+}$), multiplets of SU($3$) should rather be seen as two-dimensional objects where states are characterized by two U($1$) quantum numbers ${\bf S}^z = (s^z_1, s^z_2)$ and related by two ladder operators $({S}^{-}_1,{S}^{-}_2)$ (or $({S}^{+}_1,{S}^{+}_2)$). Note that a given couple $(s^z_1, s^z_2)$ is no longer necessarily associated to a unique state.

The multiplet structures and young tableaux of the irreps considered in this work, including $\bf{3}$, $\bf\overline{3}$, $\bf{6}$, $\bf\overline{6}$, $\bf{8}$, $\bf{10}$, $\bf\overline{10}$, $\bf{15}$, $\bf\overline{15}$, $\bf{15'}$, $\bf\overline{15'}$, $\bf{21}$, $\bf\overline{21}$, $\bf{24}$, $\bf\overline{24}$ and $\bf{27}$, are depicted in Tab.~\ref{table:su3multiplets}.

\begin{longtable}{p{0.4\columnwidth}p{0.1\columnwidth}p{0.4\columnwidth}}
	\caption{SU($3$) multiplets considered in this work. Yellow boxes display a conventional numbering of states grouped in tuples sharing the same ${\bf S}^z = (s^z_1, s^z_2)$. Red and blue arrows show how states are connected by the two lowering operators of SU($3$). The black dot in each irrep denotes the highest weight state.}\\
	\hline
	\hline\\
	\endfirsthead
	\multicolumn{3}{c}%
	{\tablename\ \thetable\ -- \textit{Continued from previous column}} \\
	\hline\vspace{0.5mm}
	\endhead
	\hline \multicolumn{3}{r}{\textit{Continued on next column}} \\
	\endfoot
	\hline\hline
	\endlastfoot
	\includegraphics[width=0.45\columnwidth]{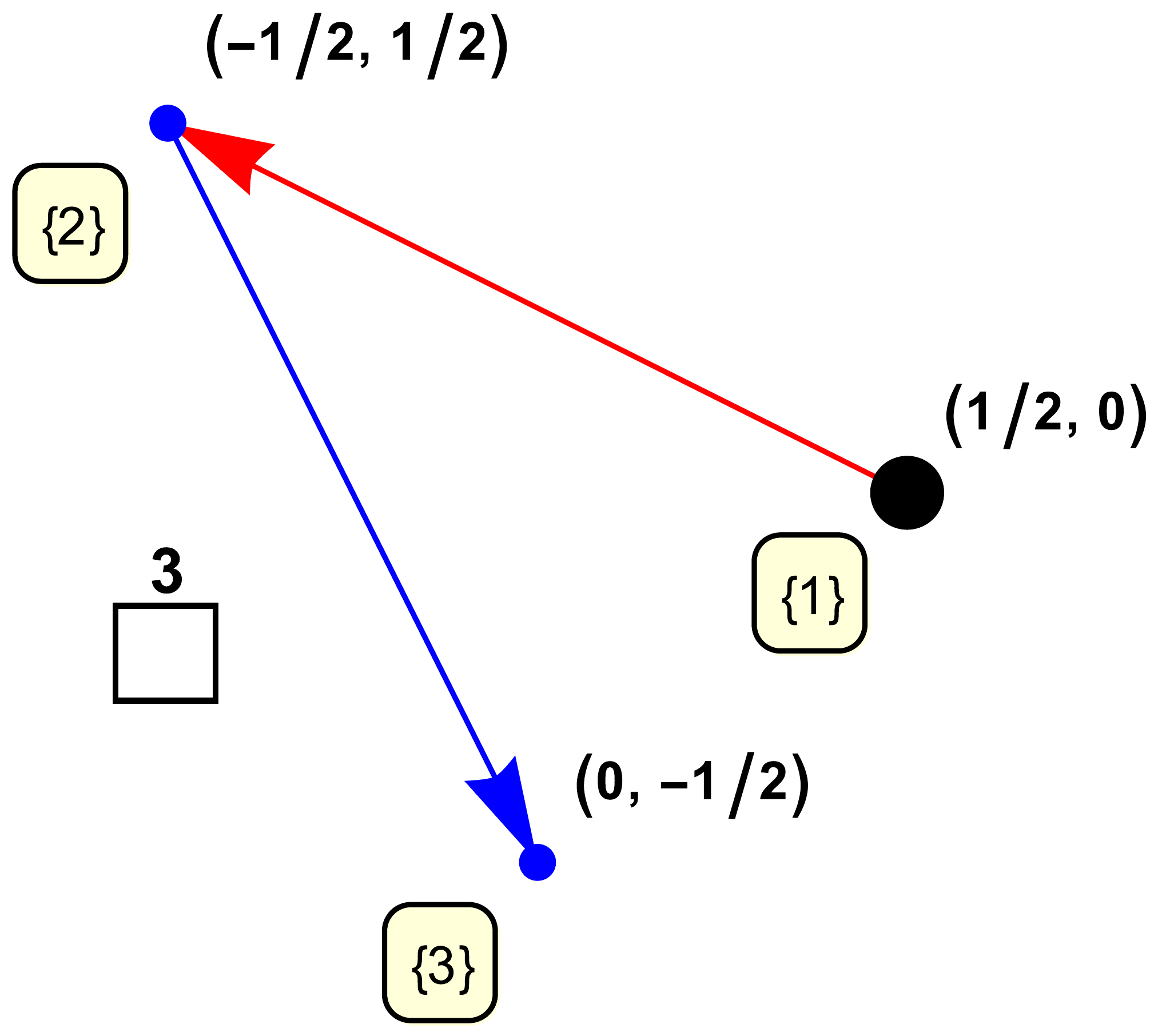}&&\includegraphics[width=0.45\columnwidth]{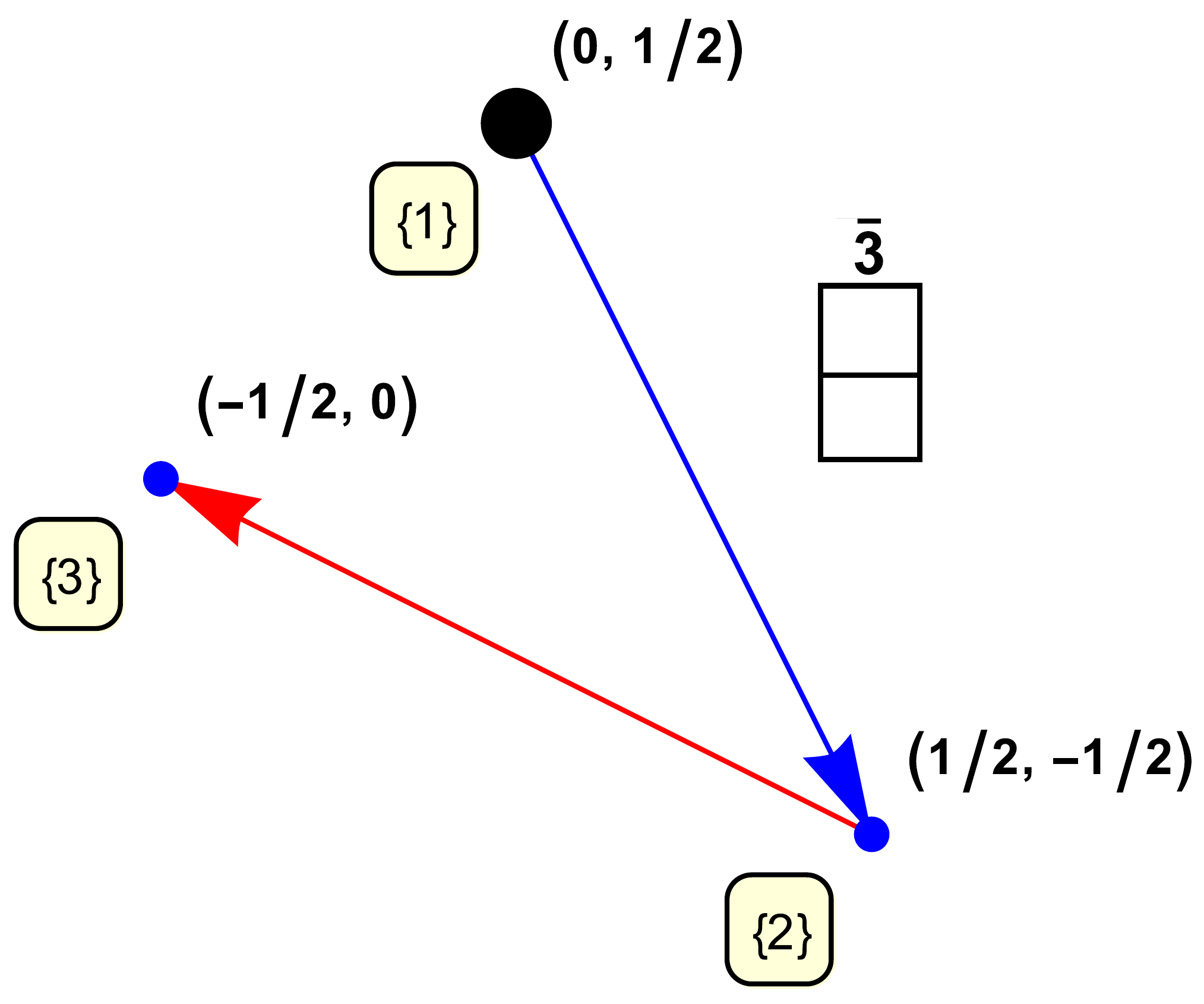} \\
	\multicolumn{3}{c}{\includegraphics[width=0.65\columnwidth]{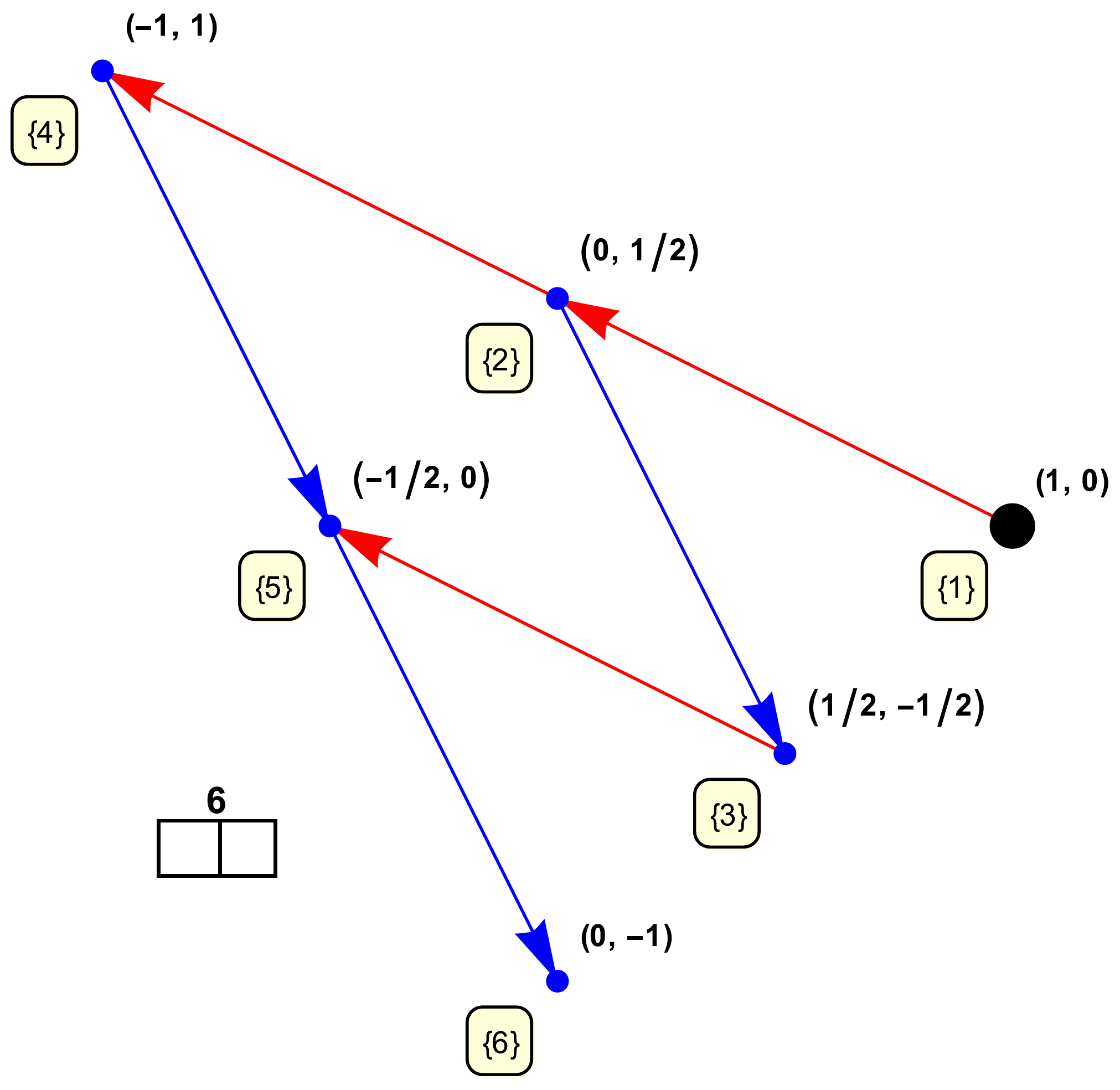}} \\ \vspace{1cm}\\
	\multicolumn{3}{c}{\includegraphics[width=0.65\columnwidth]{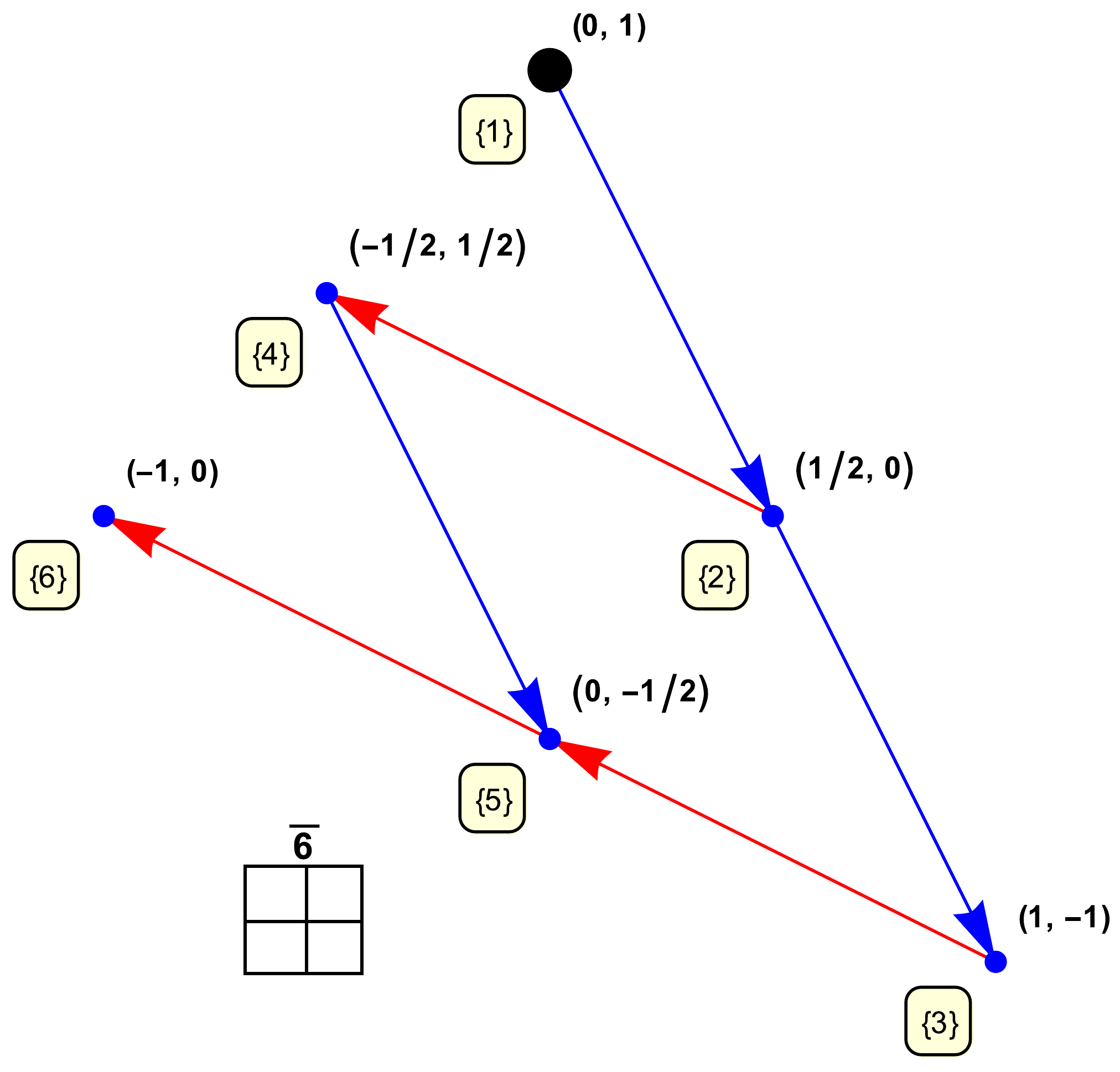}} \\ \vspace{1cm} \\
	\multicolumn{3}{c}{\includegraphics[width=0.7\columnwidth]{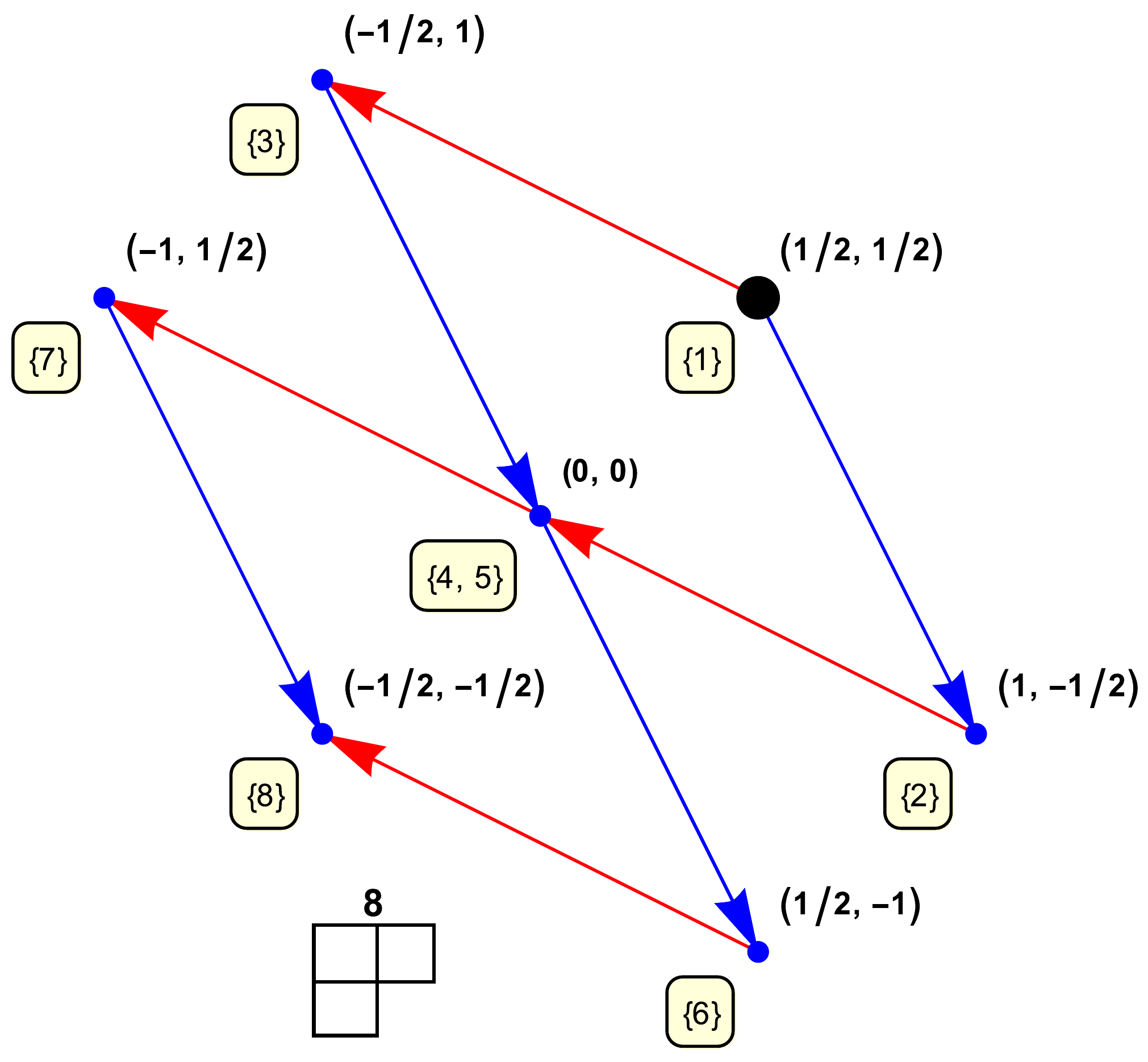}} \\	
	\multicolumn{3}{c}{\includegraphics[width=0.85\columnwidth]{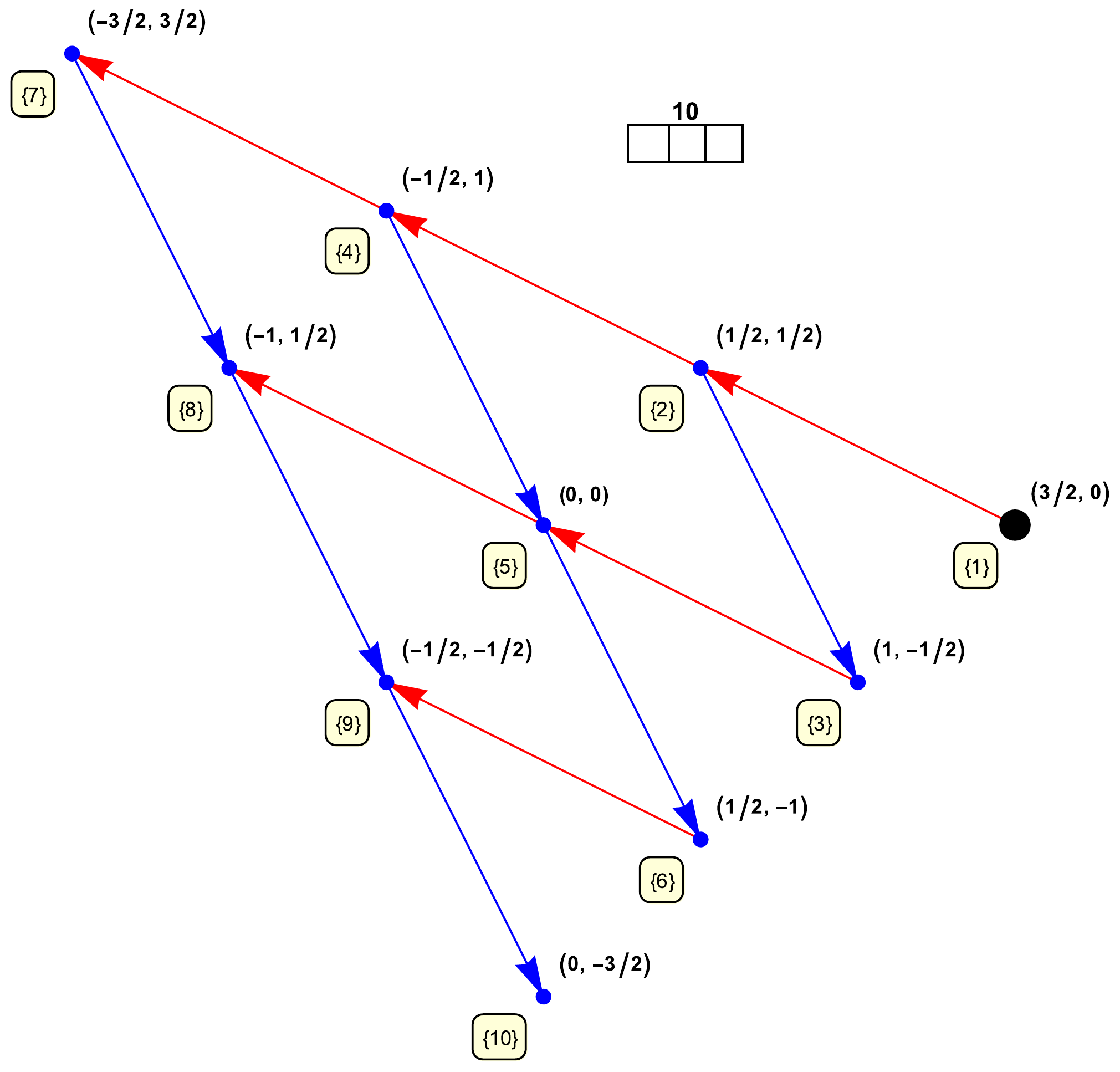}} \\
	\multicolumn{3}{c}{\includegraphics[width=0.85\columnwidth]{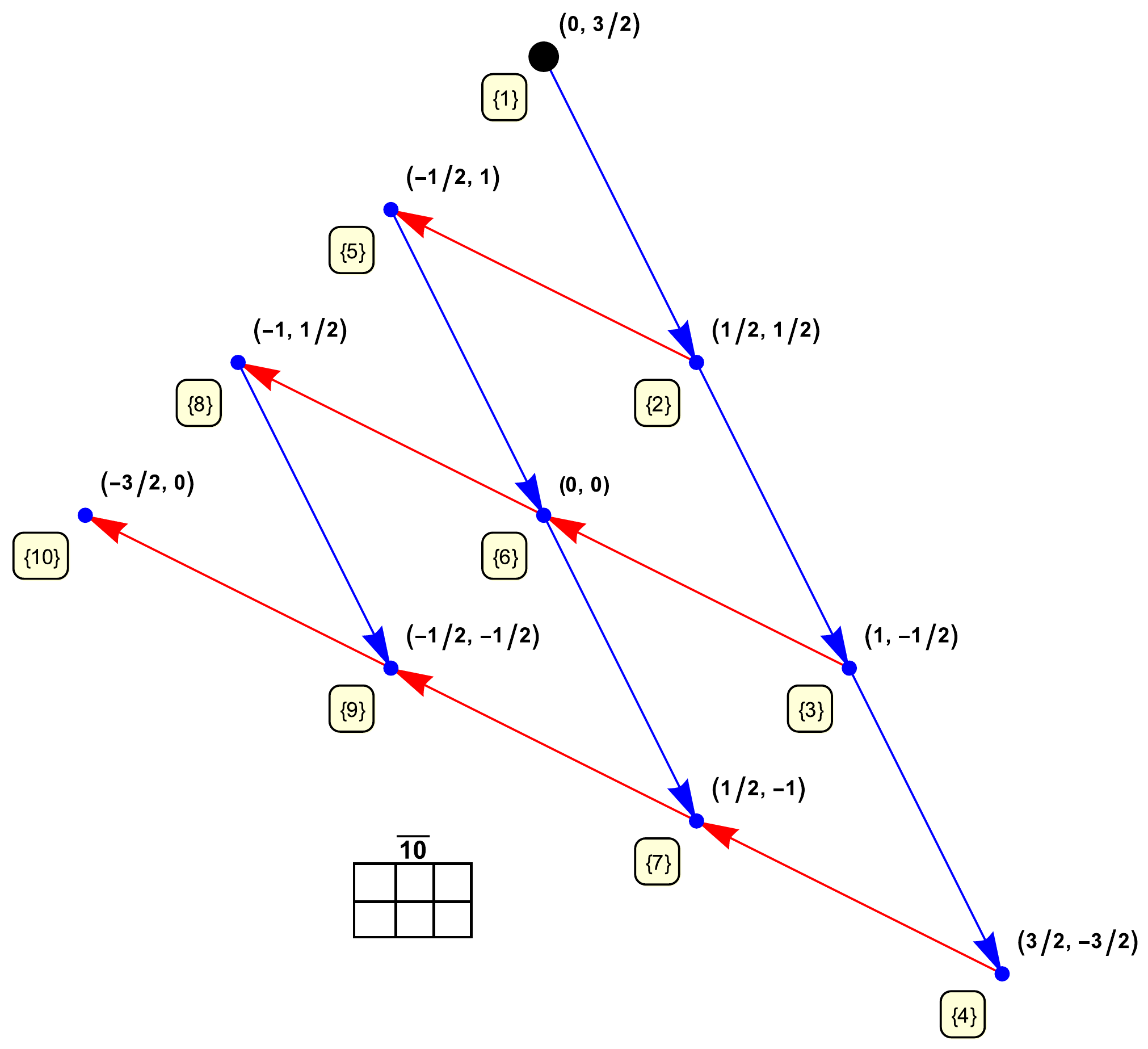}} \\
	\multicolumn{3}{c}{\includegraphics[width=0.85\columnwidth]{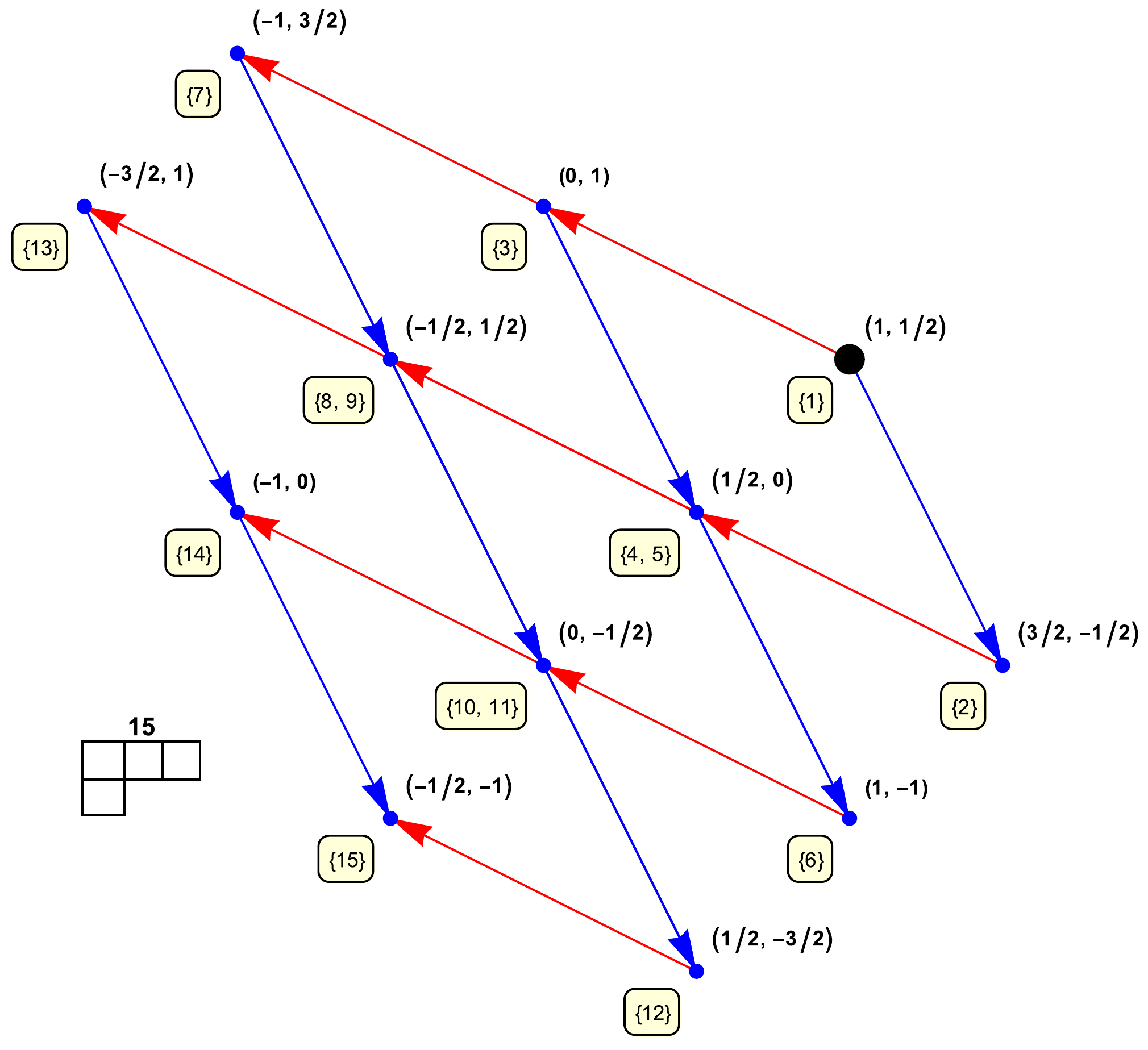}} \\
	\multicolumn{3}{c}{\includegraphics[width=0.85\columnwidth]{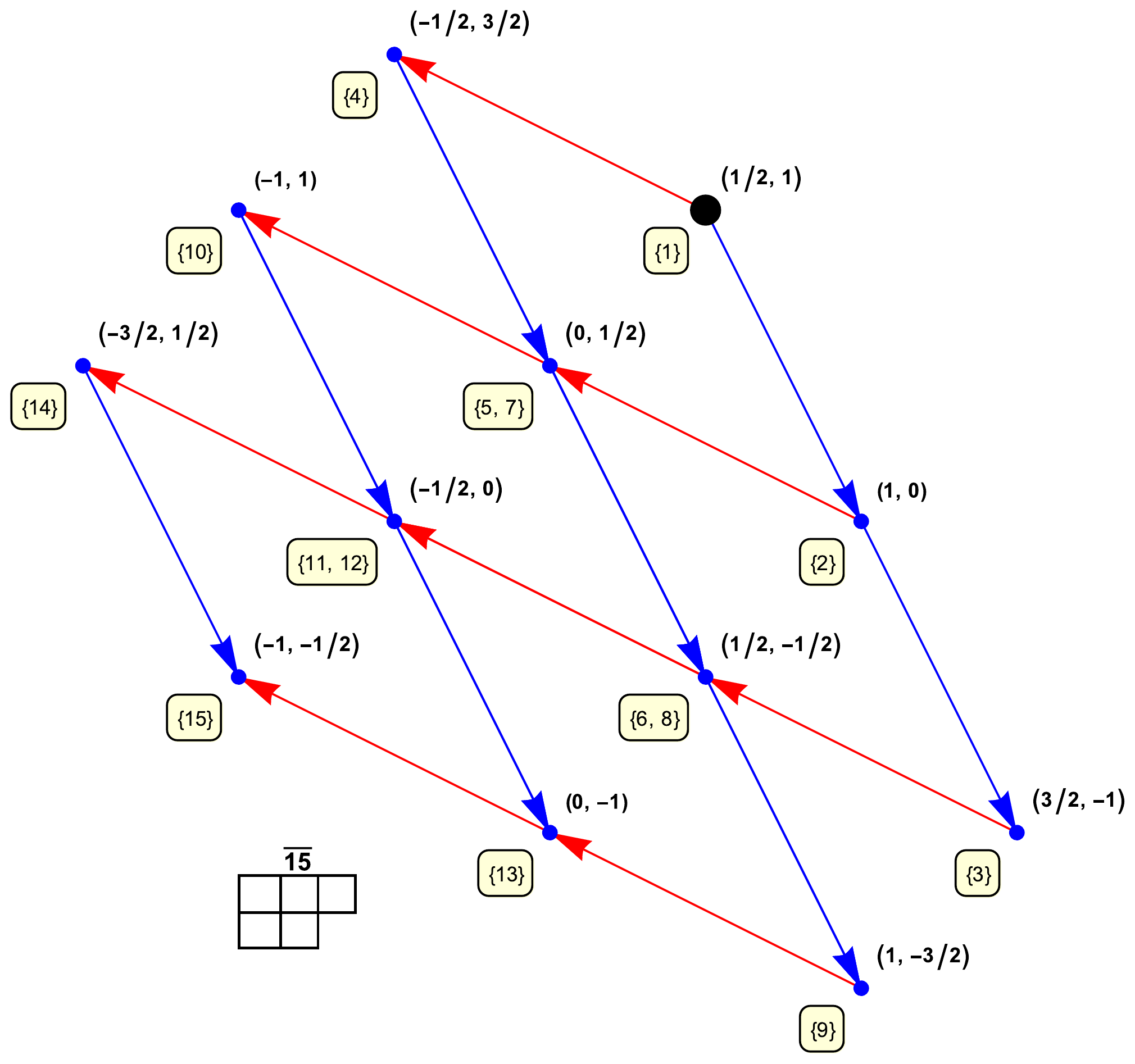}} \\
	\multicolumn{3}{c}{\includegraphics[width=\columnwidth]{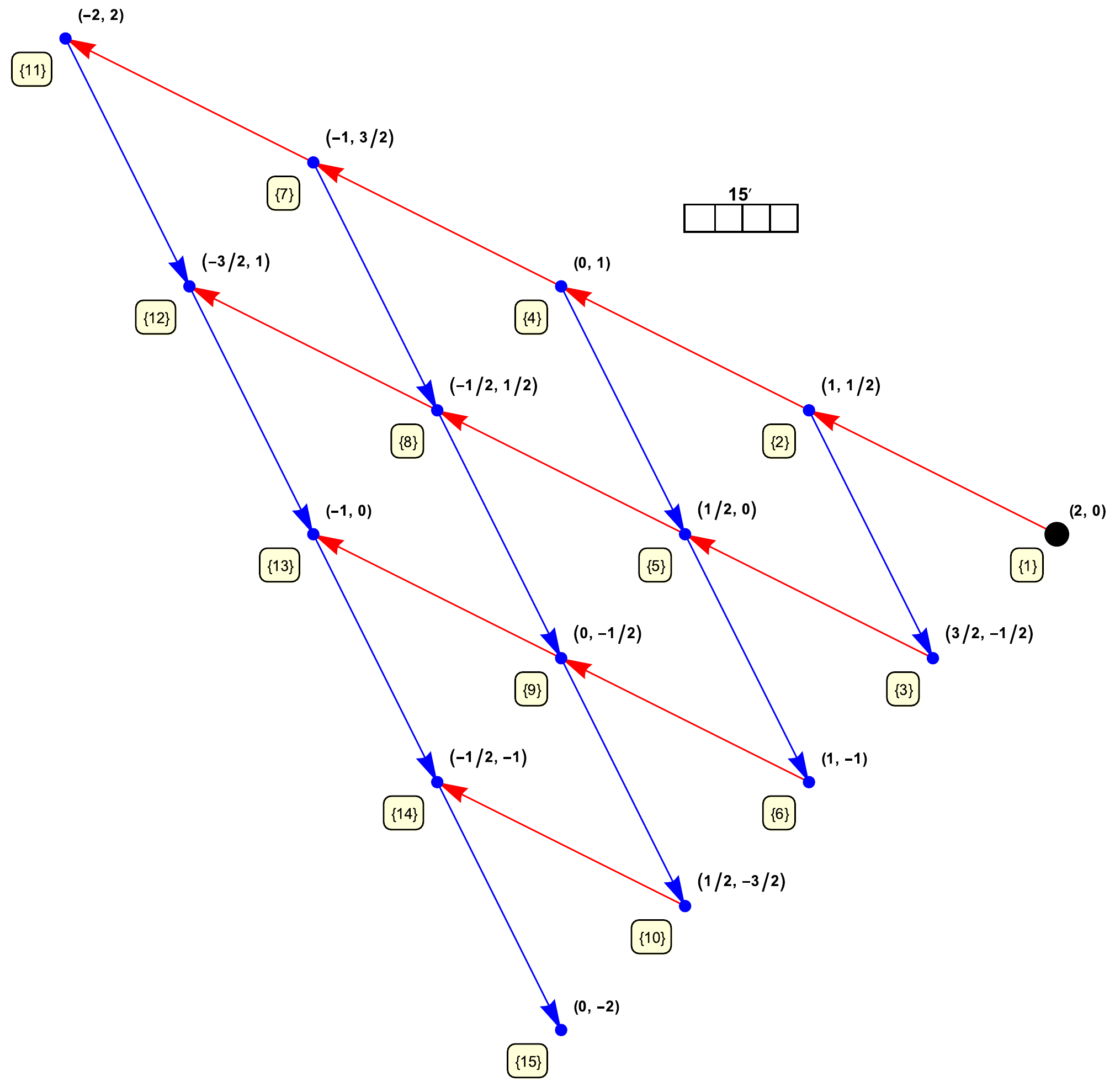}} \\  \vspace{1.5cm} \\
	\multicolumn{3}{c}{\includegraphics[width=\columnwidth]{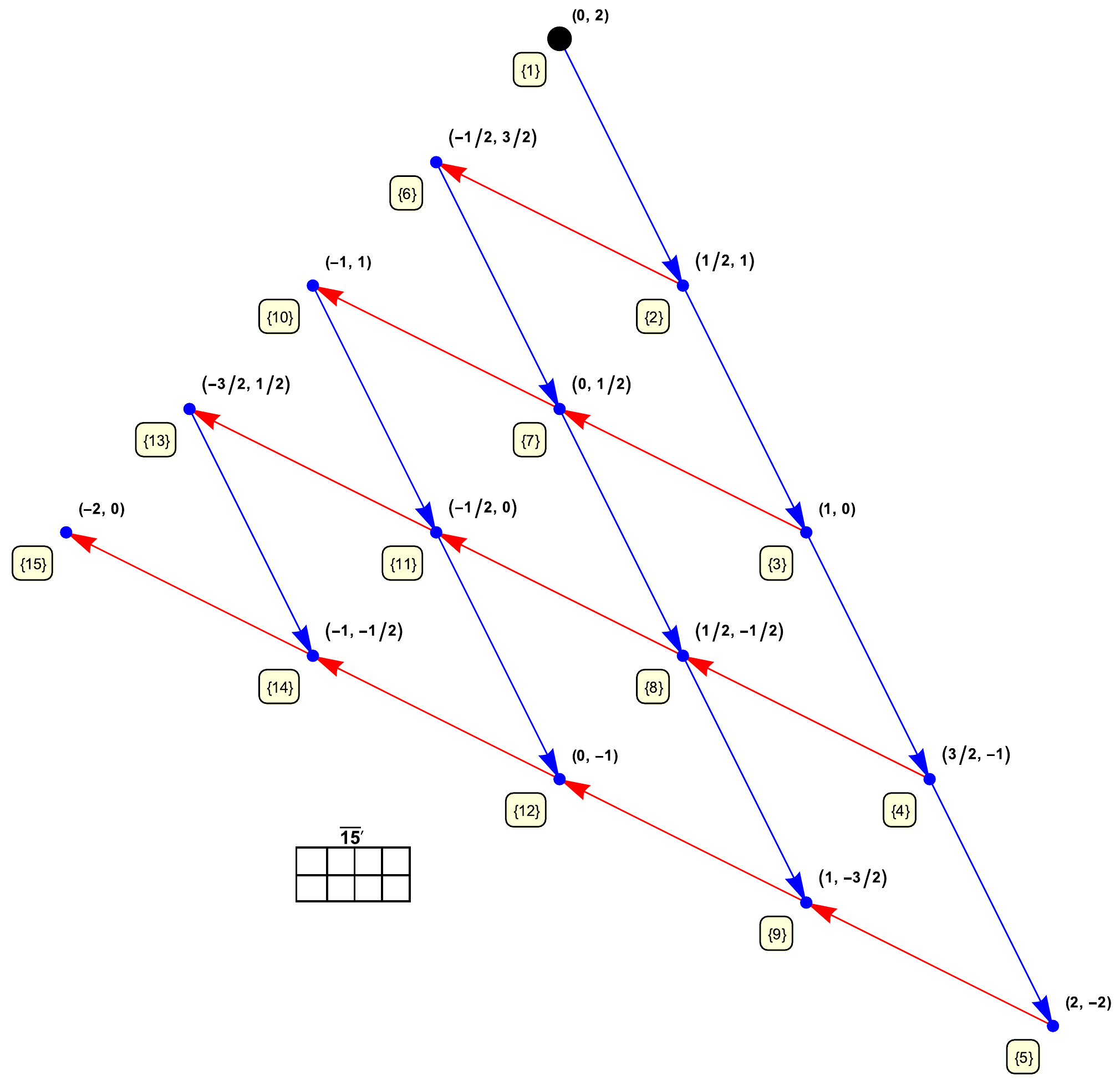}} \\
	\multicolumn{3}{c}{\includegraphics[width=\columnwidth]{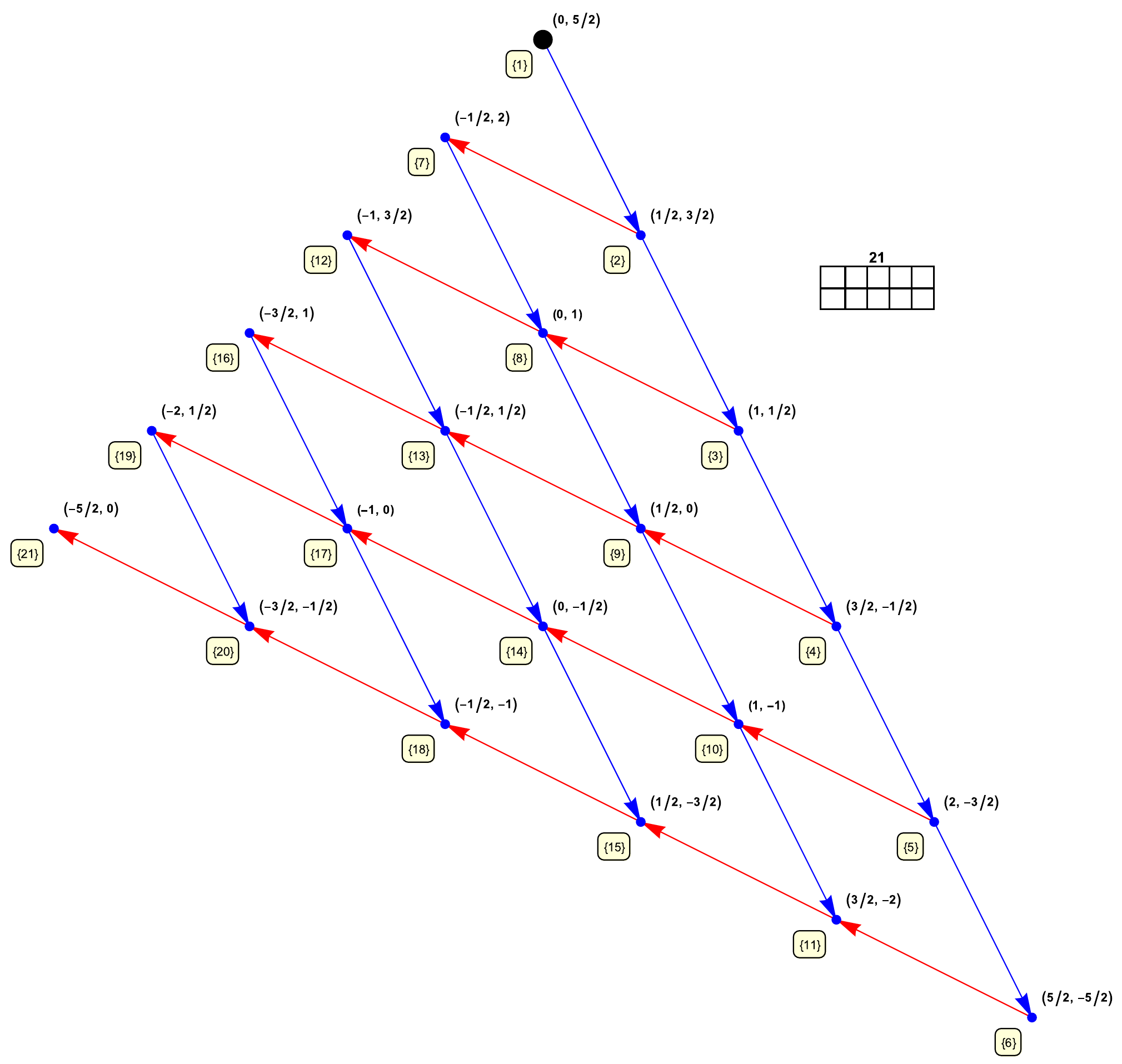}} \\  \vspace{1.5cm} \\
	\multicolumn{3}{c}{\includegraphics[width=\columnwidth]{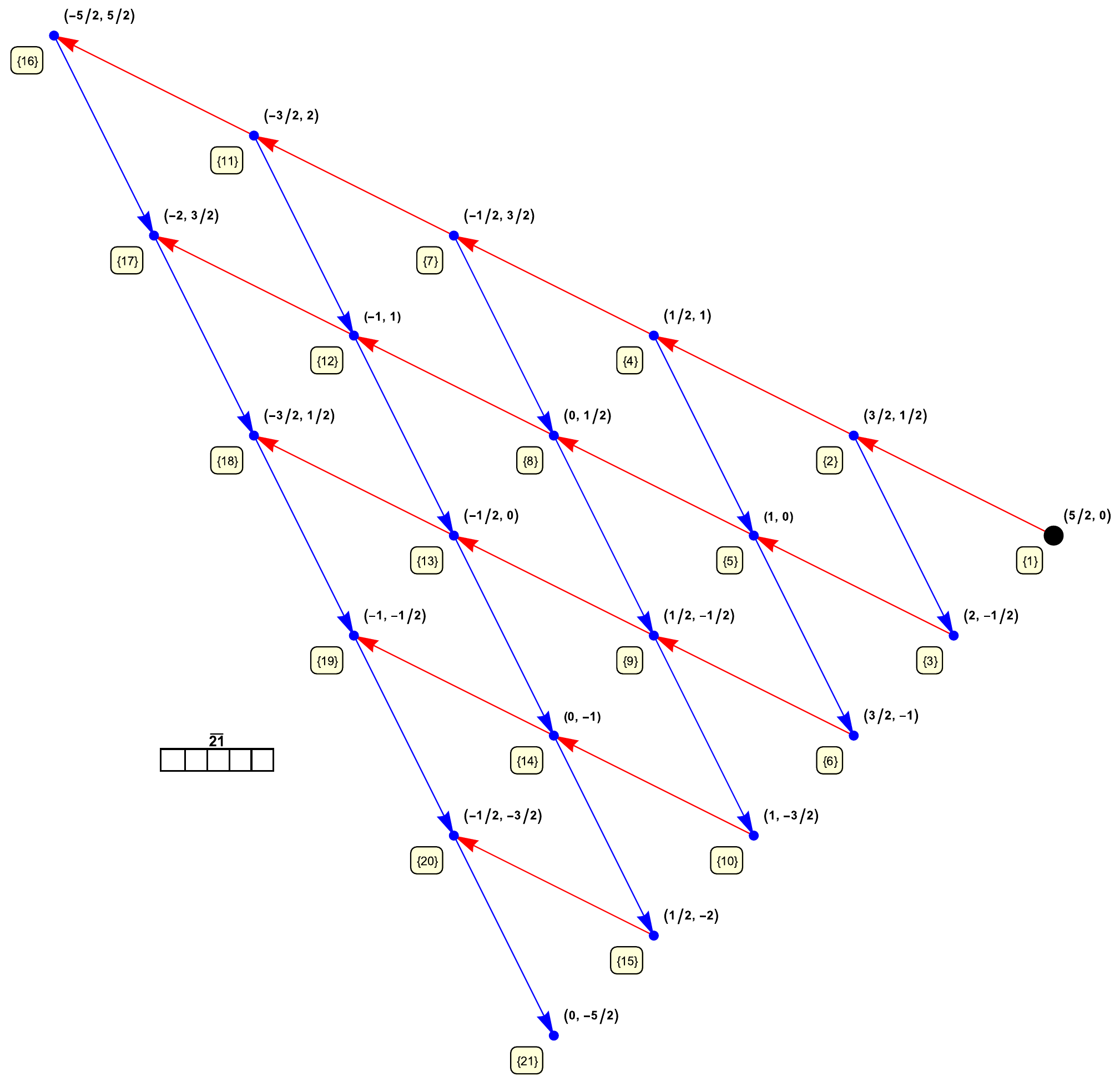}} \\
	\multicolumn{3}{c}{\includegraphics[width=\columnwidth]{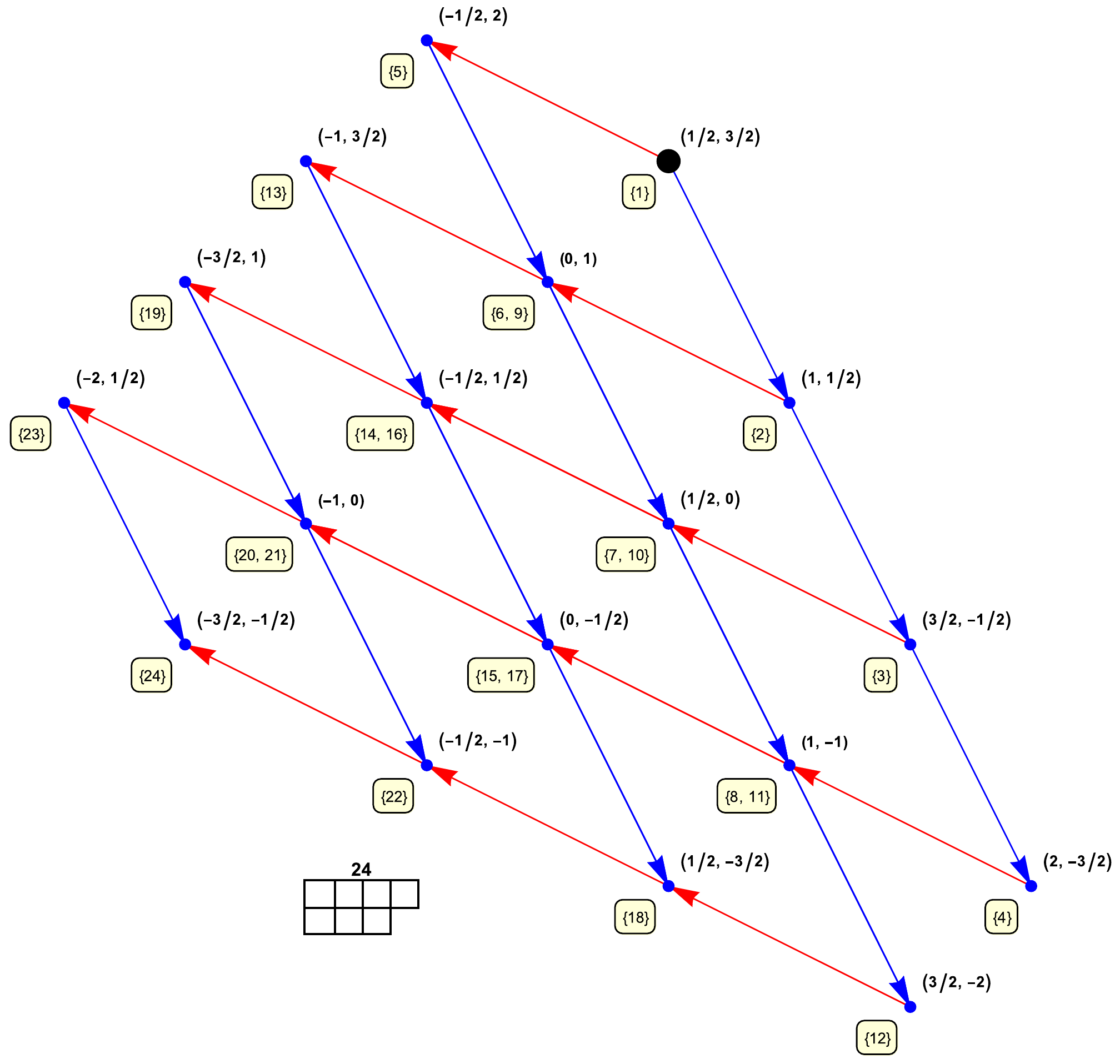}} \\  \vspace{1.5cm} \\
	\multicolumn{3}{c}{\includegraphics[width=\columnwidth]{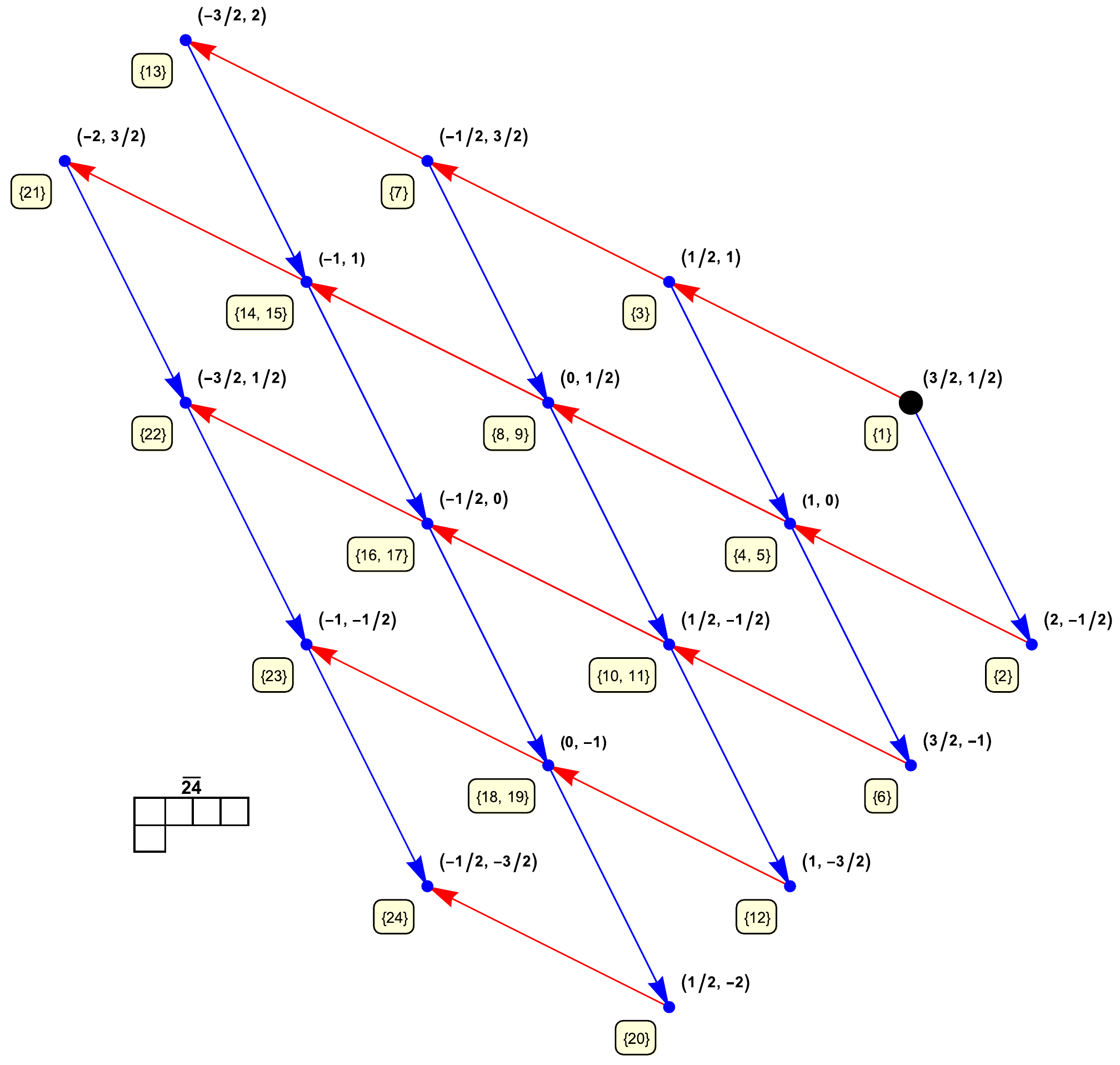}} \\
	\multicolumn{3}{c}{\includegraphics[width=\columnwidth]{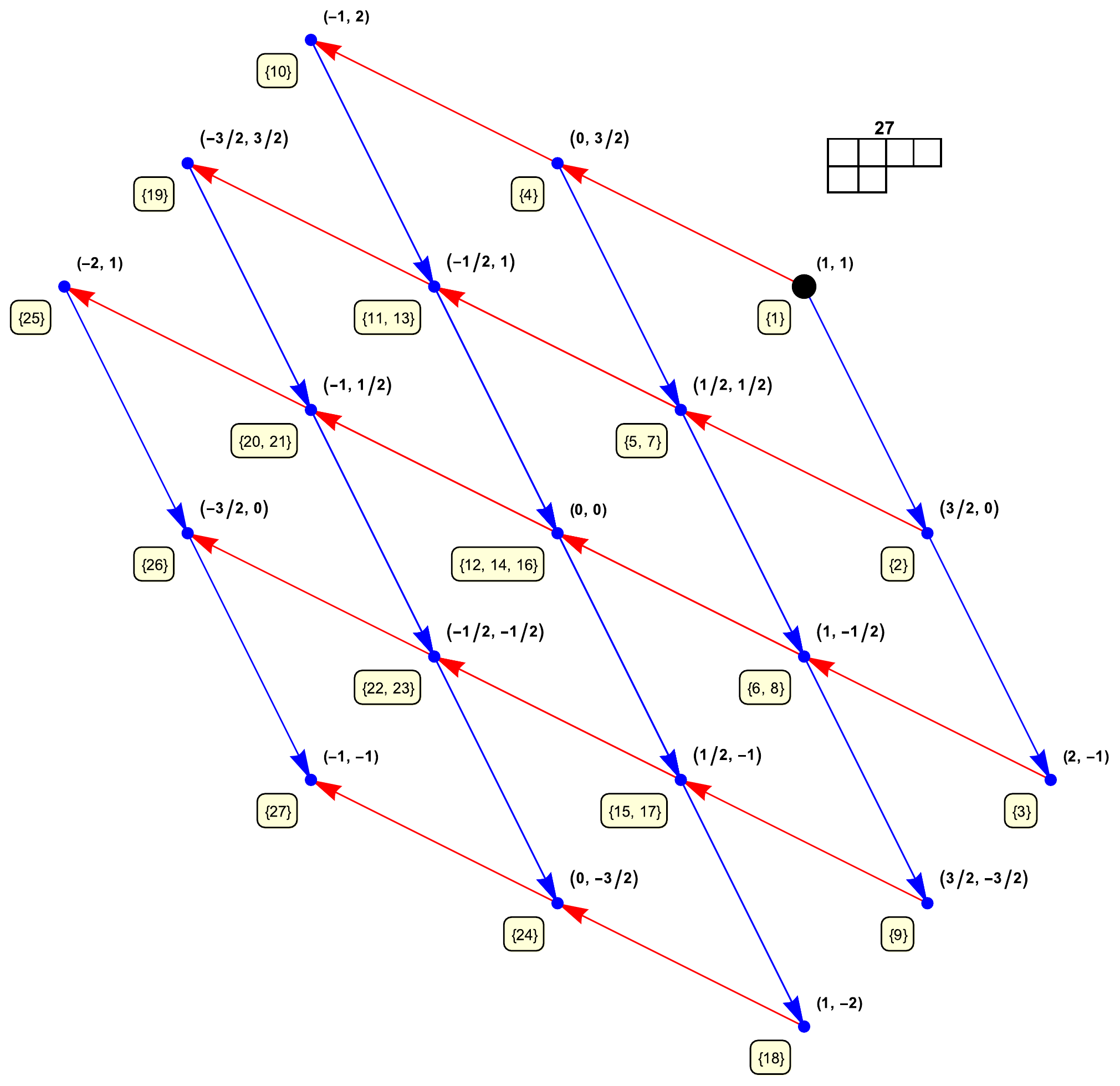}}
	\label{table:su3multiplets}
\end{longtable}

The su($3$) Lie algebra of SU($3$) group contains eight generators, denoted as ${S}^\alpha (\alpha=1,2,..,8)$, whose matrix expressions in the $\bf{3}$ and $\bf{\overline{3}}$ irreps are provided in Tab.~\ref{table:su3gen}.

\begin{table}[!htb]
	\renewcommand\arraystretch{1.5}
	\begin{tabularx}{0.45\textwidth}{ccc}
		\hline
		\hline
	generator & $\bf{3}$ & $\bf{\overline{3}}$ \\
	$S^1$ & $\left(
	\begin{array}{ccc}
	0 & 0 & 0 \\
	0 & 0 & \frac{1}{2} \\
	0 & \frac{1}{2} & 0 \\
	\end{array}
	\right)$ & $\left(
	\begin{array}{ccc}
	0 & \frac{1}{2} & 0 \\
	\frac{1}{2} & 0 & 0 \\
	0 & 0 & 0 \\
	\end{array}
	\right)$ \\
	$S^2$ & $\left(
	\begin{array}{ccc}
	0 & 0 & \frac{1}{2} \\
	0 & 0 & 0 \\
	\frac{1}{2} & 0 & 0 \\
	\end{array}
	\right)$ & $\left(
	\begin{array}{ccc}
	0 & 0 & -\frac{1}{2} \\
	0 & 0 & 0 \\
	-\frac{1}{2} & 0 & 0 \\
	\end{array}
	\right)$ \\
	$S^3$ & $\left(
	\begin{array}{ccc}
	0 & \frac{1}{2} & 0 \\
	\frac{1}{2} & 0 & 0 \\
	0 & 0 & 0 \\
	\end{array}
	\right)$ & $\left(
	\begin{array}{ccc}
	0 & 0 & 0 \\
	0 & 0 & \frac{1}{2} \\
	0 & \frac{1}{2} & 0 \\
	\end{array}
	\right)$ \\
	$S^4$ & $\left(
	\begin{array}{ccc}
	0 & 0 & 0 \\
	0 & 0 & -\frac{i}{2} \\
	0 & \frac{i}{2} & 0 \\
	\end{array}
	\right)$ & $\left(
	\begin{array}{ccc}
	0 & -\frac{i}{2} & 0 \\
	\frac{i}{2} & 0 & 0 \\
	0 & 0 & 0 \\
	\end{array}
	\right)$ \\
	$S^5$ & $\left(
	\begin{array}{ccc}
	0 & 0 & -\frac{i}{2} \\
	0 & 0 & 0 \\
	\frac{i}{2} & 0 & 0 \\
	\end{array}
	\right)$ & $\left(
	\begin{array}{ccc}
	0 & 0 & \frac{i}{2} \\
	0 & 0 & 0 \\
	-\frac{i}{2} & 0 & 0 \\
	\end{array}
	\right)$ \\
	$S^6$ & $\left(
	\begin{array}{ccc}
	0 & -\frac{i}{2} & 0 \\
	\frac{i}{2} & 0 & 0 \\
	0 & 0 & 0 \\
	\end{array}
	\right)$ & $\left(
	\begin{array}{ccc}
	0 & 0 & 0 \\
	0 & 0 & -\frac{i}{2} \\
	0 & \frac{i}{2} & 0 \\
	\end{array}
	\right)$ \\
	$S^7$ & $\left(
	\begin{array}{ccc}
	\frac{1}{2} & 0 & 0 \\
	0 & -\frac{1}{2} & 0 \\
	0 & 0 & 0 \\
	\end{array}
	\right)$ & $\left(
	\begin{array}{ccc}
	0 & 0 & 0 \\
	0 & \frac{1}{2} & 0 \\
	0 & 0 & -\frac{1}{2} \\
	\end{array}
	\right)$ \\
	$S^8$ & $\left(
	\begin{array}{ccc}
	\frac{1}{2\sqrt{3}} & 0 & 0 \\
	0 & \frac{1}{2\sqrt{3}} & 0 \\
	0 & 0 & -\frac{1}{\sqrt{3}} \\
	\end{array}
	\right)$ & $\left(
	\begin{array}{ccc}
	\frac{1}{\sqrt{3}} & 0 & 0 \\
	0 & -\frac{1}{2\sqrt{3}} & 0 \\
	0 & 0 & -\frac{1}{2\sqrt{3}} \\
	\end{array}
	\right)$ \vspace{1mm}\\
	\hline\hline
	\end{tabularx}
	\caption{Explicit expressions of su(3) generators in the $\bf{3}$ and $\bf{\overline{3}}$ representations.}
	\label{table:su3gen}
\end{table}

We note that, in irreps of SU($N$), there is some arbitrariness in defining the $N-1$ diagonal generators (so-called Cartan subalgebra). Without basis change, it is possible to linearly combine them. The two U($1$) quantum numbers shown in Tab.~\ref{table:su3multiplets} indeed correspond to the eigenvalues of ${S}^7$ and $-\frac{1}{2}{S}^7+\frac{\sqrt{3}}{2}{S}^8$. The generator for the center of SU($3$) can also be expressed in terms of the two diagonal generators as: 
\begin{equation}
Z = {\rm exp}\left( i2\pi(S^7-S^8/\sqrt{3})\right).
\end{equation}
The two-site permutation operator $P_{ij}$ used in defining the Hamiltonian, can be expressed with su($3$) generators in the irrep $\bf{3}$ as:
\begin{equation}
P_{ij}=\frac{1}{3}+2\sum_{\alpha=1}^{8}{S}_i^\alpha {S}_j^{\alpha}.
\end{equation}

\section{Exact diagonalization on various clusters}
\label{sec:ED}

For exact diagonalization, we have used the Lanczos (respectively Davidson) algorithm to compute the ground-state (respectively low-energy excitations) of our model on various finite-size clusters of $N_s$ sites. 
See Tab.~\ref{table:clusters} for a list of clusters considered with periodic boundary conditions (PBC).
Since we are looking for a quantum spin liquid state, all clusters are adequate even though they possess different momenta in their Brillouin zone. Even more, we have considered some clusters which are not perfectly square to get additional signatures: we define the eccentricity of a cluster by the ratio of the two smallest inequivalent loops of the nearest-neighbor graph around the torus, which 
is a measure of the ``two-dimensionality" of the cluster, where a value close to one is considered fully two-dimensional. In order to reduce the size of the Hilbert space, we have used all space symmetries (translation and point-group) as well as color conservation, which is equivalent to fixing the values of the two U(1) quantum numbers ${\bf S}^z$: namely, we diagonalize the Hamiltonian in a subspace with a given number of particles per color $(N_1,N_2,N_3)$ with a constraint of single-occupation, i.e. $N_1+N_2+N_3=N_s$. Using the conventions of Sec.~\ref{sec:SU3}, such a sector is equivalent to fixing the weights $(s^z_1=(N_1-N_2)/2,\, s^z_2=(N_2-N_3)/2)$.

\begin{table}[!h]
\begin{tabular}{c | c | c | c | c | c}
\hline
\hline
cluster & ${\bf t}_1$ & ${\bf t}_2$ & $N_s$ & eccentricity & point group  \\
\hline
12     & $(1,3)$     & $(4,0)$    & $12$  & $1$          & $C_{2}$           \\
15a     & $(1,4)$     & $(4,1)$    & $15$  & $1$          & $C_{2v}$           \\
15b     & $(2,-3)$     & $(3,3)$    & $15$  & $1$          & $C_{2}$           \\
18a     & $(3,3)$     & $(3,-3)$    & $18$  & $1$          & $C_{4v}$           \\
18b     & $(3,-3)$    & $(4,2)$     & $18$  & $1$          & $C_{2}$            \\
19a     & $(1,4)$     & $(4,-3)$    & $19$  & $1.2$        & $C_{2}$            \\
19b     & $(2,-5)$    & $(3,2)$     & $19$  & $1.4$        & $C_{2}$            \\
20a     & $(4,2)$     & $(2,-4)$    & $20$  & $1$          & $C_{4}$            \\
20b     & $(1,4)$     & $(5,0)$     & $20$  & $1$          & $C_{2}$            \\
20c     & $(2,-4)$    & $(3,4)$     & $20$  & $1.2$        & $C_{2}$            \\
21a     & $(2,-5)$    & $(3,3)$     & $21$  & $1.17$       & $C_{2v}$           \\
21b     & $(3,-3)$    & $(4,3)$     & $21$  & $1.17$       & $C_{2}$            \\
21c     & $(1,4)$     & $(5,-1)$    & $21$  & $1.2$        & $C_{2}$            \\
24a     & $(2,4)$     & $(4,-4)$    & $24$  & $1$          & $C_{2}$            \\
24b     & $(1,5)$     & $(5,1)$     & $24$  & $1$          & $C_{2v}$          \\
\hline
\hline
\end{tabular}
\caption{List of clusters under PBC studied with ED: name, the two unit vectors, number of sites $N_s$, eccentricity, and point-group symmetry.}
\label{table:clusters}
\end{table}

In the seminal work by Halperin~\cite{Halperin1983}, a SU(2) spin-singlet fractional quantum Hall (FQH) state was introduced for
hardcore spin-$1/2$ bosons at filling $\nu=2/3$. As stated in the main
text, the SU$(3)_1$ CSL that we are investigating is a lattice
realization of such a phase~\cite{WuTu2016}. The simplest signatures
of a FQH phase are given by the ground-state degeneracy and the
quasi-hole properties (quasi-degeneracy and momentum quantum
numbers)~\cite{Regnault2011,Bernevig2012}. Moreover, there is a simple
generalized exclusion principle~\cite{Estienne2012,Sterdyniak2013}
with $(2,1)$ clustering properties in our case: for instance, in the
spinful bosonic language, when $N_s=3p$, the three (quasi) degenerate
ground-states are given by occupations:
$(\uparrow,\downarrow,0,\uparrow,\downarrow,0,\ldots)$ and its
translations. These occupations have to be understood as a function of
$N_s$ orbitals which are obtained when folding the Brillouin
zone~\cite{Regnault2011,Bernevig2012}. This exclusion rule simply enforces that there are no more than 2 particles in 3 consecutive orbitals and that a $\downarrow$ particle must necessarily be followed by a hole.

\begin{figure}[!h]
        \centering
        \includegraphics[width=\columnwidth]{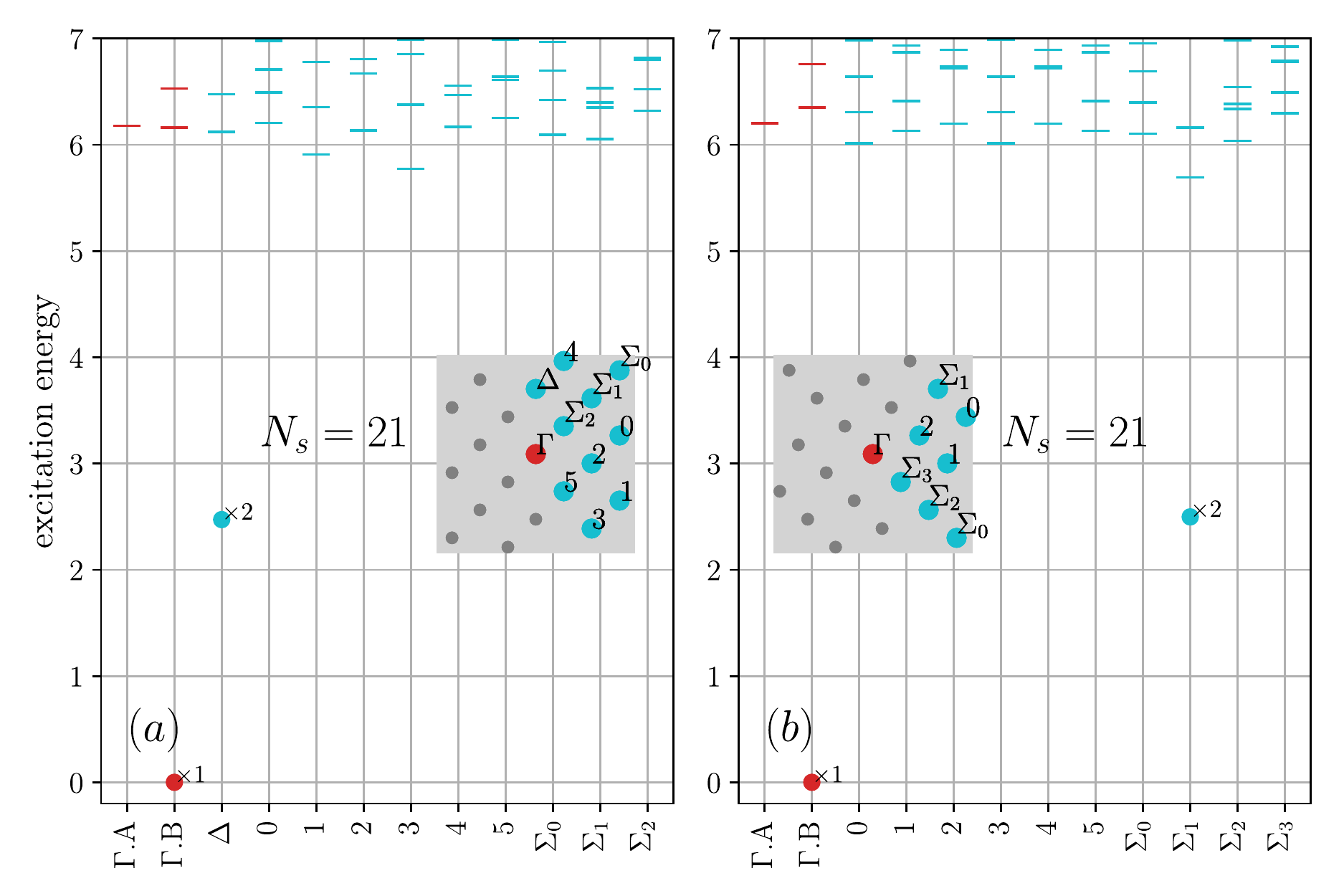}
        \caption{Low-energy spectra obtained from ED on additional $N_s=21$ clusters: (a) 21b and (b) 21a.  Each Brillouin zone is shown as inset. On these clusters, the ground-state is a global SU(3) singlet. We confirm the presence of two additional low-energy singlet states with the expected momenta,   see text.}
        \label{fig:ED21}
\end{figure}

\begin{figure}[!h]
        \centering
        \includegraphics[width=\columnwidth]{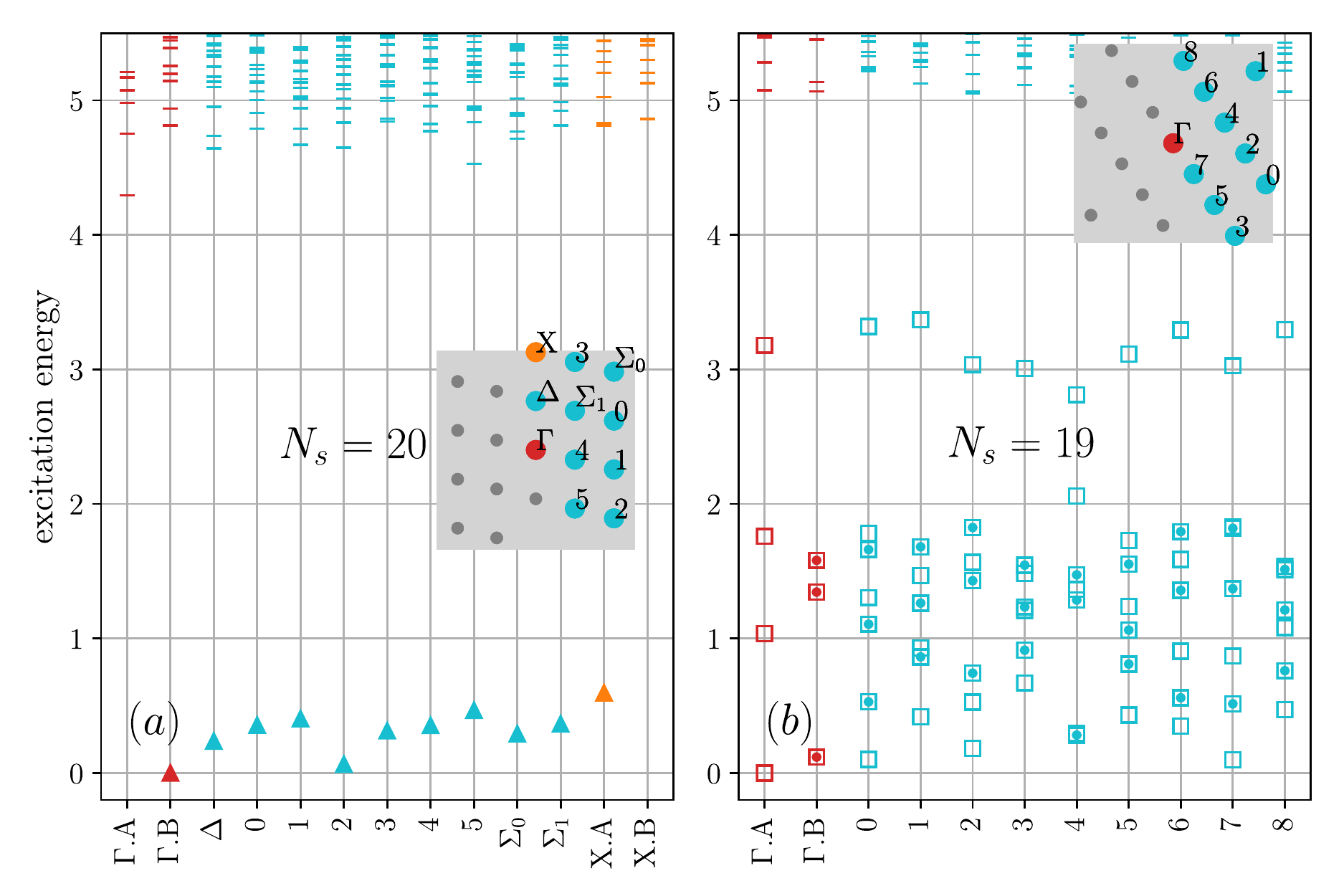}
        \caption{Low-energy spectra obtained from ED on clusters (a) 20b and (b) 19b corresponding to one or two quasi-hole excitations in the FQH language. As expected, we do observe a precise counting per momentum sector in the low-energy band, see text. In panel (a), the low-energy states (triangles) transform as $\overline{\bf{3}}$ irrep. In panel (b), the low-energy squares (respectively circles) correspond to weight sectors $(s^z_1,s^z_2)=(1/2,0)$ (respectively $(0,1)$); i.e., empty squares correpond to an irrep ${\bf 3}$ while squares filled with circles correspond to $\bf{\overline 6}$.}
        \label{fig:ED20}
\end{figure}

As a result, for all clusters $N_s=3p$ ranging from $N_s=12$ to $N_s=24$, we have confirmed that our model does indeed show a quasi three-fold degeneracy of the ground-state, and their quantum numbers are given by the above generalized exclusion principle. This is a non-trivial feature and it is different from what would be expected for a charge-density wave phase. For example, in the cluster 18a, the three states are found at momentum $\Gamma=(0,0)$, while in the cluster 18b, one state is found at $\Gamma$ and the two others at $\pm(2\pi/3,2\pi/3)$ as predicted. For $N_s=21$, in all three considered clusters, we also find three quasi-degenerate states with the correct momenta, see Fig.~1a in the main text and Fig.~\ref{fig:ED21}. 
Moreover, as shown in the main text, the ground-state energy shows a rather quick saturation with $N_s$, compatible with a gapped phase. 

Regarding the quasi-hole case, it can be obtained from clusters $N_s=3p-1$.  In such a case, the counting in the sector $(N_1,N_2,N_3)=(p,p,p-1)$ (which are three-fold degenerate due to SU(3) symmetry) is given by the generalized exclusion principle for spinful particles with $N_\uparrow=p$ and $N_\downarrow=p-1$ (number of holes being $p$). For example, for all clusters with $N_s=20$, we predict $N_s$ low-energy quasi-hole states, more precisely one (three-fold degenerate) per momentum sector using the heuristic rule~\cite{Regnault2011,Bernevig2012}, which is indeed observed in Fig.~\ref{fig:ED20}(a) (other data not shown), and each low-energy state transforms as a $\overline{\bf{3}}$ irrep.

For clusters $N_s=3p-2$, we expect that it would be best described by \emph{two} quasi-hole states since a single quasi-hole excitation is rather localized. As a result, we would observe low-energy excitations both in sectors $(N_1,N_2,N_3)=(p,p,p-2)$, equivalent to $(s^z_1,s^z_2)=(0,1)$, as well as $(p,p-1,p-1)$, equivalent to $(s^z_1,s^z_2)=(1/2,0)$, and their equivalent symmetry sectors. This is indeed what we have found: for instance in both clusters $N_s=19$, the counting in weight sector $(0,1)$ predicts 3 states per momentum and the one in sector $(1/2,0)$ leads to 7 states per momentum, which is indeed observed in Fig.~\ref{fig:ED20}(b) (other data not shown). Reconstructing the SU(3) irreps from the states quantum numbers, one then finds four ${\bf 3}$ and three ${\bf{\overline 6}}$ multiplets in each momentum sector, as predicted. 

\begin{figure}[!h]
        \centering
        \includegraphics[width=0.9\columnwidth]{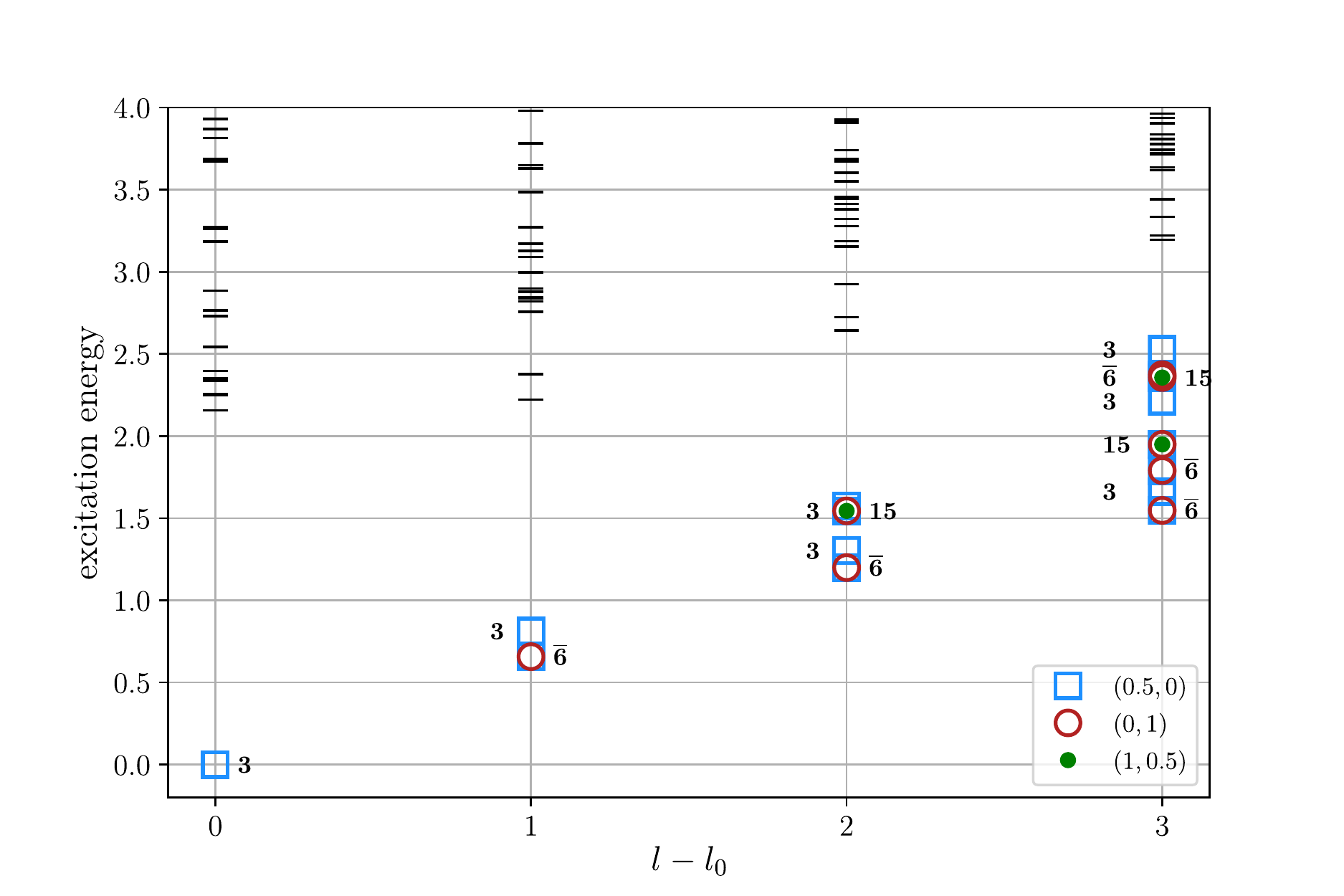}
        \caption{Low-energy spectrum obtained from ED on a $4\times 4$ ($N_s=16=3p+1$) cluster with open boundary conditions with respect to the angular momentum. Using the knowledge of the weights $(s^z_1,s^z_2)$ of the low-energy states, one can reconstruct the SU(3) irreps which are in perfect agreement with the expected counting for a single chiral branch, see main text and Tab.~\ref{table:su3_1}.}
        \label{fig:ED16obc}
\end{figure}

As an additional numerical evidence for the existence of the SU$(3)_1$ CSL, we have computed the low-energy spectrum of a $4\times 4$ cluster with open boundary conditions which should host a single chiral branch at its edge. Since the number of sites $N_s=16=3p+1$, the ground-state transforms as a ${\bf 3}$ irrep. As shown in Fig.~\ref{fig:ED16obc}, the low-energy spectrum perfectly agrees with the Virasoro levels of the SU$(3)_1$ CFT (see main text and Tab.~\ref{table:su3_1}). Note that the levels are plotted vs the relative angular momentum since the ground-state is found in a non-zero angular momentum sector with $\l_0=1$, similar to the momentum shift found in the PEPS calculation, see main text.

\begin{figure*}[!ht]
        \centering
        \includegraphics[width=0.9\textwidth]{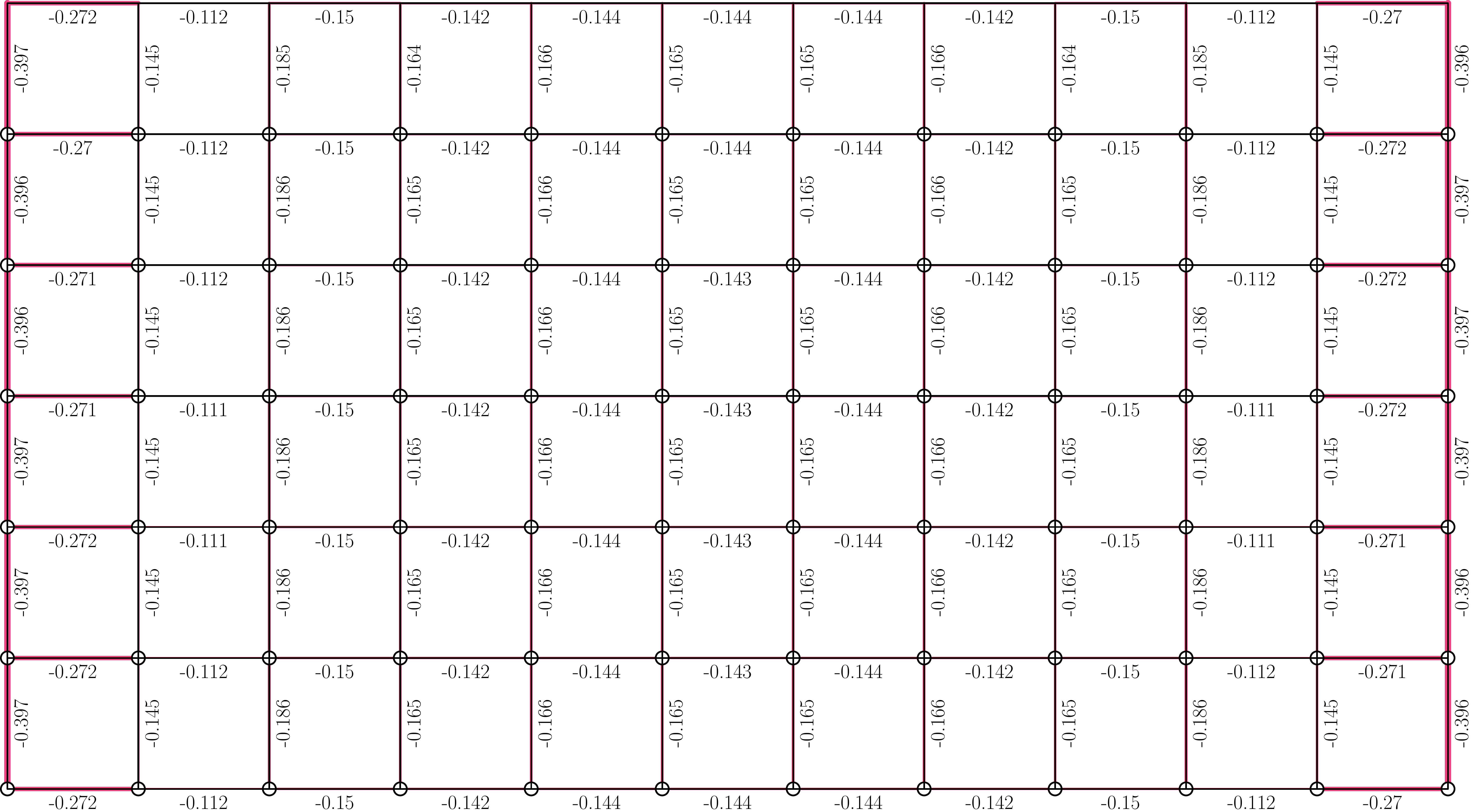}
        \caption{Local bond strengths $\langle P_{ij} \rangle$ obtained from DMRG simulation on a $12\times 6$ cylinder with open/periodic boundary conditions in the long/short directions.}
        \label{fig:SM_DMRG}
\end{figure*}

\section{Relevant details of DMRG method}
\label{sec:DMRG}

For DMRG, we have computed the ground-state wavefunction on various cylinders $N_s=N_h \times N_v$ (with open/periodic boundary conditions in the long/short direction). We have used explicitly the two U(1) quantum numbers to ease convergence. Using up to $m=4\, 000$ states, we can obtain reliable energies (discarded weight below $5\text{e-}5$) up to $N_v=6$. In order to stabilize a global SU(3) singlet ground-state, we have chosen the system size $N_s$ multiple of 3, more specifically we took $N_h=3p$ as the integer closest to $2N_v$.

By computing the total energy for cylinders $N_h\times N_v$ and $(N_h+3)\times N_v$, we can obtain an accurate estimate of the ground-state energy density (per site) by subtraction, providing the data plotted in Fig.~2b in the main article. Quite remarkably, there is a very fast convergence since all data for $N_v=4$, $5$ or $6$ are compatible with a ground-state energy density $e_0=-2.05(1)$, very close to the ED estimate.

In Fig.~\ref{fig:SM_DMRG}, we plot the bond strengths $\langle P_{ij} \rangle$ on nearest-neighbor bonds which do not show any modulation at all. Moreover, we have measured the local Cartan $(s_1^z,s_2^z)$ average values and found that they are vanishing (below $1\text{e-}6$). All these measurements are indicative of a featureless phase. 

Since our model possesses one SU(3) fermion per unit cell (equivalent to 1/3 filling), a trivial featureless gapped ground-state is impossible according to the Lieb-Schultz-Mattis theorem for SU(N) spin systems and its generalization to two dimensions~\cite{Lieb1961,Oshikawa2000,Hastings2005,Totsuka2017}. Therefore, our ED and DMRG data are suggestive of a gapped topological phase.

\section{SU(3) symmetric PEPS on the square lattice}
\label{sec:PEPSansatz}

Here we present details about the symmetric PEPS construction, following the same spirit of Ref.~\onlinecite{Mambrini2016, Poilblanc2017b, Chen2018b}.

To construct a faithful PEPS representation of a chiral spin liquid wave function, we could encode the symmetry property of the desired wave function into local tensors. On the microscopic lattice scale, the symmetries that we need to take into account are: (1) the wave function $|\psi\rangle$ is invariant under global SU($3$) rotations, i.e., it is a SU($3$) singlet; (2) under one-site translation and $\pi/2$ lattice rotation, $|\psi\rangle$ is invariant up to a phase; (3) under lattice reflection P or time-reversal action T, $|\psi\rangle$ is transformed into its complex conjugate $|\psi\rangle\rightarrow|\bar{\psi}\rangle$ (also up to a possible phase), but is invariant under their combination PT. These symmetry requirements can be fulfilled by taking a suitable unit-cell of tensors, where these tensors satisfy certain symmetry constraints. 

To implement global SU($3$) symmetry, the PEPS wave function can take a form where a virtual SU($3$) singlet formed by two virtual spins, denoted as $|\Omega\rangle$, is put on every bond, and the on-site tensor $\mathcal{P}$ does a projection from four virtual spins on every site into the physical spin. Using translation symmetry, we can put the same virtual singlet on every bond, and the same local projector on every site, so as to work directly in the thermodynamic limit. 
The wave function then takes a simple form:
\begin{equation}
\label{eq:wavefunction}
|\psi\rangle = \prod_i\left(\sum_{lrud}^{s}\mathcal{P}^{s}_{lrud}|s\rangle\langle lrud|\right)_i|\Omega\rangle\otimes\cdots\otimes|\Omega\rangle,
\end{equation}
where $i$ stands for site index, and $lrud$, $s$ is for left, right, up, down virtual, and physical spin on every site, respectively. The lattice point group symmetry imposes strong constraints on these local tensors, namely, up to a phase, the local projector $\mathcal{P}$ should be invariant under $\pi/2$ lattice rotation, and becomes complex conjugate under reflection, and the virtual singlet $|\Omega\rangle$ should be invariant under reflection.

One noticeable difference between SU($3$) (more generally SU($N$)) and SU(2) group is that SU($2$) group is self-conjugate such that one can form a singlet with two spins carrying the same irrep. Such a property is absent with general SU($3$) spins, and we need to combine two spins carrying irreps with opposite $\mathbb{Z}_3$ charges to form the singlet. The virtual space we use in this work, $\mathcal{V}=\bf{3}\oplus\bf{\overline{3}}\oplus 1$ with bond dimension $D=7$, satisfies this SU(3) symmetry requirement, and allows us to construct the virtual singlet:
\begin{equation}
\label{eq:virtualSinglet}
\begin{split}
|\Omega\rangle &= |{\bf 3},1\rangle \otimes|{\bf\overline 3},3\rangle - |{\bf 3},2\rangle \otimes|{\bf\overline 3},2\rangle +|{\bf 3},3\rangle \otimes|{\bf\overline 3},1\rangle\\
 & + |{\bf\overline 3},1\rangle \otimes|{\bf 3},3\rangle - |{\bf\overline 3},2\rangle \otimes|{\bf 3},2\rangle + |{\bf\overline 3},3\rangle \otimes|{\bf 3},1\rangle \\
 & + |{\bf 1},1\rangle \otimes|{\bf 1},1\rangle,
\end{split}
\end{equation}
where the labeling of basis for each irrep follows Tab.~\ref{table:su3multiplets}. This (unnormalized) maximally entangled virtual singlet is indeed symmetric under reflection. Unlike the SU($2$) case, where using on-site unitary transformation on one sublattice, one can transform bond singlet into an identity matrix without changing the on-site projector, the same trick cannot be applied to SU($3$) virtual singlet $|\Omega\rangle$.
Nevertheless, we can absorb the two neighboring bond matrices into the on-site projector $\mathcal{P}$, e.g. the right and down one, forming the tensor $\mathcal{A}$, without enlarging the unit cell. This strategy is taken in the numerical calculation in this work.

To systematically construct the on-site projector, we first did a classification of the rank-5 tensor. According to the fusion rules of SU($3$), we expand $\mathcal{V}^{\otimes4}$ onto the irreps of SU($3$) (see Tab.~\ref{table:tensors1}). Considering the physical variable $\bf 3$ (see second row of Tab.~\ref{table:tensors1}), relevant occupation numbers ($n_{\rm occ}$) are then determined. For each occupation number channel, the highest weight states (corresponding to ${\bf S}^z = (1/2, 0)$) are expressed in the tensor product basis of $\mathcal{V}^{\otimes4}$. A point group analysis ($C_{\rm4v}$) is then performed (see Tab.~\ref{table:tensors2}), and highest weight states are symmetrized accordingly. Lower weight states (namely ${\bf S}^z = (-1/2, 1/2)$ and ${\bf S}^z = (0, -1/2)$) are determined using lowering operators expressed in the tensor product basis of $\mathcal{V}^{\otimes4}$.

\begin{table*}[!htb]
	\renewcommand\arraystretch{1.5}
$\begin{array}{cc}
	\hline\hline\vspace{-0.2cm}\\ 
	{\bf 0} 			& {\it 1}\times{ \{ 0,0,4\}} + {\it 4}\times { \{ 0,3,1\}} + {\it 12}\times { \{ 1,1,2\}} +{\it 12}\times { \{ 2,2,0\}} +{\it 4}\times { \{ 3,0,1\}}  \\
	{\bf 3} 			& {\it 6}\times { \{ 0,2,2\}} +{\it 4}\times { \{ 1,0,3\}} +{\it 12}\times { \{ 1,3,0\}} +{\it 24}\times { \{ 2,1,1\}} +{\it 3}\times { \{ 4,0,0\}}  \\
	{\bf\overline{3}}   & {\it 4}\times { \{ 0,1,3\}} +{\it 3}\times { \{ 0,4,0\}} +{\it 24}\times { \{ 1,2,1\}} +{\it 6}\times { \{ 2,0,2\}} +{\it 12}\times { \{ 3,1,0\}}  \\
	{\bf 6} 			& {\it 2}\times { \{ 0,4,0\}} +{\it 12}\times { \{ 1,2,1\}} +{\it 6}\times { \{ 2,0,2\}} +{\it 12}\times { \{ 3,1,0\}}  \\
	{\bf\overline{6}} 	& {\it 6}\times { \{ 0,2,2\}} +{\it 12}\times { \{ 1,3,0\}} +{\it 12}\times { \{ 2,1,1\}} +{\it 2}\times { \{ 4,0,0\}}  \\
	{\bf 8} 			& {\it 8}\times { \{ 0,3,1\}} +{\it 12}\times { \{ 1,1,2\}} +{\it 24}\times { \{ 2,2,0\}} +{\it 8}\times { \{ 3,0,1\}}  \\
	{\bf 10} 			& {\it 6}\times { \{ 2,2,0\}} +{\it 4}\times { \{ 3,0,1\}}  \\
	{\bf\overline{10}} 	& {\it 4}\times { \{ 0,3,1\}} +{\it 6}\times { \{ 2,2,0\}}  \\
	{\bf 15} 			& {\it 8}\times { \{ 1,3,0\}} +{\it 12}\times { \{ 2,1,1\}} +{\it 3}\times { \{ 4,0,0\}}  \\
	{\bf\overline{15}} 	& {\it 3}\times { \{ 0,4,0\}} +{\it 12}\times { \{ 1,2,1\}} +{\it 8}\times { \{ 3,1,0\}}  \\
	{\bf 15'} 			& {\it 1}\times{ \{ 4,0,0\}}  \\	
	{\bf\overline{15'}} & {\it 1}\times{ \{ 0,4,0\}}  \\
	{\bf 24} 			& {\it 4}\times { \{ 1,3,0\}}  \\
	{\bf\overline{24}} 	& {\it 4}\times { \{ 3,1,0\}}  \\
	{\bf 27} 			& {\it 6}\times { \{ 2,2,0\}}  \\\hline\hline
\end{array}$
\caption{Decomposition of $\left ({\bf 3}\oplus{\bf\overline{3}}\oplus{\bf 1} \right )^{\otimes4}$ onto the irreps of SU($3$) (first column). Triplets of integers in brackets denote the occupation numbers of species ${\bf 3}$, ${\bf\overline{3}}$ and ${\bf 1}$. Italic integers are the multiplicities and hence the number of tensors generated in each occupation number channel. The second row, i.e. the linear maps (named projectors) onto the physical irrep $\bf 3$ is relevant to this work.}
\label{table:tensors1}
\end{table*}

\begin{table}[!htb]
	\begin{tabularx}{0.498\textwidth}{p{0.12\linewidth}p{0.38\linewidth}p{0.12\linewidth}p{0.38\linewidth}}
		\hline
		\hline
		{\small $n_{\rm occ}$}   & {\small Multiplicities / Pg}&{\small $n_{\rm occ}$}  & {\small Multiplicities / Pg} \\
		\hline
		\{4,0,0\} &
		$\begin{array}{ll}
		1 & \mathcal{A}_2 \\
		1 & \mathcal{E}^{(1)} \\
		1 & \mathcal{E}^{(2)} \\
		\end{array}$
		& \{1,0,3\} &
		$\begin{array}{ll}
		1 & \mathcal{A}_1 \\
		1 & \mathcal{B}_1 \\
		1 & \mathcal{E}^{(1)} \\
		1 & \mathcal{E}^{(2)} \\
		\end{array}$
		\\ \hline
		\{0,2,2\} &
		$\begin{array}{ll}
		1 & \mathcal{A}_2 \\
		1 & \mathcal{B}_1 \\
		2 & \mathcal{E}^{(1)} \\
		2 & \mathcal{E}^{(2)} \\
		\end{array}$
		& \{1,3,0\} &
		$\begin{array}{ll}
		1 & \mathcal{A}_1 \\
		2 & \mathcal{A}_2 \\
		1 & \mathcal{B}_1 \\
		2 & \mathcal{B}_2 \\
		3 & \mathcal{E}^{(1)} \\
		3 & \mathcal{E}^{(2)} \\
		\end{array}$
		\\ \hline
		\{2,1,1\} &
	$	\begin{array}{ll}
		3 & \mathcal{A}_1 \\
		3 & \mathcal{A}_2 \\
		3 & \mathcal{B}_1 \\
		3 & \mathcal{B}_2 \\
		6 & \mathcal{E}^{(1)} \\
		6 & \mathcal{E}^{(2)} \\
		\end{array}$
		&  & \\
		\hline
		\hline
	\end{tabularx}
	\caption{$C_{\rm4v}$ point group (Pg) analysis in each relevant occupation number channel for all the linear maps (projectors) onto the physical irrep $\bf 3$ (see second line of Tab.~\ref{table:tensors1}).}
\label{table:tensors2}
\end{table}

\begin{figure}[!htb]
\centering
	\includegraphics[width=0.99\columnwidth]{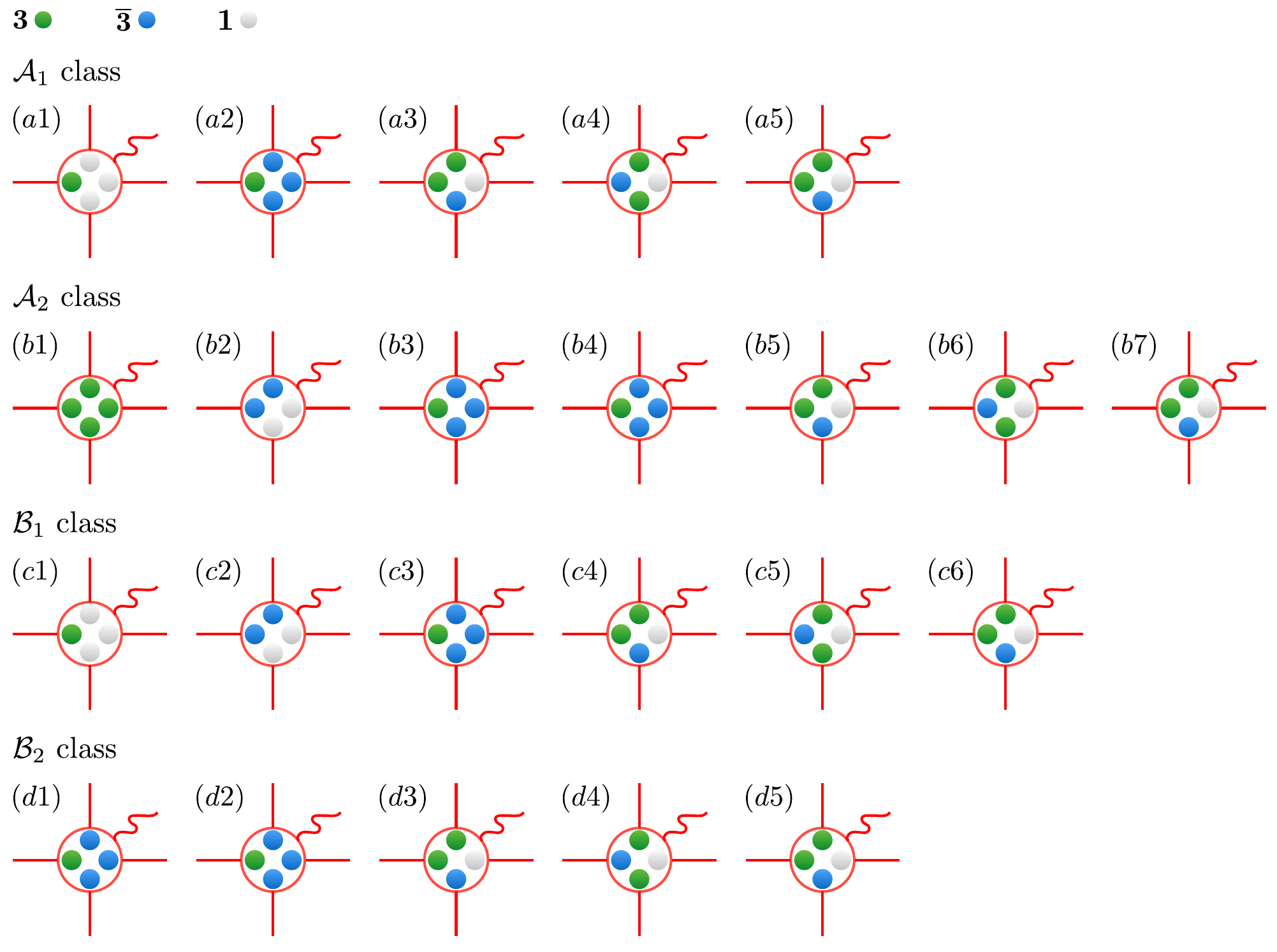}
\caption{Illustration for the classification of local projectors (i.e. linear maps). For bond dimension $D=7$ with virtual space $\mathcal{V}=\bf{3}\oplus\bf\overline{3}\oplus\bf{1}$, the local projectors
have been classified according to the irreps of the point group $C_{\rm4v}$. Class $\mathcal{A}_1/\mathcal{A}_2/\mathcal{B}_1/\mathcal{B}_2$ has $5/7/6/5$ linearly independent tensors, whose typical configurations are shown in $(a1)-(a5)$, $(b1)-(b7)$, $(c1)-(c6)$, $(d1)-(d5)$, separately. Other configurations of the elementary tensors can be obtained through lattice rotation and reflection, along with the corresponding characters. Each elementary tensor in every class has a fixed occupation number $n_{\rm occ}$, which can be directly read out. For some configurations, the number of linearly independent tensors is larger than one, due to multiple SU(3) fusion channels (see e.g.  $(a3)$ and $(a5)$).} 
\label{fig:classification}
\end{figure}

As a result of the classification, the local projectors are now classified according to irreps of the square lattice point group $C_{4\rm v}$, denoted as $\mathcal{A}_1, \mathcal{A}_2, \mathcal{B}_1, \mathcal{B}_2$ and $\mathcal{E}$. One can then construct the on-site projector $\mathcal{P}$ by linearly combining different classes of tensors, such that it is invariant under $\pi/2$ lattice rotation but becomes its complex conjugate upon reflection (up to an irrelevant phase). One choice we considered is: 
\begin{equation}
\label{eq:pepsa}
\mathcal{P} = \mathcal{B}_1 + i\mathcal{B}_2 = \sum_{a=1}^{N_1} \lambda_1^a\mathcal{B}_1^a + i\sum_{b=1}^{N_2} \lambda_2^b\mathcal{B}_2^b,
\end{equation}
where $\{\lambda_1^a,\lambda_2^b\}$ are real coefficients, as mentioned in the main text. Here $N_1=6$, $N_2=5$ is the number of tensors in the $\mathcal{B}_1$ and $\mathcal{B}_2$ classes, respectively. We note that, one could also use $\mathcal{A}_1$ and $\mathcal{A}_2$ classes to build chiral PEPS~\cite{Chen2018b}, whose energy turns out to be significantly higher than Eq.~\eqref{eq:pepsa} (data not shown). Thus we do not examine the detailed property of the later.

The expressions for the classes of tensor considered in this work are provided in Sec.~\ref{sec:tensorExpression}. See also Fig.~\ref{fig:classification} for a pictorial illustration.

\section{CTMRG method and variational optimization}
\label{sec:CTMRG}

For completeness, here we briefly describe the specific CTMRG method we used in this work, which follows Ref.~\onlinecite{Corboz2014} and is further simplified in Ref.~\onlinecite{Fishman2018}.

\begin{figure}[!hbt]
	\centering
	\includegraphics[width=0.99\columnwidth]{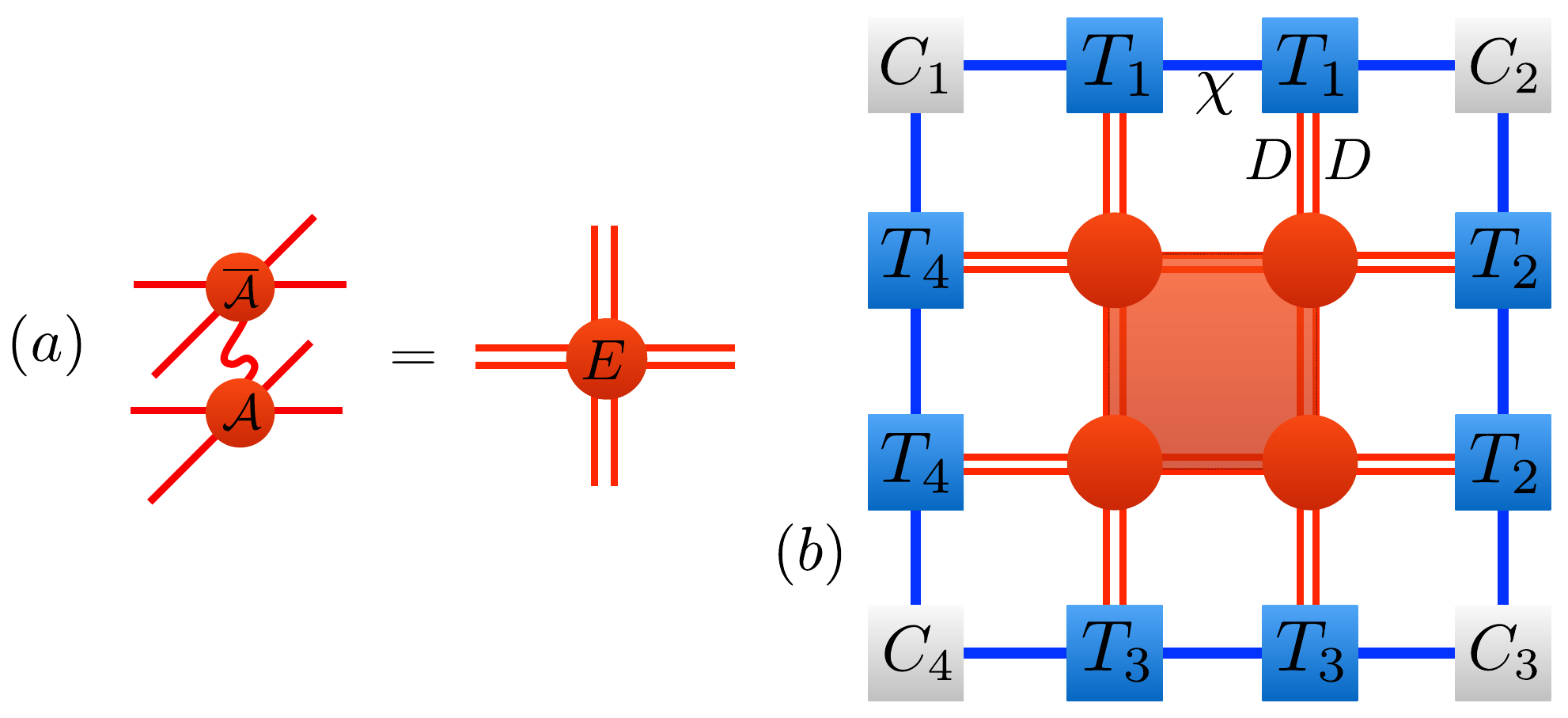}
	\caption{(a) The double tensor $E$ is obtained by contracting tensor $\mathcal{A}$ and its complex conjugate over the physical index. (b) The environment tensors $\{C_i, T_i\} (i=1,2,3,4)$ surrounding the $2\times2$ central region, computed with CTMRG method.}
\label{fig:CTMRG}
\end{figure}


In our setting, for tensor network of the wave function norm, the unit cell contains only one tensor, denoted as $E$ (see Fig.~\ref{fig:CTMRG}(a)), which is obtained by contracting tensor $\mathcal{A}$ and its complex conjugate over the physical index. The CTMRG method allows us to approximately contract the whole network on the infinite plane by computing the effective environment tensors surrounding the unit cell. In our case, the environment tensors are composed by corner tensors and edge tensors $\{C_i, T_i\} (i=1,2,3,4)$, see Fig.~\ref{fig:CTMRG}(b) for graphic notation. The accuracy of CTMRG method is controlled by the environment bond dimension, denoted as $\chi$, and typically we choose $\chi=kD^2 (k\in \mathbb{N^{+}})$. In the CTMRG procedure, we dynamically increase $\chi$ by a small amount to keep the complete SU($3$) multiplet structure. To further speed up the CTMRG procedure, we have explicitly kept track of the first U($1$) quantum number of the SU($3$) multiplets~\cite{Bauer2011}.

We note that, although the wave function has certain lattice symmetries, we do not use them in the CTMRG procedure, since after absorbing the bond singlet into on-site projector to construct tensor $\mathcal{A}$, the tensor $\mathcal{A}$ is not invariant under $\pi/2$ lattice rotation. As a result, the four corner (edge) tensors $\{C_i\} (i=1,2,3,4)$ ($\{T_i\} (i=1,2,3,4)$) are not necessarily the same. Nevertheless, we have checked that the physical observables, e.g., correlation lengths, along the horizontal and vertical directions are the same, as expected.

For a given set of variational parameters $\{\lambda_1^a, \lambda_2^b\}$, we can now compute the energy density with the environment tensors, simply by inserting the identity operator or local Hamiltonian terms in the central region, see Fig.~\ref{fig:CTMRG}(b). Energy gradient can then be easily obtained by finite difference method, which is feasible due to the significantly reduced number of variational parameters (compare to general PEPS ansatz). The conjugate-gradient method~\cite{NoceWrig06} is then utilized to find the variational optimal parameters. In practice, this optimization procedure is carried out with $\chi=D^2$. Then we evaluate the energy density of the optimized ansatz with several larger $\chi=kD^2 (k=2,...,6)$ and eventually extrapolate to the $\chi\to\infty$ limit.

\section{Additional data for entanglement spectrum}
\label{sec:ES}

The entanglement property of PEPS can be most easily characterized by studying the entanglement spectrum on finite width cylinders, which is defined to be minus log of the spectrum of reduced density matrix (RDM) of subsystem~\cite{Li2008}, say the left half of the cylinder. For PEPS on an infinitely long cylinder, the RDM can be constructed from the leading eigenvector of the transfer operator through the relation:
\begin{equation}
\rho=U\sqrt{\sigma_L}\sigma_R^T\sqrt{\sigma_L}U^\dag,
\end{equation}
where U is an isometry relating the physical degrees of freedom to the virtual ones~\cite{Cirac2011}, and we have adopted the convention that the first index of $\sigma_{L,R}$ is in the bra layer. This RDM further shares the same spectrum as $\rho=\sigma_L\sigma_R^T$, which we diagonalize to get information about edge properties. 

\begin{figure}[!htb]
\centering
	\includegraphics[width=0.9\columnwidth]{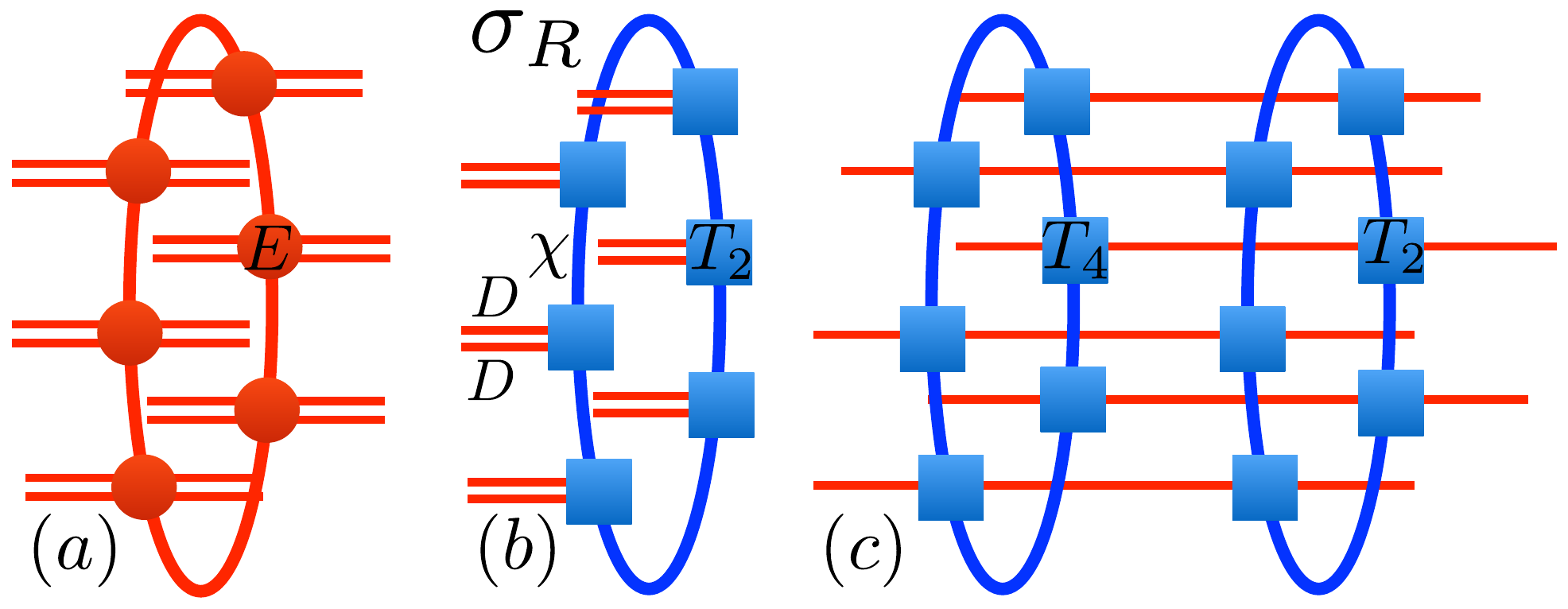}\\
\caption{Using CTMRG environment tensors to construct RDM for the left half. (a) shows the transfer operator on a width $N_v=6$ cylinder, whose right leading eigenvector $\sigma_R$ is approximated by a ring of $T_2$ tensors, shown in (b). Similarly for the left leading eigenvector $\sigma_L$. The RDM is then obtained by contracting a ring of $T_4$ and $T_2$ tensors, shown in (c).
}
\label{fig:RDM}
\end{figure}

As mentioned in the main text, for our case with bond dimension $D=7$, it is not feasible to compute $\sigma_{L,R}$ exactly, except for cylinders with small width. Instead, we can use the environment tensors computed from CTMRG method to approximate the $\sigma_{L,R}$, see Fig.~\ref{fig:RDM} for illustration. This is justified by the fact that CTMRG method is essentially approximating the fixed-point of certain transfer operator with matrix product state formed by environment tensors. One advantage of this approach is that, using $\sigma_{L,R}$ constructed in this way one can find RDM in all different charge sectors simultaneously, while with exact contraction one has to find them separately.

\begin{figure*}[htb]
\centering
	\subfloat{\includegraphics[width=0.655\columnwidth]{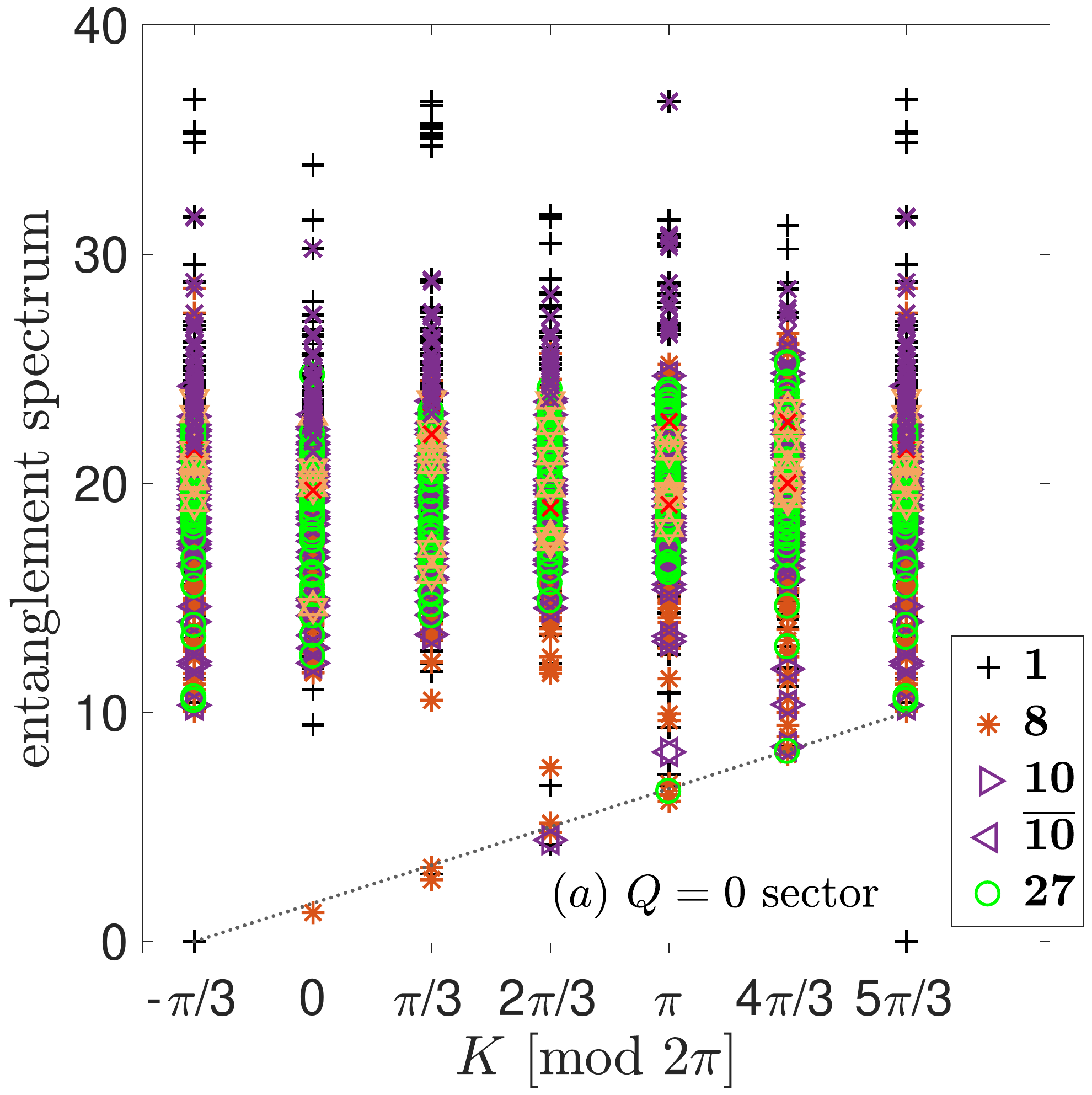}}
	\subfloat{\includegraphics[width=0.66\columnwidth]{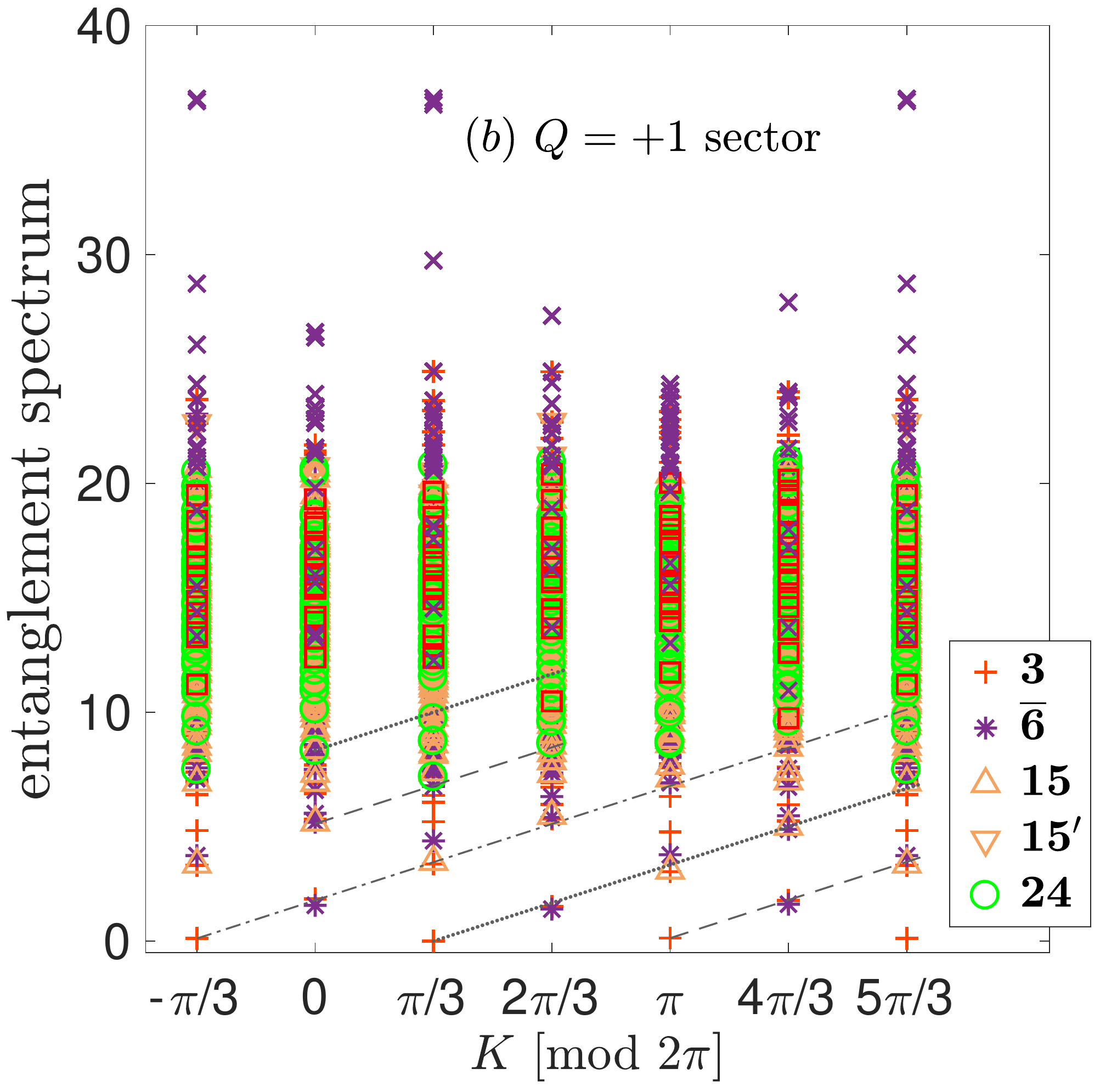}}
	\subfloat{\includegraphics[width=0.66\columnwidth]{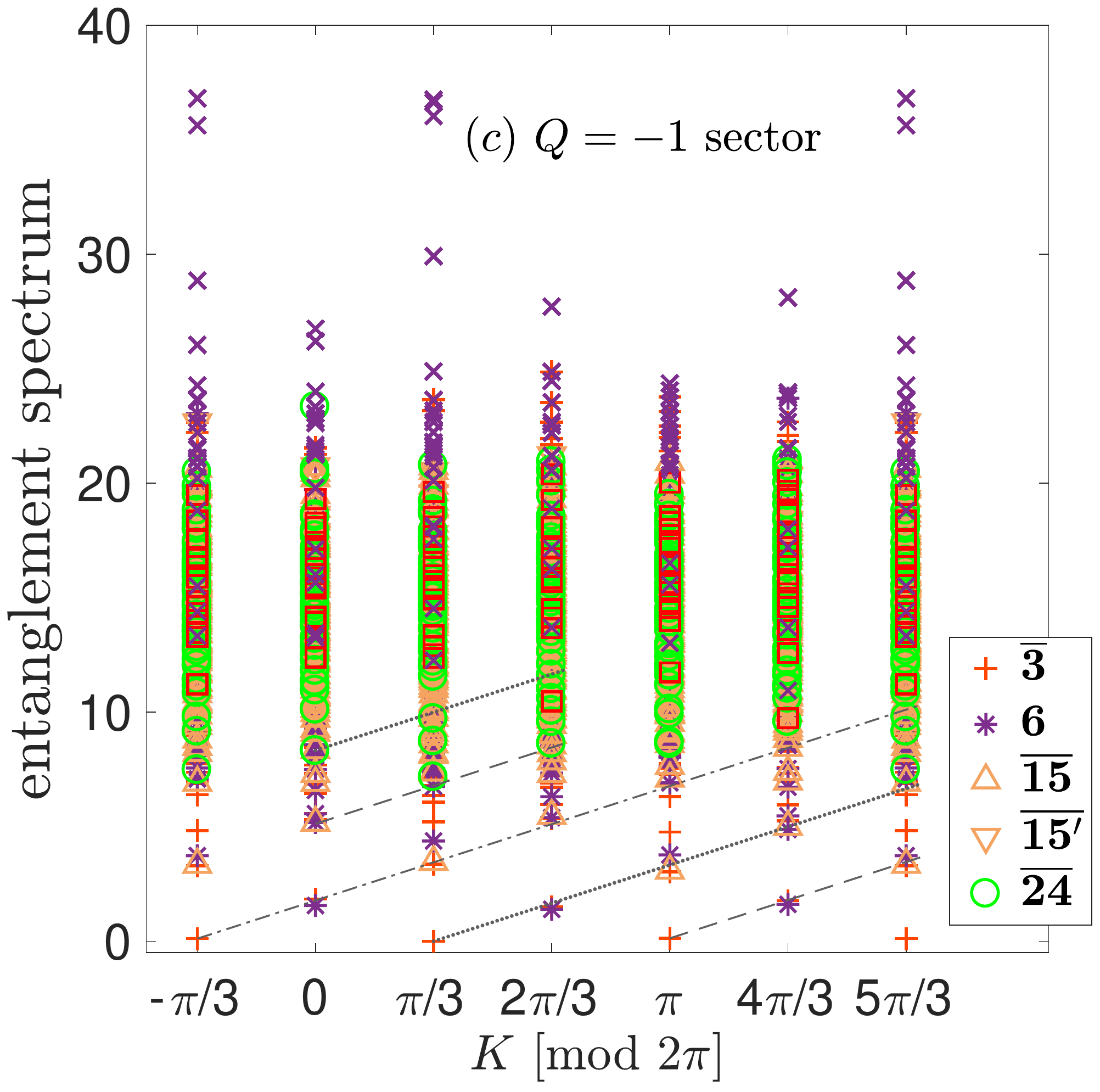}}
\caption{Complete entanglement spectrum on a $N_v=6$ infinitely long cylinder, computed with $\chi=343$. Linear dispersing modes can be clearly seen in the low energy spectrum, with one branch for the charge $Q=0$ sector, indicated as dotted line in (a), and three branches for charge $Q=\pm 1$ sectors, shown as dotted, dashed, dash-dotted line in (b) and (c), separately. The spectra in the charge $Q=+1$ sector is degenerate with the $Q=-1$ sector, except very small difference at highest energy level. See text for further descriptions.}
\label{fig:fullSpectrum}
\end{figure*}

\begin{figure*}[!htb]
\centering
	\subfloat{\includegraphics[width=0.49\columnwidth]{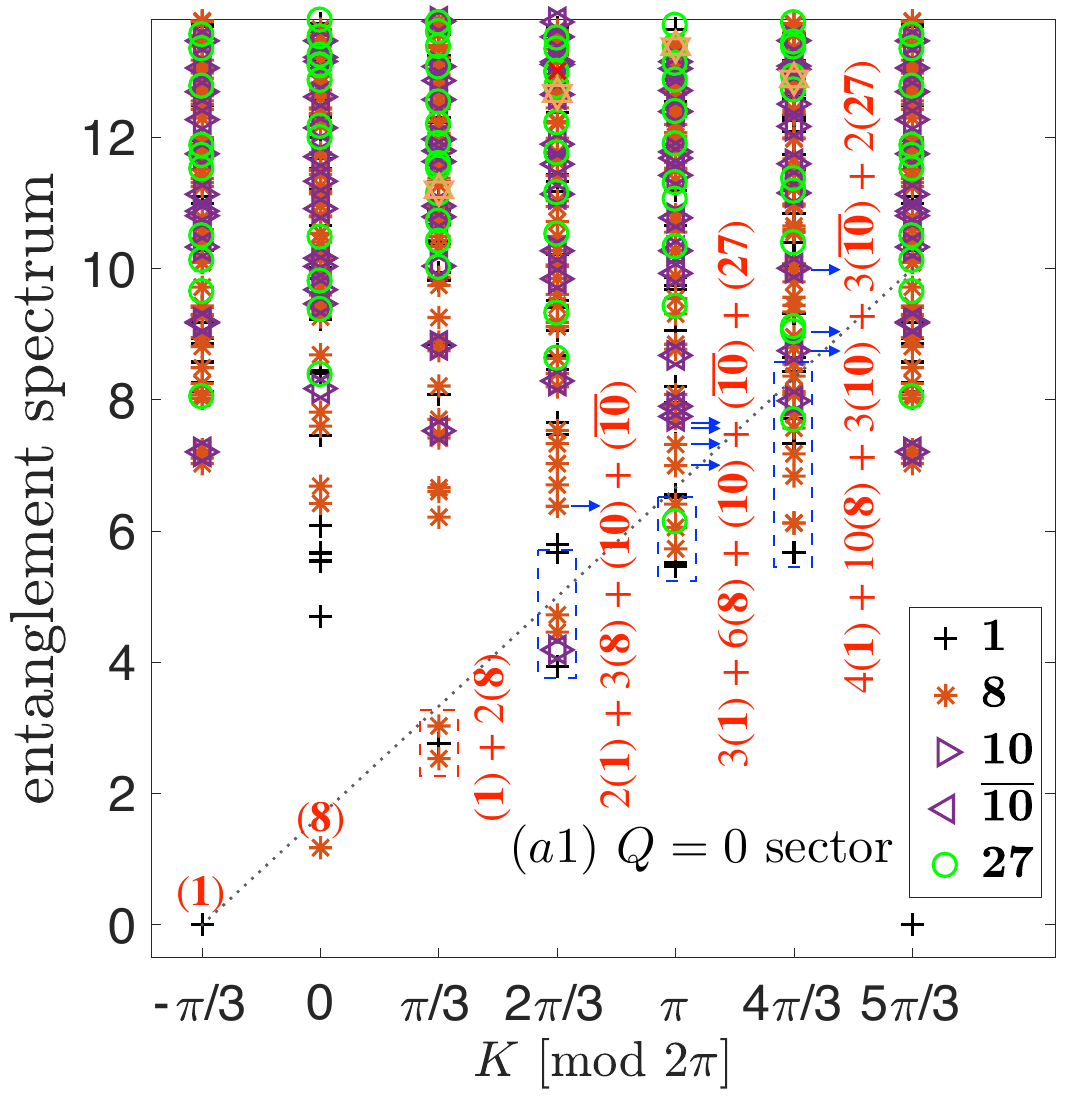}}
	\subfloat{\includegraphics[width=0.49\columnwidth]{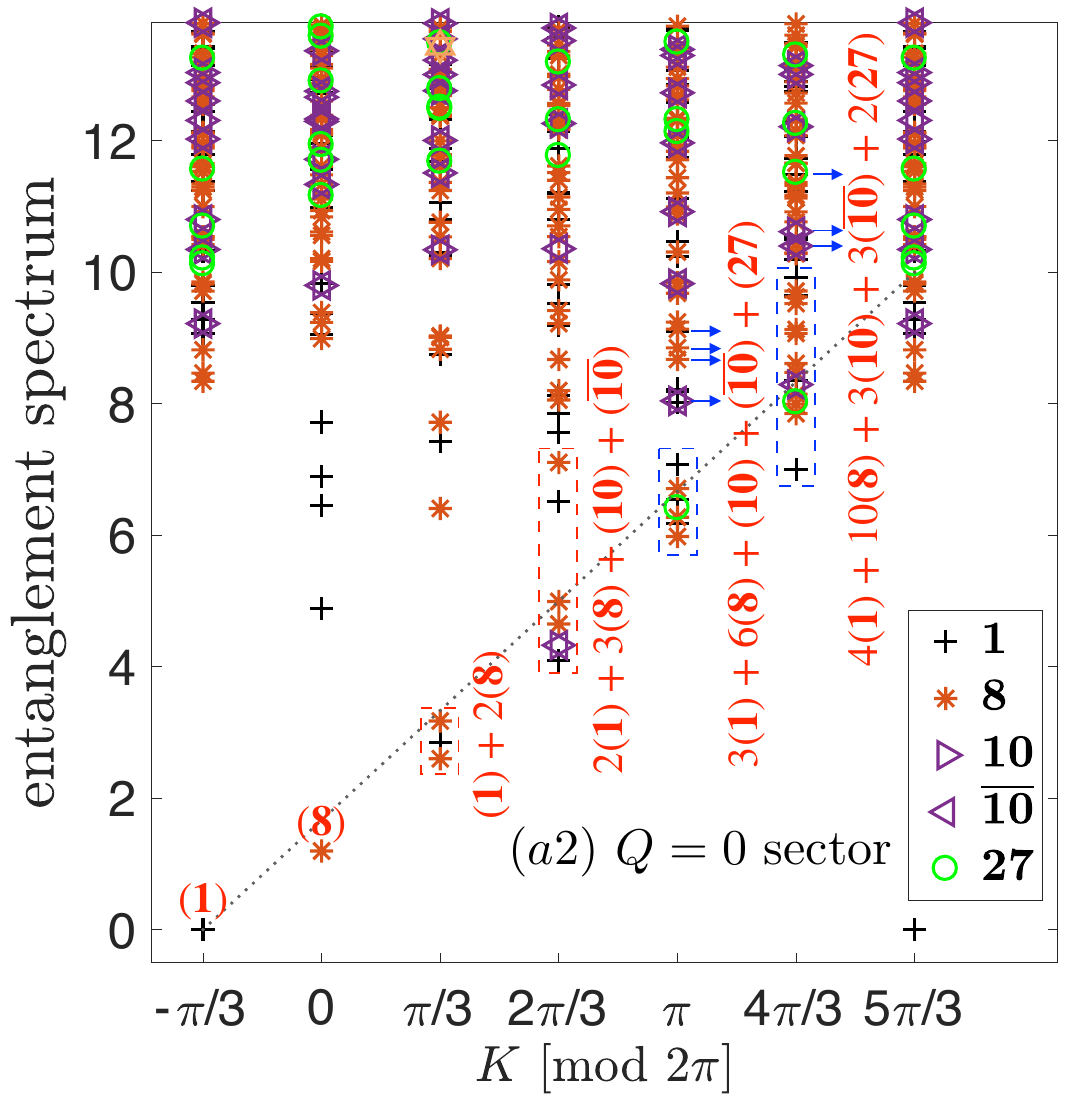}}
	\subfloat{\includegraphics[width=0.49\columnwidth]{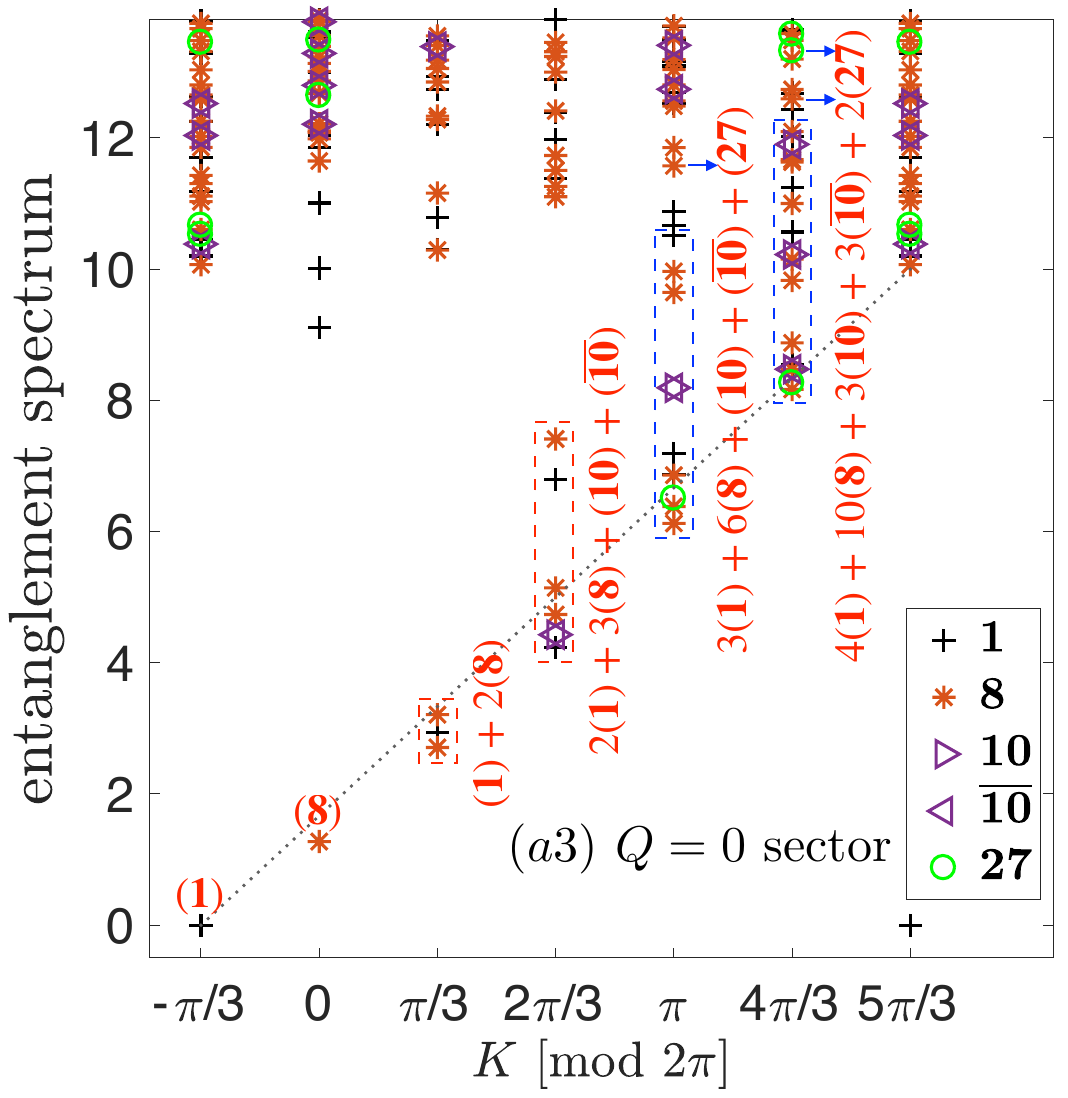}}
	\subfloat{\includegraphics[width=0.49\columnwidth]{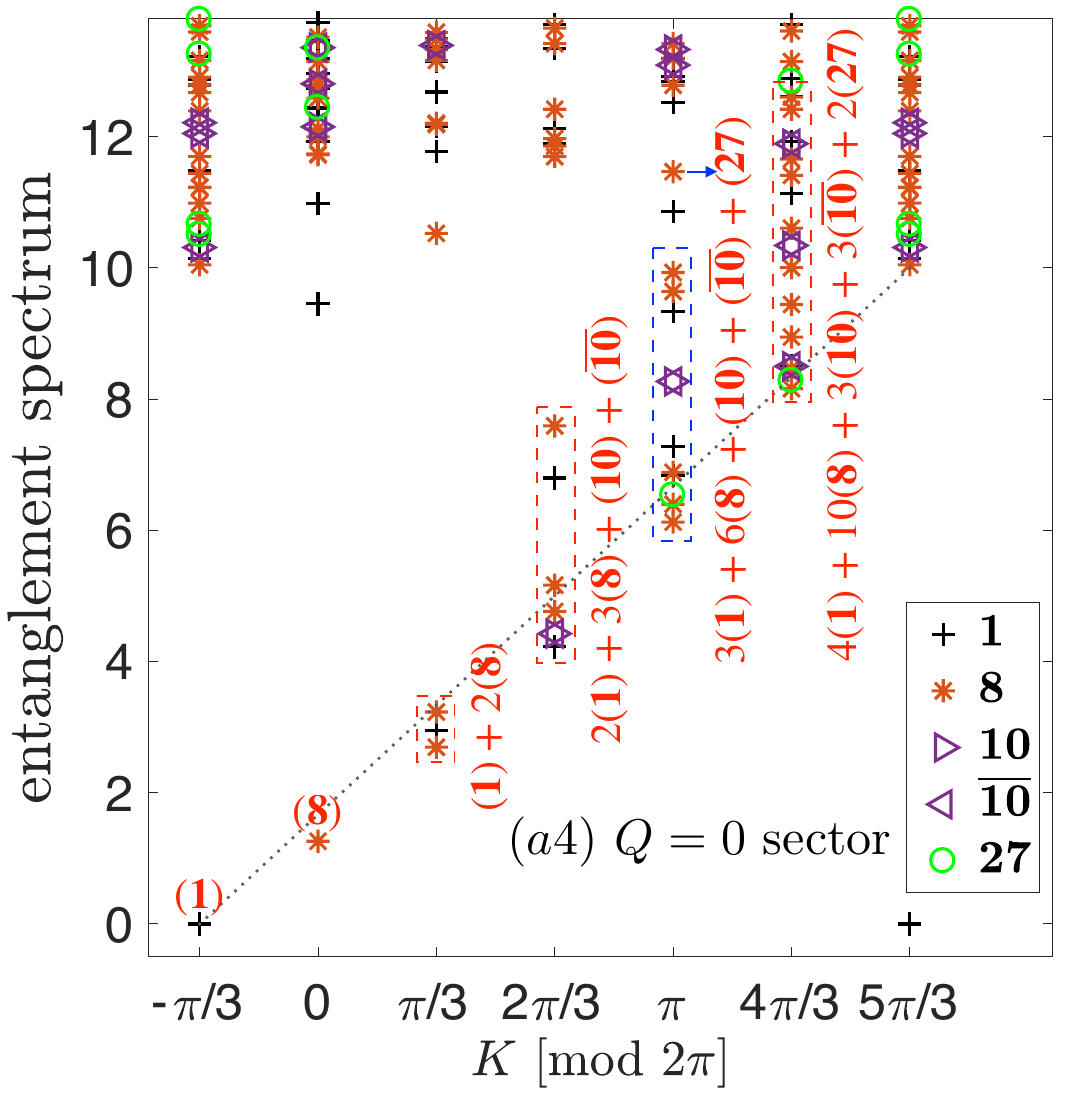}}\\
	\subfloat{\includegraphics[width=0.49\columnwidth]{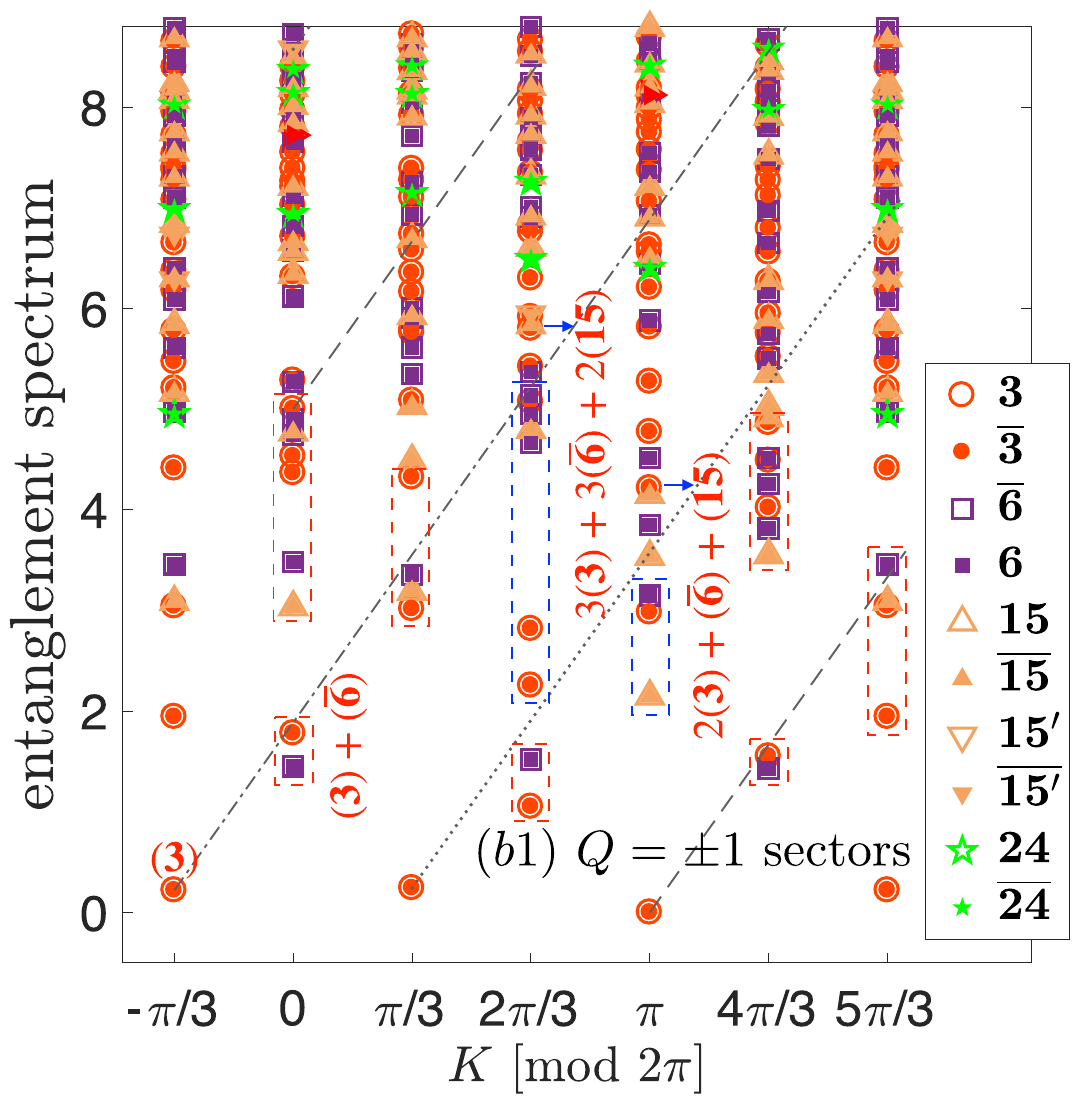}}
	\subfloat{\includegraphics[width=0.49\columnwidth]{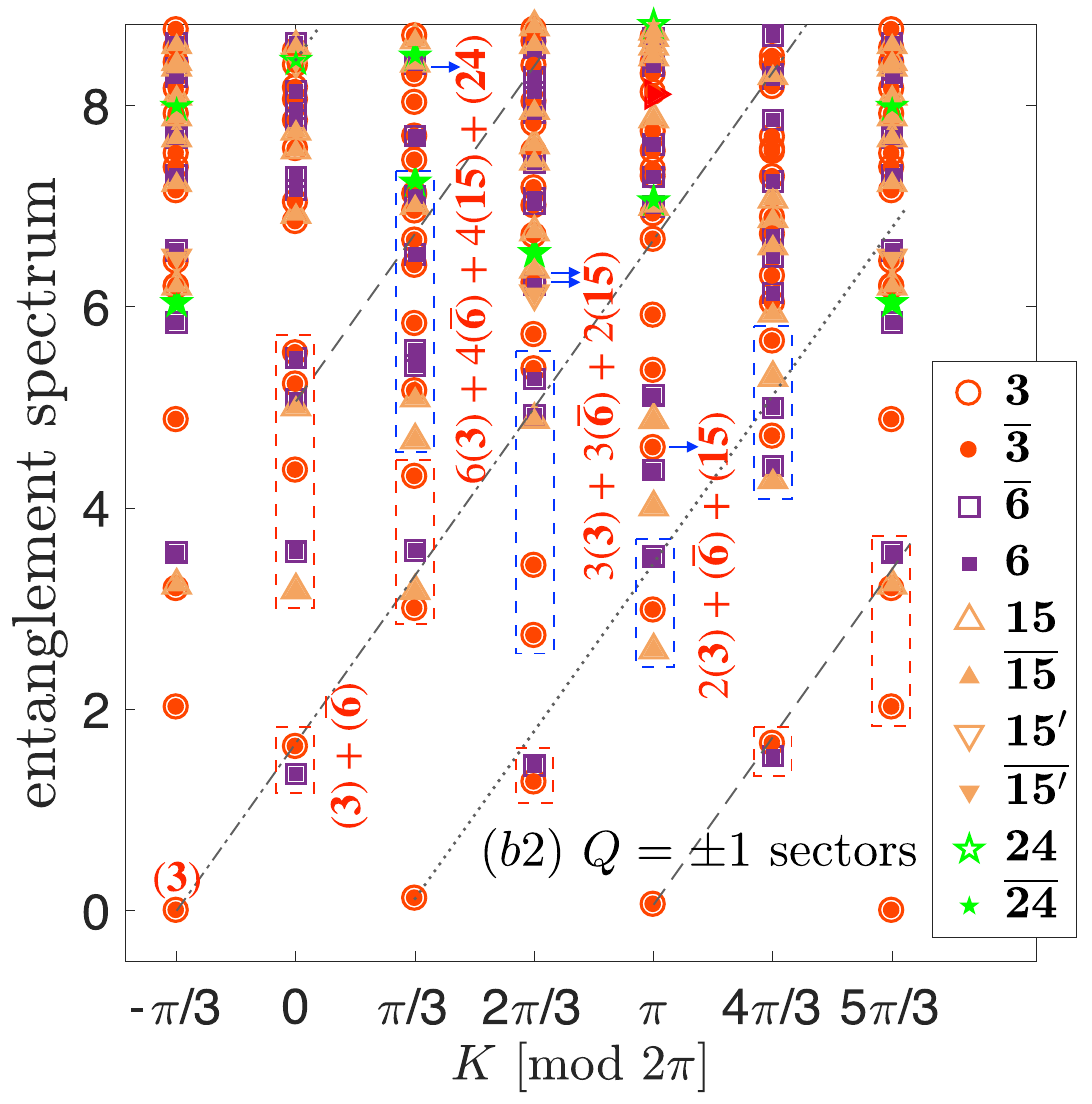}}
	\subfloat{\includegraphics[width=0.49\columnwidth]{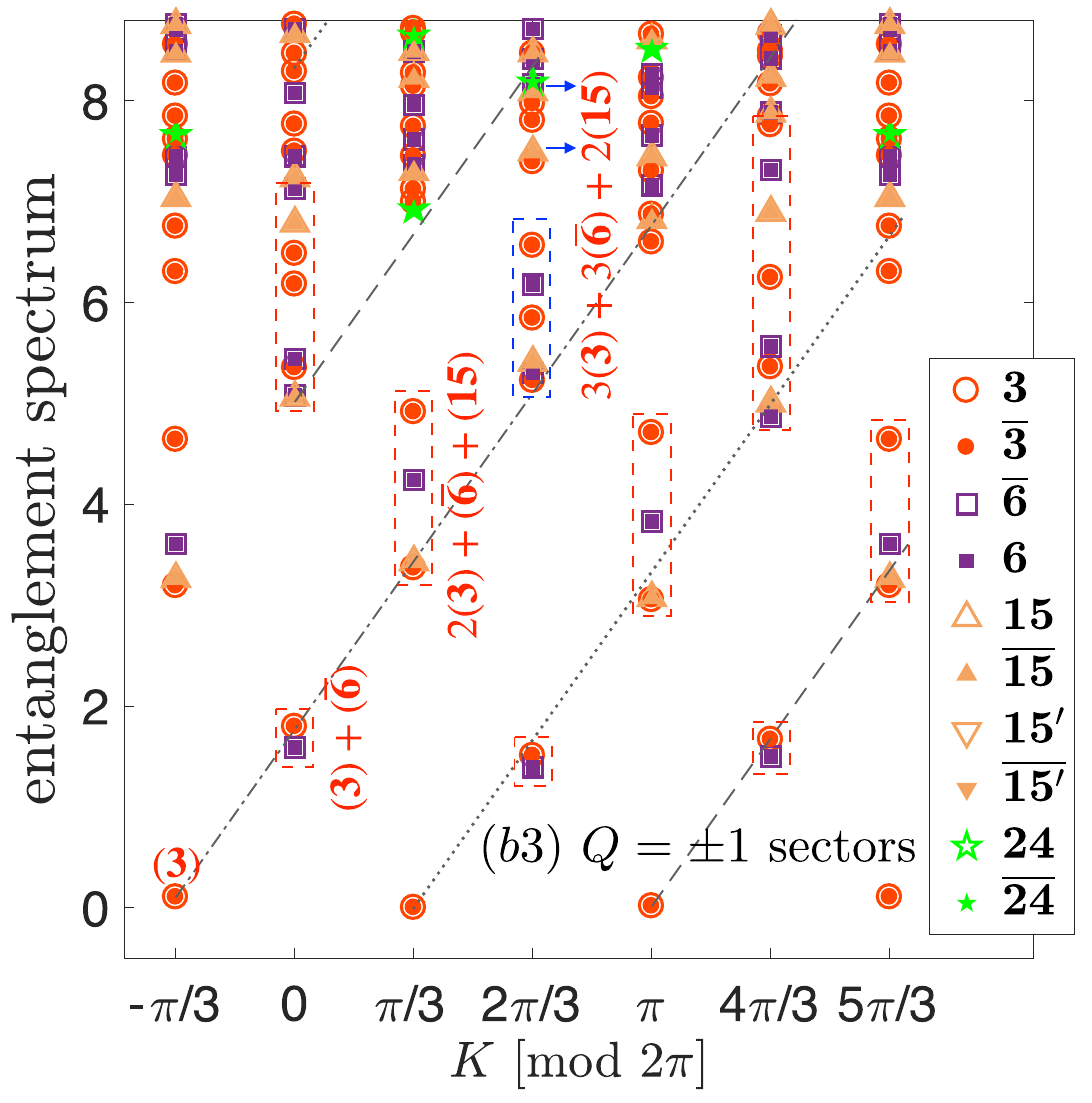}}
	\subfloat{\includegraphics[width=0.49\columnwidth]{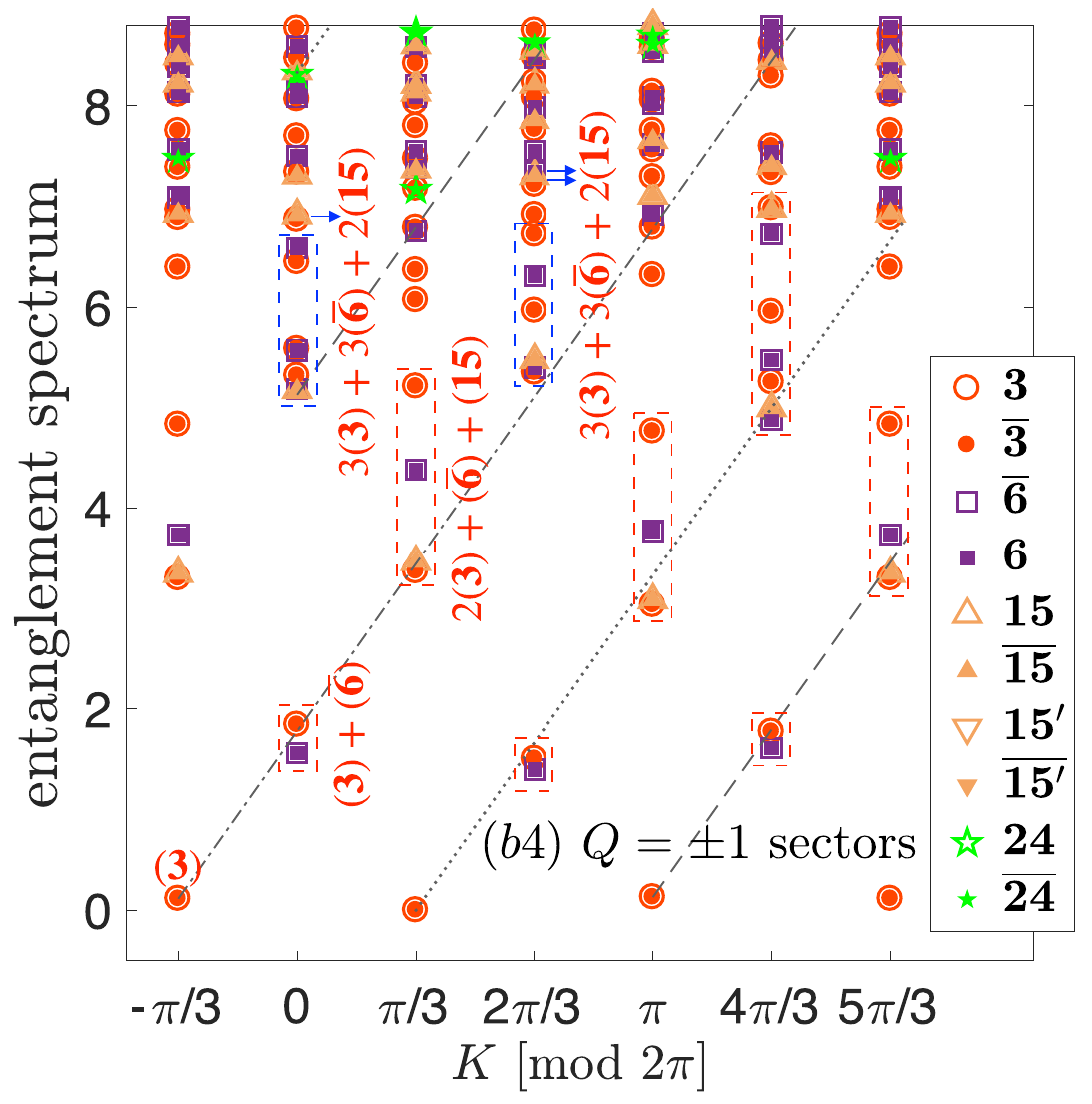}}
\caption{Finite $\chi$ effect on the low energy spectrum of ES. (a1)-a(4) ((b1)-(b4)) show ES in the charge $Q=0$ ($Q=\pm 1$) sector(s), computed with $\chi=49, 98, 245, 343$, respectively. In (b1)-(b4), open symbols are for the $Q=+1$ sector while filled symbols are for the $Q=-1$ sector. In each sector, the (in)complete Virasoro levels have been indicated by (blue) red boxes when necessary, and the missing levels can be found in the higher energy spectrum, marked as blue arrows. Their contents are shown in red vertically (see also Tab.~\ref{table:su3_1}). (Since the $Q=+1$ sector is degenerate with the $Q=-1$ sector, only Virasoro levels in the former sector are marked out.) With increasing $\chi$, more Virasoro levels become complete, which is evidently seen in the charge $Q=0$ sector. This trend is not monotonic in charged sectors, due to the large number of branches -- three -- on a relatively small width ($N_v=6$) cylinder, which mix each other in the higher energy spectrum. Nevertheless, the CFT spectrum in all sectors get more separated from the high energy continuum with increasing $\chi$.}
\label{fig:finite_chi_effect}
\end{figure*}

\begin{figure}[!htb]
\centering
	\includegraphics[width=0.99\columnwidth]{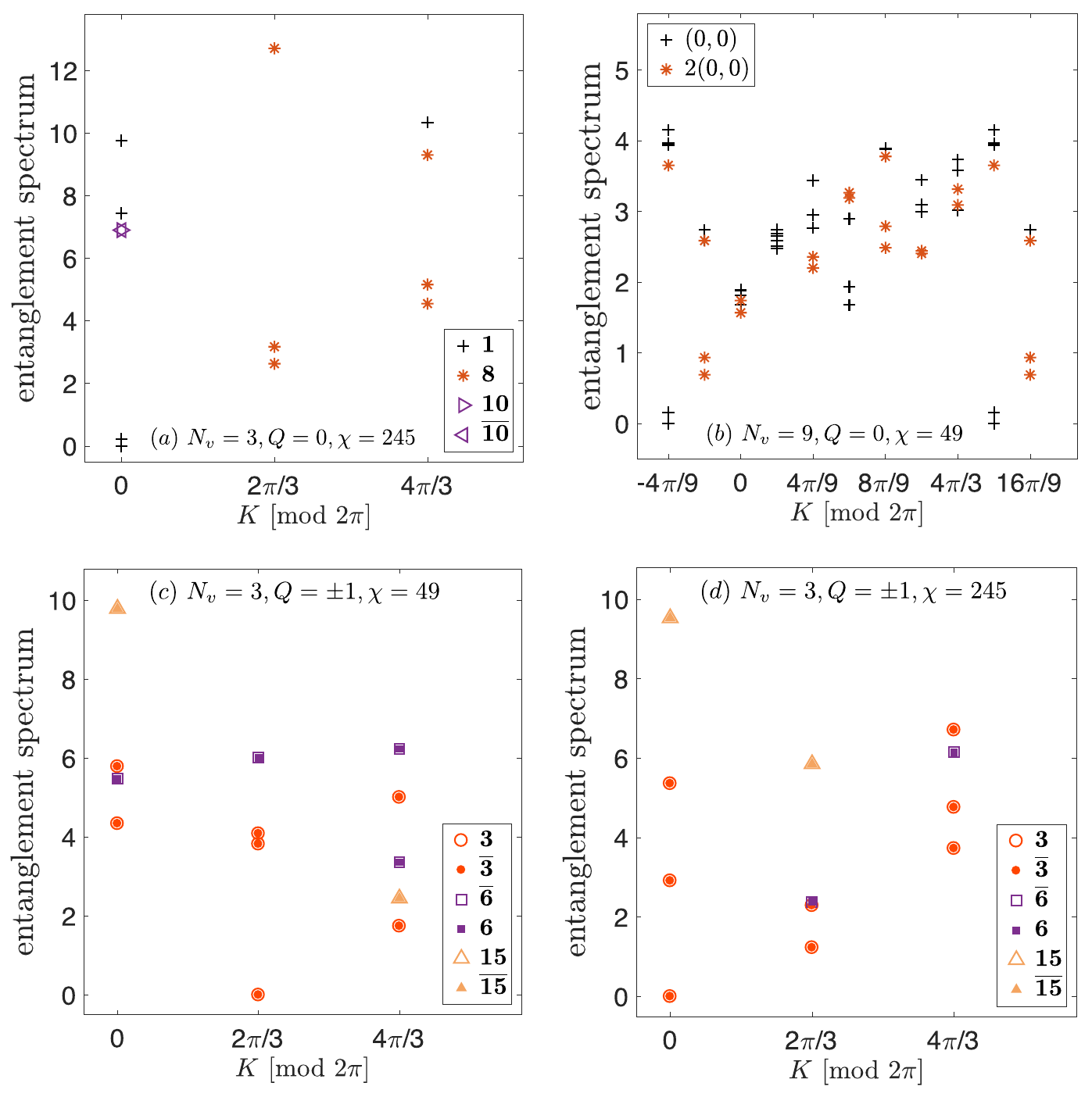}\\
\caption{Entanglement spectrum on $N_v=3$ and $N_v=9$ cylinders. (a) and (b) shows ES in the charge $Q=0$ sector. For $N_v=9$, only low energy spectrum in ${\bf S}^z=(0,0)$ sector is computed, where the ground state is supposed to be. (c), (d) shows ES in charge $Q=\pm 1$ sectors for $N_v=3$ with different $\chi$.}
\label{fig:ES_3_9}
\end{figure}

The RDM has both translation symmetry and SU($3$) symmetry, allowing us to block diagonalize it with momentum quantum number $K$, the two U($1$) quantum number, and $\mathbb{Z}_3$ quantum number (charge $Q$). A typical result of full diagonalization is shown in Fig.~\ref{fig:fullSpectrum}, where linear dispersing chiral modes can be seen in the low energy spectrum. The degeneracy between the charge $Q=+1$ and $Q=-1$ sector can be identified with Fig.~\ref{fig:fullSpectrum}(b) and (c), where same energy levels with same momenta but conjugated SU($3$) irreps appear in the $Q=+1$ and $Q=-1$ sector separately, confirming the degeneracy mentioned in the main text.
In all three $\mathbb{Z}_3$ sectors, an entanglement gap~\cite{Li2008} separating the chiral mode from the high energy continuum can be identified, although the magnitude of the gap in charge $Q=0$ sector is much larger than the gap in charge $Q=\pm 1$ sectors. In the $Q=0$ sector, the chiral branch starts at momentum $K_0=-\pi/3$, while the three quasi-degenerate chiral branches in the $Q=\pm 1$ sectors start at momenta $K_{\pm 1}=-\pi/3, \pi/3$ and $\pi$, individually.

Further examining the level counting of the chiral modes confirms that they satisfy SU$(3)_1$ Wess-Zumino-Witten (WZW) conformal field theory (CFT) prediction~\cite{Bouwknegt1996,Franscesco1997}. See Tab.~\ref{table:su3_1} for a list of the tower of states in SU$(3)_1$ WZW CFT. It is interesting to see that the Virasoro level contents also exhibit degeneracy between the charge $Q=+1$ and $Q=-1$ sectors, which is perfectly recovered by the numerically computed entanglement spectrum. Due to this degeneracy, in the following, we have plotted the $Q=+1$ sector and $Q=-1$ sector together, using open symbols and filled symbols respectively, to emphasize this symmetry (see Fig.~\ref{fig:finite_chi_effect} and Fig.~\ref{fig:ES_3_9}).

\begin{table}[htb]
\begin{center}
\resizebox{1\columnwidth}{!}{%
    \begin{tabular}{@{} ccc @{}}
    \hline
    \hline
    $n \backslash {\rm sector}$ & $\bf{1}$ & $\bf{3}$\\
    \hline
0 & ($\bf{1}$) & ($\bf{3}$)
\vspace{1mm}\\
1 & ($\bf{8}$) & ($\bf{3}$)+($\bf\overline{6}$)
\vspace{1mm}\\
2 & ($\bf{1}$)+2($\bf{8}$) & 2($\bf{3}$)+($\bf\overline{6}$)+($\bf{15}$)
\vspace{1mm}\\
3 & 2($\bf{1}$)+3($\bf{8}$)+($\bf{10}$)+($\bf\overline{10}$) & 3($\bf{3}$)+3($\bf\overline{6}$)+2($\bf{15}$)
\vspace{1mm}\\
4 & 3($\bf{1}$)+6($\bf{8}$)+($\bf{10}$)+($\bf\overline{10}$)+($\bf{27}$) & 6($\bf{3}$)+4($\bf\overline{6}$)+4($\bf{15}$)+($\bf{24}$)
\vspace{1mm}\\
5 & 4($\bf{1}$)+10($\bf{8}$)+3($\bf{10}$)+3($\bf\overline{10}$)+2($\bf{27}$) & 9($\bf{3}$)+8($\bf\overline{6}$)+7($\bf{15}$)+2($\bf{24}$)+($\bf{15'}$)
\vspace{1mm}\\
    \hline
    \hline
    \end{tabular}
     }
\caption{Tower of states in charge $Q=0$ and $1$ sectors of SU$(3)_1$ WZW model, which are characterized by primary fields $S={\bf 1}, {\bf 3}$ and conformal weight $0, 1/3$, separately. Each line corresponds to a Virasoro level indexed by $n$. For each sector and each level, the (quasi-) degenerate states can be grouped in terms of exact SU(3) multiplets like $n_1 ({\bf 1}) + n_8 ({\bf 8}) +\cdots$ (meaning $n_1$ singlets, $n_8$ octets, etc). The $Q=-1$ sector, characterized by primary field $S=\bf\overline{3}$ and conformal weight $1/3$, has the same structure as $Q=1$ sector but with conjugate SU(3) irrep, therefore not shown in the table. The Young tableau for each SU(3) irrep can be found in Tab.~\ref{table:su3multiplets}.}
\label{table:su3_1}
\end{center}
\end{table}

It should be noted, when using CTMRG environment tensors to construct approximate RDM, the environment bond dimension $\chi$ is the only tuning parameter. Since $\chi$ controls the accuracy of CTMRG procedure, we expect that the level contents of chiral CFT mode becomes more complete with increasing $\chi$. This finite $\chi$ effect is shown in Fig.~\ref{fig:finite_chi_effect}. Certain features of the ES, e.g., momentum shift in all sectors and three branches in charged sectors, are present for all different $\chi$ we have considered. This is reasonable since the low energy spectrum converges first with increasing $\chi$, and suggests that these features are intrinsic properties of the optimized PEPS wave function, i.e., not artifacts of the approximation.

The momentum shift observed in ES on $N_v=6$ cylinder has dramatic consequence for ES on $N_v=3$ and $N_v=9$ cylinders, shown in Fig.~\ref{fig:ES_3_9}.
In the charge $Q=0$ sector for both $N_v=3$ and $9$, a linear dispersing mode can be vaguely identified. However the content of each level is doubled, i.e., two singlets for $n=0$ level, two $\bf{8}$ for $n=1$ level, since the finite momentum of ground state $K_0=-\pi/3$ is incommensurate with $N_v=3,9$. This scenario is further confirmed by exact contraction for $N_v=3$ case (data not shown). In the charged sectors with $N_v=3$, the lowest level can appear at different momenta, which typically depends on $\chi$, see Fig.~\ref{fig:ES_3_9}(c) and (d). This is in agreement with the three quasi-degenerate branches with different momenta $K_{\pm1}=-\pi/3,\pi/3$ and $\pi$ in $N_v=6$ case.

\begin{figure}[!htb]
\centering
	\includegraphics[width=0.9\columnwidth]{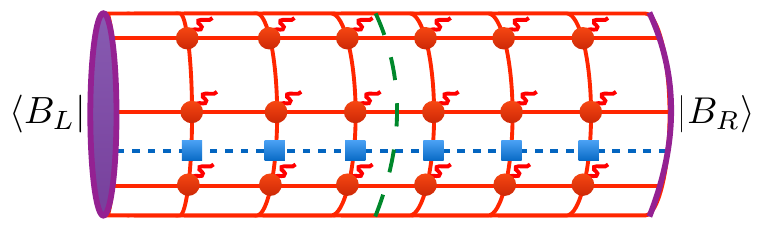}\\
\caption{On infinitely long cylinders, topologically quasi-degenerate states can be constructed by choosing virtual boundary state $|B_L\rangle, |B_R\rangle$ belonging to fixed $\mathbb{Z}_3$ charge sector, with or without nontrivial $\mathbb{Z}_3$ flux insertion (shown as blue squares).}
\label{fig:topoSec}
\end{figure}

Finally, we close this section by briefly discussing the ES in the flux sector. In this work we have mainly focused on ES in the topological sectors without flux, and succeeded in finding signatures for chiral topological order. The $\mathbb{Z}_3$ gauge symmetry implies that we can also construct topological sectors with flux, see Fig.~\ref{fig:topoSec} for illustration. However, previous studies in the SU($2$) case~\cite{Poilblanc2016, Hackenbroich2018} suggest the ES in flux sectors does not follow a simple CFT description. Therefore, we do not explore it here but leave it to further study.

\section{topological excitations and correlations in symmetric PEPS}
\label{sec:excitations}

As discussed in the main text, $\mathbb{Z}_3$ gauge symmetry, generated by $Z (Z^3=\mathbb{I}_D)$, implies topological excitations on infinite plane, whose type can be labeled by the group element and group irreps. Here we will not describe the full details of the theory, but refer to Ref.~\cite{Duivenvoorden2017} for the interested reader. 

\begin{figure}[!htb]
\centering
	\includegraphics[width=0.95\columnwidth]{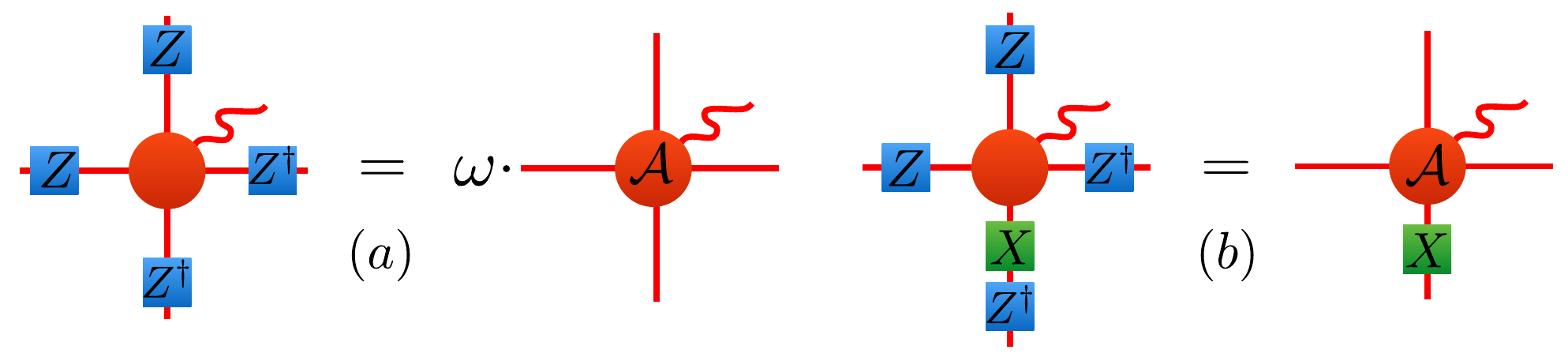}\\
\caption{Creating spinon excitation by acting on the virtual level. Since the original tensor $\mathcal{A}$ carries charge 1, shown in (a), acting an operator $X$ on the virtual level will create a local excitation carrying charge 0, shown in (b). See text for further details.}
\label{fig:spinon}
\end{figure}

Spinon, one of the topologically nontrivial elementary excitations, can be created by modifying a local tensor such that it belongs to different irreps of $\mathbb{Z}_3$. This can be achieved by acting on the virtual level of the local tensor $\mathcal{A}$ with an operator $X$, see Fig.~\ref{fig:spinon}. Its anti-particle can then be similarly created with an operator $X^2$ also acting on the virtual index.

Apart from basic algebraic relation of $X$ and $Z$: 
\begin{equation}
XZ = \omega ZX,
\end{equation}
with $\omega={\rm e}^{i2\pi/3}$, which generalizes the anti-commutation relation between Pauli matrix $\sigma_x$ and $\sigma_z$ to $\mathbb{Z}_3$, the choice of $X$ is not unique, due to the internal SU(3) symmetry in the ansatz. The specific $X$ we use to compute spinon-antispinon correlation function is:
\begin{equation}
X = \begin{pmatrix} 0 & 0 & 0 & 1 & 0 & 0 & 0\\
0 & 0 & 0 & 0 & 1 & 0 & 0\\
0 & 0 & 0 & 0 & 0 & 1 & 0\\
0 & 0 & 0 & 0 & 0 & 0 & 1\\
0 & 0 & 0 & 0 & 0 & 0 & 1\\
0 & 0 & 0 & 0 & 0 & 0 & 1\\
1 & 1 & 1 & 0 & 0 & 0 & 0 \end{pmatrix}.
\end{equation}
In principle, the operator X can be put on any of the four virtual indices of $\mathcal{A}$. However, unlike the Pauli matrix $\sigma_x$ in the $\mathbb{Z}_2$ case,  $X$ in the $\mathbb{Z}_3$ case cannot be chosen to be symmetric. Thus the order of indices matters when computing anyonic correlation functions using $X$. In practice, we always put $X$ or $X^2$ in the ket layer, with the first index contracted with down index of local tensor $\mathcal{A}$.

\begin{figure}[!htb]
\centering
	\includegraphics[width=0.7\columnwidth]{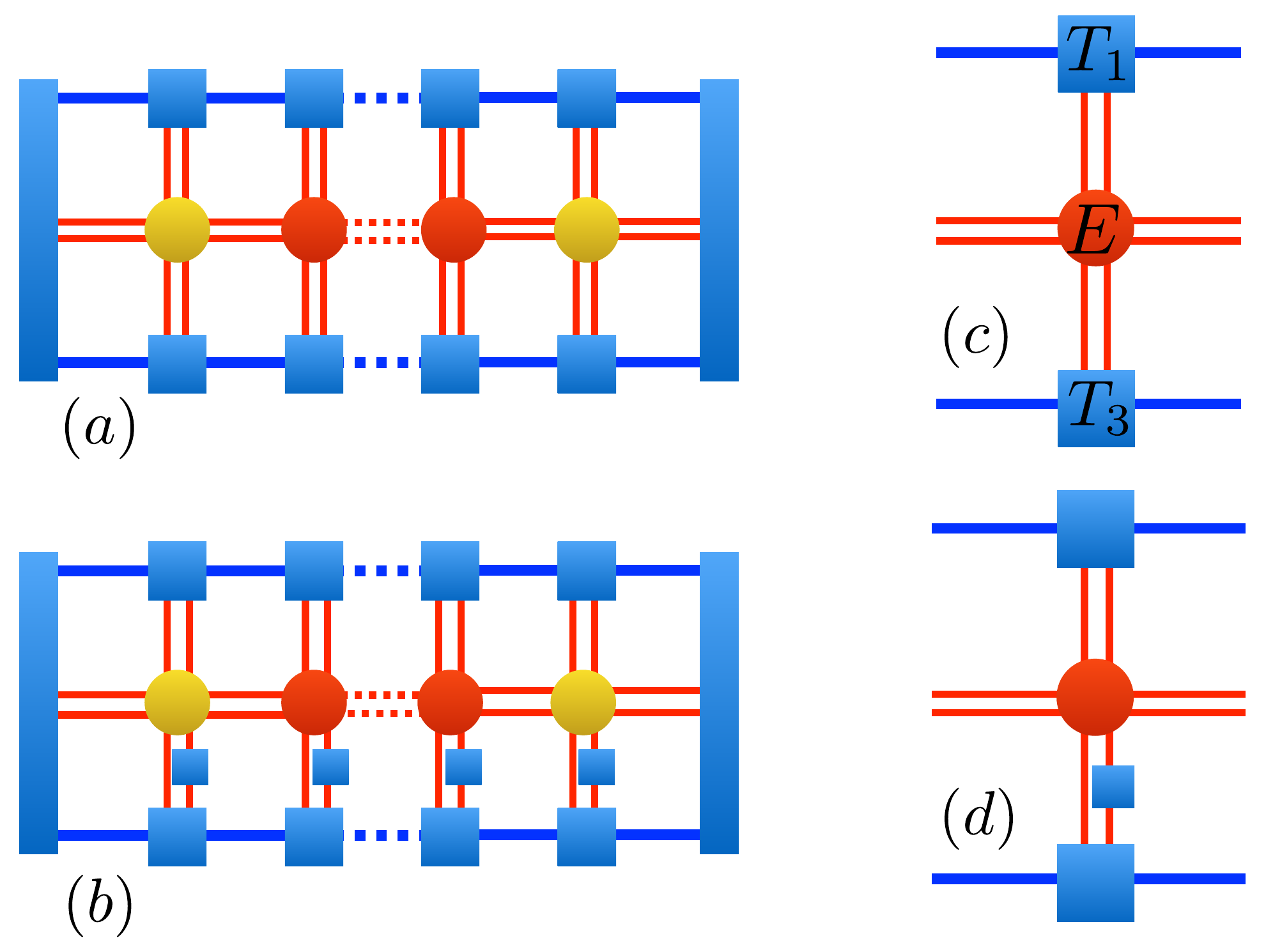}\\
\caption{Computing correlation functions and correlation lengths using CTMRG environment tensors. (a) shows computing real-space correlations using CTMRG environment tensors, for topologically trivial excitations created by, e.g. spin operator acting on the physical index, and topologically nontrivial excitations, e.g. spinons. (b) is for real-space vison or parafermion correlations. In both (a) and (b) the yellow dots represent the local excitations. (c), (d) stands for transfer matrix without or with flux, respectively. In (b) and (d), the blue squares represent the gauge group element $Z$ or $Z^2$, acting on the virtual indices of ket layer.}
\label{fig:correlation_sm}
\end{figure}

Besides spinons, the $\mathbb{Z}_3$ gauge symmetry also allows us to construct vison excitations, which are end points of $Z$ or $Z^2$ string operator acting on the virtual level. Their bound states, so-called parafermions, can be created by attaching spinons to the end points of virtual string.

With these topological excitations at hand, the calculation of their correlation functions is straightforward, shown in Fig.~\ref{fig:correlation_sm}(a) and (b). The corresponding transfer matrix can also be constructed, without or with flux, see Fig.~\ref{fig:correlation_sm}(c) and (d).

\section{Nonzero elements of symmetric tensors}
\label{sec:tensorExpression}

For the sake of completeness, we now present the resulting tensors from classification. According to the irreps of $C_{\rm 4v}$ group, the on-site projectors can be classified into four real classes, and two complex classes. Since only the real classes are used, we list their nonzero elements below, denoted as $\mathcal{A}_1$, $\mathcal{B}_1$, $\mathcal{A}_2$, $\mathcal{B}_2$. For virtual space $\mathcal{V}=\bf{3}\oplus{\bf\overline{3}}\oplus{\bf 1}$, we relabel the basis as $|{\bf 3},1\rangle\equiv|0\rangle$, $|{\bf 3},2\rangle\equiv|1\rangle$, $|{\bf 3},3\rangle\equiv|2\rangle$, $|{\bf\overline{3}},1\rangle\equiv|3\rangle$, $|{\bf\overline{3}},2\rangle\equiv|4\rangle$, $|{\bf\overline{3}},3\rangle\equiv|5\rangle$, $|{\bf1},1\rangle\equiv|6\rangle$. See also Tab.~\ref{table:su3multiplets} for the U($1$) quantum numbers of each basis in both physical and virtual spaces.

The expressions of the three components of each real tensors are provided in tables~\ref{table:S2}-~\ref{table:S22}, where the tensor indices are in $[{\rm up, left, down, right}]$ order.

\clearpage

\begin{table}[!htb]
	\renewcommand\arraystretch{1.5}
	\caption{$\mathcal{A}_2^{(1)}$, $n_{\rm occ}=\{4,0,0\}$}
	\label{table:S2}
	\scriptsize
\begin{tabularx}{0.498\textwidth}{p{0.33\linewidth} p{0.33\linewidth} p{0.33\linewidth}}
		\hline
		\hline
		{\small ${\bf S}^z = (1/2, 0)$}  & {\small ${\bf S}^z = (-1/2, 1/2)$} & {\small ${\bf S}^z = (0, -1/2)$} \\
		\hline
		$
		\begin{array}{l}
		t_{0,0,1,2}=-\frac{1}{2 \sqrt{2}} \\
		t_{0,0,2,1}=\frac{1}{2 \sqrt{2}} \\
		t_{0,1,2,0}=-\frac{1}{2 \sqrt{2}} \\
		t_{0,2,1,0}=\frac{1}{2 \sqrt{2}} \\
		t_{1,0,0,2}=\frac{1}{2 \sqrt{2}} \\
		t_{1,2,0,0}=-\frac{1}{2 \sqrt{2}} \\
		t_{2,0,0,1}=-\frac{1}{2 \sqrt{2}} \\
		t_{2,1,0,0}=\frac{1}{2 \sqrt{2}} \\
		\end{array}
		$&
		$
		\begin{array}{l}
		t_{0,1,1,2}=-\frac{1}{2 \sqrt{2}} \\
		t_{0,2,1,1}=\frac{1}{2 \sqrt{2}} \\
		t_{1,0,2,1}=\frac{1}{2 \sqrt{2}} \\
		t_{1,1,0,2}=\frac{1}{2 \sqrt{2}} \\
		t_{1,1,2,0}=-\frac{1}{2 \sqrt{2}} \\
		t_{1,2,0,1}=-\frac{1}{2 \sqrt{2}} \\
		t_{2,0,1,1}=-\frac{1}{2 \sqrt{2}} \\
		t_{2,1,1,0}=\frac{1}{2 \sqrt{2}} \\
		\end{array}
		$&
		$
		\begin{array}{l}
		t_{0,1,2,2}=-\frac{1}{2 \sqrt{2}} \\
		t_{0,2,2,1}=\frac{1}{2 \sqrt{2}} \\
		t_{1,0,2,2}=\frac{1}{2 \sqrt{2}} \\
		t_{1,2,2,0}=-\frac{1}{2 \sqrt{2}} \\
		t_{2,0,1,2}=-\frac{1}{2 \sqrt{2}} \\
		t_{2,1,0,2}=\frac{1}{2 \sqrt{2}} \\
		t_{2,2,0,1}=-\frac{1}{2 \sqrt{2}} \\
		t_{2,2,1,0}=\frac{1}{2 \sqrt{2}} \\
		\end{array}
		$\vspace{1mm}\\
		\hline
		\hline
	\end{tabularx}
\end{table}

\begin{table}[!htb]
	\renewcommand\arraystretch{1.5}
	\caption{$\mathcal{A}_1^{(1)}$ and $\mathcal{B}_1^{(1)}$, $n_{\rm occ}=\{1,0,3\}$}
	\label{table:S3}
	\scriptsize
	\begin{tabularx}{0.498\textwidth}{p{0.33\linewidth} p{0.33\linewidth} p{0.33\linewidth}}
		\hline
		\hline
		{\small ${\bf S}^z = (1/2, 0)$}  & {\small ${\bf S}^z = (-1/2, 1/2)$} & {\small ${\bf S}^z = (0, -1/2)$} \\
		\hline
		$
		\begin{array}{l}
		t_{0,6,6,6}=\frac{1}{2} \\
		t_{6,0,6,6}=\frac{1}{2} \\
		t_{6,6,0,6}=\frac{1}{2} \\
		t_{6,6,6,0}=\frac{1}{2} \\
		\end{array}
		$&
		$
		\begin{array}{l}
		t_{1,6,6,6}=\frac{1}{2} \\
		t_{6,1,6,6}=\frac{1}{2} \\
		t_{6,6,1,6}=\frac{1}{2} \\
		t_{6,6,6,1}=\frac{1}{2} \\
		\end{array}
		$&
		$
		\begin{array}{l}
		t_{2,6,6,6}=\frac{1}{2} \\
		t_{6,2,6,6}=\frac{1}{2} \\
		t_{6,6,2,6}=\frac{1}{2} \\
		t_{6,6,6,2}=\frac{1}{2} \\
		\end{array}
		$
		\vspace{1mm}\\
		\hline
		\hline
	\end{tabularx}
	$\phantom{a}$\vspace{0.5mm}
	\begin{tabularx}{0.498\textwidth}{p{0.33\linewidth} p{0.33\linewidth} p{0.33\linewidth}}
	\hline
	\hline
	{\small ${\bf S}^z = (1/2, 0)$}  & {\small ${\bf S}^z = (-1/2, 1/2)$} & {\small ${\bf S}^z = (0, -1/2)$} \\
	\hline
	$
	\begin{array}{l}
	t_{0,6,6,6}=\frac{1}{2} \\
	t_{6,0,6,6}=-\frac{1}{2} \\
	t_{6,6,0,6}=\frac{1}{2} \\
	t_{6,6,6,0}=-\frac{1}{2} \\
	\end{array}
	$&
	$
	\begin{array}{l}
	t_{1,6,6,6}=\frac{1}{2} \\
	t_{6,1,6,6}=-\frac{1}{2} \\
	t_{6,6,1,6}=\frac{1}{2} \\
	t_{6,6,6,1}=-\frac{1}{2} \\
	\end{array}
	$&
	$
	\begin{array}{l}
	t_{2,6,6,6}=\frac{1}{2} \\
	t_{6,2,6,6}=-\frac{1}{2} \\
	t_{6,6,2,6}=\frac{1}{2} \\
	t_{6,6,6,2}=-\frac{1}{2} \\
	\end{array}
	$
	\vspace{1mm}\\
	\hline
	\hline
\end{tabularx}	
\end{table}

\begin{table}[!htb]
	\renewcommand\arraystretch{1.5}
	\caption{$\mathcal{A}_2^{(1)}$ and $\mathcal{B}_1^{(1)}$, $n_{\rm occ}=\{0,2,2\}$}
	\label{table:S4}
	\scriptsize
	\begin{tabularx}{0.498\textwidth}{p{0.33\linewidth} p{0.33\linewidth} p{0.33\linewidth}}
		\hline
		\hline
		{\small ${\bf S}^z = (1/2, 0)$}  & {\small ${\bf S}^z = (-1/2, 1/2)$} & {\small ${\bf S}^z = (0, -1/2)$} \\
		\hline
		$
		\begin{array}{l}
		t_{3,4,6,6}=\frac{1}{2 \sqrt{2}} \\
		t_{3,6,6,4}=-\frac{1}{2 \sqrt{2}} \\
		t_{4,3,6,6}=-\frac{1}{2 \sqrt{2}} \\
		t_{4,6,6,3}=\frac{1}{2 \sqrt{2}} \\
		t_{6,3,4,6}=\frac{1}{2 \sqrt{2}} \\
		t_{6,4,3,6}=-\frac{1}{2 \sqrt{2}} \\
		t_{6,6,3,4}=\frac{1}{2 \sqrt{2}} \\
		t_{6,6,4,3}=-\frac{1}{2 \sqrt{2}} \\
		\end{array}
		$&
		$
		\begin{array}{l}
		t_{3,5,6,6}=\frac{1}{2 \sqrt{2}} \\
		t_{3,6,6,5}=-\frac{1}{2 \sqrt{2}} \\
		t_{5,3,6,6}=-\frac{1}{2 \sqrt{2}} \\
		t_{5,6,6,3}=\frac{1}{2 \sqrt{2}} \\
		t_{6,3,5,6}=\frac{1}{2 \sqrt{2}} \\
		t_{6,5,3,6}=-\frac{1}{2 \sqrt{2}} \\
		t_{6,6,3,5}=\frac{1}{2 \sqrt{2}} \\
		t_{6,6,5,3}=-\frac{1}{2 \sqrt{2}} \\
		\end{array}
		$&
		$
		\begin{array}{l}
		t_{4,5,6,6}=\frac{1}{2 \sqrt{2}} \\
		t_{4,6,6,5}=-\frac{1}{2 \sqrt{2}} \\
		t_{5,4,6,6}=-\frac{1}{2 \sqrt{2}} \\
		t_{5,6,6,4}=\frac{1}{2 \sqrt{2}} \\
		t_{6,4,5,6}=\frac{1}{2 \sqrt{2}} \\
		t_{6,5,4,6}=-\frac{1}{2 \sqrt{2}} \\
		t_{6,6,4,5}=\frac{1}{2 \sqrt{2}} \\
		t_{6,6,5,4}=-\frac{1}{2 \sqrt{2}} \\
		\end{array}
		$\vspace{1mm}\\
		\hline
		\hline
	\end{tabularx}
	$\phantom{a}$\vspace{0.5mm}
	\begin{tabularx}{0.498\textwidth}{p{0.33\linewidth} p{0.33\linewidth} p{0.33\linewidth}}
	\hline
	\hline
	{\small ${\bf S}^z = (1/2, 0)$}  & {\small ${\bf S}^z = (-1/2, 1/2)$} & {\small ${\bf S}^z = (0, -1/2)$} \\
	\hline
	$
	\begin{array}{l}
	t_{3,4,6,6}=-\frac{1}{2 \sqrt{2}} \\
	t_{3,6,6,4}=-\frac{1}{2 \sqrt{2}} \\
	t_{4,3,6,6}=\frac{1}{2 \sqrt{2}} \\
	t_{4,6,6,3}=\frac{1}{2 \sqrt{2}} \\
	t_{6,3,4,6}=\frac{1}{2 \sqrt{2}} \\
	t_{6,4,3,6}=-\frac{1}{2 \sqrt{2}} \\
	t_{6,6,3,4}=-\frac{1}{2 \sqrt{2}} \\
	t_{6,6,4,3}=\frac{1}{2 \sqrt{2}} \\
	\end{array}
	$&
	$
	\begin{array}{l}
	t_{3,5,6,6}=-\frac{1}{2 \sqrt{2}} \\
	t_{3,6,6,5}=-\frac{1}{2 \sqrt{2}} \\
	t_{5,3,6,6}=\frac{1}{2 \sqrt{2}} \\
	t_{5,6,6,3}=\frac{1}{2 \sqrt{2}} \\
	t_{6,3,5,6}=\frac{1}{2 \sqrt{2}} \\
	t_{6,5,3,6}=-\frac{1}{2 \sqrt{2}} \\
	t_{6,6,3,5}=-\frac{1}{2 \sqrt{2}} \\
	t_{6,6,5,3}=\frac{1}{2 \sqrt{2}} \\
	\end{array}
	$&
	$
	\begin{array}{l}
	t_{4,5,6,6}=-\frac{1}{2 \sqrt{2}} \\
	t_{4,6,6,5}=-\frac{1}{2 \sqrt{2}} \\
	t_{5,4,6,6}=\frac{1}{2 \sqrt{2}} \\
	t_{5,6,6,4}=\frac{1}{2 \sqrt{2}} \\
	t_{6,4,5,6}=\frac{1}{2 \sqrt{2}} \\
	t_{6,5,4,6}=-\frac{1}{2 \sqrt{2}} \\
	t_{6,6,4,5}=-\frac{1}{2 \sqrt{2}} \\
	t_{6,6,5,4}=\frac{1}{2 \sqrt{2}} \\
	\end{array}
	$\vspace{1mm}\\
	\hline
	\hline
\end{tabularx}	
\end{table}

\begin{table}[htb]
	\renewcommand\arraystretch{1.5}
	\caption{$\mathcal{A}_1^{(1)}$, $n_{\rm occ}=\{1,3,0\}$}
	\label{table:S5}
	\scriptsize
	\begin{tabularx}{0.498\textwidth}{p{0.33\linewidth} p{0.33\linewidth} p{0.33\linewidth}}
		\hline
		\hline
		{\small ${\bf S}^z = (1/2, 0)$}  & {\small ${\bf S}^z = (-1/2, 1/2)$} & {\small ${\bf S}^z = (0, -1/2)$} \\
\hline
		\hline
	\end{tabularx}
\end{table}

\clearpage

\begin{table}[!htb]
	\renewcommand\arraystretch{1.5}
	\caption{$\mathcal{A}_2^{(1)}$, $n_{\rm occ}=\{1,3,0\}$}
	\label{table:S6}
	\scriptsize
	\begin{tabularx}{0.498\textwidth}{p{0.33\linewidth} p{0.33\linewidth} p{0.33\linewidth}}
		\hline
		\hline
		{\small ${\bf S}^z = (1/2, 0)$}  & {\small ${\bf S}^z = (-1/2, 1/2)$} & {\small ${\bf S}^z = (0, -1/2)$} \\
\hline
		\hline
	\end{tabularx}
\end{table}

\begin{table}[!htb]
	\renewcommand\arraystretch{1.5}
	\caption{$\mathcal{A}_2^{(2)}$, $n_{\rm occ}=\{1,3,0\}$}
	\label{table:S7}
	\scriptsize
	\begin{tabularx}{0.498\textwidth}{p{0.33\linewidth} p{0.33\linewidth} p{0.33\linewidth}}
		\hline
		\hline
		{\small ${\bf S}^z = (1/2, 0)$}  & {\small ${\bf S}^z = (-1/2, 1/2)$} & {\small ${\bf S}^z = (0, -1/2)$} \\
\hline
		\hline
	\end{tabularx}
\end{table}

\clearpage

\begin{table}[!htb]
	\renewcommand\arraystretch{1.5}
	\caption{$\mathcal{B}_1^{(1)}$, $n_{\rm occ}=\{1,3,0\}$}
	\label{table:S8}
	\scriptsize
	\begin{tabularx}{0.498\textwidth}{p{0.33\linewidth} p{0.33\linewidth} p{0.33\linewidth}}
		\hline
		\hline
		{\small ${\bf S}^z = (1/2, 0)$}  & {\small ${\bf S}^z = (-1/2, 1/2)$} & {\small ${\bf S}^z = (0, -1/2)$} \\
\hline
		\hline
	\end{tabularx}
\end{table}

\begin{table}[!htb]
	\renewcommand\arraystretch{1.5}
	\caption{$\mathcal{B}_2^{(1)}$, $n_{\rm occ}=\{1,3,0\}$}
	\label{table:S9}
	\scriptsize
	\begin{tabularx}{0.498\textwidth}{p{0.33\linewidth} p{0.33\linewidth} p{0.33\linewidth}}
		\hline
		\hline
		{\small ${\bf S}^z = (1/2, 0)$}  & {\small ${\bf S}^z = (-1/2, 1/2)$} & {\small ${\bf S}^z = (0, -1/2)$} \\
\hline
		\hline
	\end{tabularx}
\end{table}
\clearpage
\begin{table}[!htb]
	\renewcommand\arraystretch{1.5}
	\caption{$\mathcal{B}_2^{(2)}$, $n_{\rm occ}=\{1,3,0\}$}
	\label{table:S10}
	\scriptsize
	\begin{tabularx}{0.498\textwidth}{p{0.33\linewidth} p{0.33\linewidth} p{0.33\linewidth}}
		\hline
		\hline
		{\small ${\bf S}^z = (1/2, 0)$}  & {\small ${\bf S}^z = (-1/2, 1/2)$} & {\small ${\bf S}^z = (0, -1/2)$} \\
\hline
		\hline
	\end{tabularx}
\end{table}

\begin{table}[!htb]
	\renewcommand\arraystretch{1.5}
	\caption{$\mathcal{A}_1^{(1)}$, $n_{\rm occ}=\{2,1,1\}$}
	\label{table:S11}
	\scriptsize
	\begin{tabularx}{0.498\textwidth}{p{0.33\linewidth} p{0.33\linewidth} p{0.33\linewidth}}
		\hline
		\hline
		{\small ${\bf S}^z = (1/2, 0)$}  & {\small ${\bf S}^z = (-1/2, 1/2)$} & {\small ${\bf S}^z = (0, -1/2)$} \\
\hline
		\hline
	\end{tabularx}
\end{table}
\clearpage
\begin{table}[!htb]
	\renewcommand\arraystretch{1.5}
	\caption{$\mathcal{A}_1^{(2)}$, $n_{\rm occ}=\{2,1,1\}$}
	\label{table:S12}
	\scriptsize
	\begin{tabularx}{0.498\textwidth}{p{0.33\linewidth} p{0.33\linewidth} p{0.33\linewidth}}
		\hline
		\hline
		{\small ${\bf S}^z = (1/2, 0)$}  & {\small ${\bf S}^z = (-1/2, 1/2)$} & {\small ${\bf S}^z = (0, -1/2)$} \\
\hline
		\hline
	\end{tabularx}
\end{table}

\begin{table}[!htb]
	\renewcommand\arraystretch{1.5}
	\caption{$\mathcal{A}_1^{(3)}$, $n_{\rm occ}=\{2,1,1\}$}
	\label{table:S13}
	\scriptsize
	\begin{tabularx}{0.498\textwidth}{p{0.33\linewidth} p{0.33\linewidth} p{0.33\linewidth}}
		\hline
		\hline
		{\small ${\bf S}^z = (1/2, 0)$}  & {\small ${\bf S}^z = (-1/2, 1/2)$} & {\small ${\bf S}^z = (0, -1/2)$} \\
\hline
		\hline
	\end{tabularx}
\end{table}
\clearpage
\begin{table}[!htb]
	\renewcommand\arraystretch{1.5}
	\caption{$\mathcal{A}_2^{(1)}$, $n_{\rm occ}=\{2,1,1\}$}
	\label{table:S14}
	\scriptsize
	\begin{tabularx}{0.498\textwidth}{p{0.33\linewidth} p{0.33\linewidth} p{0.33\linewidth}}
		\hline
		\hline
		{\small ${\bf S}^z = (1/2, 0)$}  & {\small ${\bf S}^z = (-1/2, 1/2)$} & {\small ${\bf S}^z = (0, -1/2)$} \\
\hline
		\hline
	\end{tabularx}
\end{table}

\begin{table}[!htb]
	\renewcommand\arraystretch{1.5}
	\caption{$\mathcal{A}_2^{(2)}$, $n_{\rm occ}=\{2,1,1\}$}
	\label{table:S15}
	\scriptsize
	\begin{tabularx}{0.498\textwidth}{p{0.33\linewidth} p{0.33\linewidth} p{0.33\linewidth}}
		\hline
		\hline
		{\small ${\bf S}^z = (1/2, 0)$}  & {\small ${\bf S}^z = (-1/2, 1/2)$} & {\small ${\bf S}^z = (0, -1/2)$} \\
		\hline
		$
		\begin{array}{l}
		t_{0,3,2,6}=-\frac{1}{4} \\
		t_{0,4,1,6}=\frac{1}{4} \\
		t_{0,6,1,4}=-\frac{1}{4} \\
		t_{0,6,2,3}=\frac{1}{4} \\
		t_{1,4,0,6}=-\frac{1}{4} \\
		t_{1,6,0,4}=\frac{1}{4} \\
		t_{2,3,0,6}=\frac{1}{4} \\
		t_{2,6,0,3}=-\frac{1}{4} \\
		t_{3,0,6,2}=\frac{1}{4} \\
		t_{3,2,6,0}=-\frac{1}{4} \\
		t_{4,0,6,1}=-\frac{1}{4} \\
		t_{4,1,6,0}=\frac{1}{4} \\
		t_{6,0,3,2}=-\frac{1}{4} \\
		t_{6,0,4,1}=\frac{1}{4} \\
		t_{6,1,4,0}=-\frac{1}{4} \\
		t_{6,2,3,0}=\frac{1}{4} \\
		\end{array}
		$&
		$
		\begin{array}{l}
		t_{0,5,1,6}=\frac{1}{4} \\
		t_{0,6,1,5}=-\frac{1}{4} \\
		t_{1,3,2,6}=-\frac{1}{4} \\
		t_{1,5,0,6}=-\frac{1}{4} \\
		t_{1,6,0,5}=\frac{1}{4} \\
		t_{1,6,2,3}=\frac{1}{4} \\
		t_{2,3,1,6}=\frac{1}{4} \\
		t_{2,6,1,3}=-\frac{1}{4} \\
		t_{3,1,6,2}=\frac{1}{4} \\
		t_{3,2,6,1}=-\frac{1}{4} \\
		t_{5,0,6,1}=-\frac{1}{4} \\
		t_{5,1,6,0}=\frac{1}{4} \\
		t_{6,0,5,1}=\frac{1}{4} \\
		t_{6,1,3,2}=-\frac{1}{4} \\
		t_{6,1,5,0}=-\frac{1}{4} \\
		t_{6,2,3,1}=\frac{1}{4} \\
		\end{array}
		$&
		$
		\begin{array}{l}
		t_{0,5,2,6}=\frac{1}{4} \\
		t_{0,6,2,5}=-\frac{1}{4} \\
		t_{1,4,2,6}=-\frac{1}{4} \\
		t_{1,6,2,4}=\frac{1}{4} \\
		t_{2,4,1,6}=\frac{1}{4} \\
		t_{2,5,0,6}=-\frac{1}{4} \\
		t_{2,6,0,5}=\frac{1}{4} \\
		t_{2,6,1,4}=-\frac{1}{4} \\
		t_{4,1,6,2}=\frac{1}{4} \\
		t_{4,2,6,1}=-\frac{1}{4} \\
		t_{5,0,6,2}=-\frac{1}{4} \\
		t_{5,2,6,0}=\frac{1}{4} \\
		t_{6,0,5,2}=\frac{1}{4} \\
		t_{6,1,4,2}=-\frac{1}{4} \\
		t_{6,2,4,1}=\frac{1}{4} \\
		t_{6,2,5,0}=-\frac{1}{4} \\
		\end{array}
		$\vspace{1mm}\\
		\hline
		\hline
	\end{tabularx}
\end{table}

\clearpage
\begin{table}[!htb]
	\renewcommand\arraystretch{1.5}
	\caption{$\mathcal{A}_2^{(3)}$, $n_{\rm occ}=\{2,1,1\}$}
	\label{table:S16}
	\scriptsize
	\begin{tabularx}{0.498\textwidth}{p{0.33\linewidth} p{0.33\linewidth} p{0.33\linewidth}}
		\hline
		\hline
		{\small ${\bf S}^z = (1/2, 0)$}  & {\small ${\bf S}^z = (-1/2, 1/2)$} & {\small ${\bf S}^z = (0, -1/2)$} \\
\hline
		\hline
	\end{tabularx}
\end{table}

\begin{table}[!htb]
	\renewcommand\arraystretch{1.5}
	\caption{$\mathcal{B}_1^{(1)}$, $n_{\rm occ}=\{2,1,1\}$}
	\label{table:S17}
	\scriptsize
	\begin{tabularx}{0.498\textwidth}{p{0.33\linewidth} p{0.33\linewidth} p{0.33\linewidth}}
		\hline
		\hline
		{\small ${\bf S}^z = (1/2, 0)$}  & {\small ${\bf S}^z = (-1/2, 1/2)$} & {\small ${\bf S}^z = (0, -1/2)$} \\
\hline
		\hline
	\end{tabularx}
\end{table}
\clearpage
\begin{table}[!htb]
	\renewcommand\arraystretch{1.5}
	\caption{$\mathcal{B}_1^{(2)}$, $n_{\rm occ}=\{2,1,1\}$}
	\label{table:S18}
	\scriptsize
	\begin{tabularx}{0.498\textwidth}{p{0.33\linewidth} p{0.33\linewidth} p{0.33\linewidth}}
		\hline
		\hline
		{\small ${\bf S}^z = (1/2, 0)$}  & {\small ${\bf S}^z = (-1/2, 1/2)$} & {\small ${\bf S}^z = (0, -1/2)$} \\
\hline
		\hline
	\end{tabularx}
\end{table}

\begin{table}[!htb]
	\renewcommand\arraystretch{1.5}
	\caption{$\mathcal{B}_1^{(3)}$, $n_{\rm occ}=\{2,1,1\}$}
	\label{table:S19}
	\scriptsize
	\begin{tabularx}{0.498\textwidth}{p{0.33\linewidth} p{0.33\linewidth} p{0.33\linewidth}}
		\hline
		\hline
		{\small ${\bf S}^z = (1/2, 0)$}  & {\small ${\bf S}^z = (-1/2, 1/2)$} & {\small ${\bf S}^z = (0, -1/2)$} \\
\hline
		\hline
	\end{tabularx}
\end{table}
\clearpage
\begin{table}[!htb]
	\renewcommand\arraystretch{1.5}
	\caption{$\mathcal{B}_2^{(1)}$, $n_{\rm occ}=\{2,1,1\}$}
	\label{table:S20}
	\scriptsize
	\begin{tabularx}{0.498\textwidth}{p{0.33\linewidth} p{0.33\linewidth} p{0.33\linewidth}}
		\hline
		\hline
		{\small ${\bf S}^z = (1/2, 0)$}  & {\small ${\bf S}^z = (-1/2, 1/2)$} & {\small ${\bf S}^z = (0, -1/2)$} \\
\hline
		\hline
	\end{tabularx}
\end{table}

\begin{table}[!htb]
	\renewcommand\arraystretch{1.5}
	\caption{$\mathcal{B}_2^{(2)}$, $n_{\rm occ}=\{2,1,1\}$}
	\label{table:S21}
	\scriptsize
	\begin{tabularx}{0.498\textwidth}{p{0.33\linewidth} p{0.33\linewidth} p{0.33\linewidth}}
		\hline
		\hline
		{\small ${\bf S}^z = (1/2, 0)$}  & {\small ${\bf S}^z = (-1/2, 1/2)$} & {\small ${\bf S}^z = (0, -1/2)$} \\
		\hline
		$
		\begin{array}{l}
		t_{0,3,2,6}=-\frac{1}{4} \\
		t_{0,4,1,6}=\frac{1}{4} \\
		t_{0,6,1,4}=-\frac{1}{4} \\
		t_{0,6,2,3}=\frac{1}{4} \\
		t_{1,4,0,6}=-\frac{1}{4} \\
		t_{1,6,0,4}=\frac{1}{4} \\
		t_{2,3,0,6}=\frac{1}{4} \\
		t_{2,6,0,3}=-\frac{1}{4} \\
		t_{3,0,6,2}=-\frac{1}{4} \\
		t_{3,2,6,0}=\frac{1}{4} \\
		t_{4,0,6,1}=\frac{1}{4} \\
		t_{4,1,6,0}=-\frac{1}{4} \\
		t_{6,0,3,2}=\frac{1}{4} \\
		t_{6,0,4,1}=-\frac{1}{4} \\
		t_{6,1,4,0}=\frac{1}{4} \\
		t_{6,2,3,0}=-\frac{1}{4} \\
		\end{array}
		$&
		$
		\begin{array}{l}
		t_{0,5,1,6}=\frac{1}{4} \\
		t_{0,6,1,5}=-\frac{1}{4} \\
		t_{1,3,2,6}=-\frac{1}{4} \\
		t_{1,5,0,6}=-\frac{1}{4} \\
		t_{1,6,0,5}=\frac{1}{4} \\
		t_{1,6,2,3}=\frac{1}{4} \\
		t_{2,3,1,6}=\frac{1}{4} \\
		t_{2,6,1,3}=-\frac{1}{4} \\
		t_{3,1,6,2}=-\frac{1}{4} \\
		t_{3,2,6,1}=\frac{1}{4} \\
		t_{5,0,6,1}=\frac{1}{4} \\
		t_{5,1,6,0}=-\frac{1}{4} \\
		t_{6,0,5,1}=-\frac{1}{4} \\
		t_{6,1,3,2}=\frac{1}{4} \\
		t_{6,1,5,0}=\frac{1}{4} \\
		t_{6,2,3,1}=-\frac{1}{4} \\
		\end{array}
		$&
		$
		\begin{array}{l}
		t_{0,5,2,6}=\frac{1}{4} \\
		t_{0,6,2,5}=-\frac{1}{4} \\
		t_{1,4,2,6}=-\frac{1}{4} \\
		t_{1,6,2,4}=\frac{1}{4} \\
		t_{2,4,1,6}=\frac{1}{4} \\
		t_{2,5,0,6}=-\frac{1}{4} \\
		t_{2,6,0,5}=\frac{1}{4} \\
		t_{2,6,1,4}=-\frac{1}{4} \\
		t_{4,1,6,2}=-\frac{1}{4} \\
		t_{4,2,6,1}=\frac{1}{4} \\
		t_{5,0,6,2}=\frac{1}{4} \\
		t_{5,2,6,0}=-\frac{1}{4} \\
		t_{6,0,5,2}=-\frac{1}{4} \\
		t_{6,1,4,2}=\frac{1}{4} \\
		t_{6,2,4,1}=-\frac{1}{4} \\
		t_{6,2,5,0}=\frac{1}{4} \\
		\end{array}
		$\vspace{1mm}\\
		\hline
		\hline
	\end{tabularx}
\end{table}

\begin{table}[!htb]
	\renewcommand\arraystretch{1.5}
	\caption{$\mathcal{B}_2^{(3)}$, $n_{\rm occ}=\{2,1,1\}$}
	\label{table:S22}
	\scriptsize
	\begin{tabularx}{0.498\textwidth}{p{0.33\linewidth} p{0.33\linewidth} p{0.33\linewidth}}
		\hline
		\hline
		{\small ${\bf S}^z = (1/2, 0)$}  & {\small ${\bf S}^z = (-1/2, 1/2)$} & {\small ${\bf S}^z = (0, -1/2)$} \\
\hline
		\hline
	\end{tabularx}
\end{table}
\clearpage

\bibliography{bibliography}